\pdfoutput=1
\documentclass{jfm}
\usepackage{graphicx}
\usepackage{grffile} 
\usepackage{epstopdf, epsfig}
\usepackage{caption, subcaption} 
\usepackage{amsmath}

\usepackage{xcolor}
\usepackage{latexsym}
\usepackage{hyperref}

\usepackage[demo,abs]{overpic}
\usepackage[export]{adjustbox}

\shorttitle{ From linear to nonlinear dynamics}
\shortauthor{W. Zhang, M. Wei}

\title{A theoretical framework for Koopman analyses of fluid flows, part 2: from linear to nonlinear dynamics}

\author{Wei Zhang\aff{1},
 Mingjun Wei\aff{1}
 \corresp{\email{mjwei@ksu.edu}}
}

\affiliation{\aff{1}Mechanical and Nuclear Engineering, Kansas State University,
1701B Platt St, Manhattan, KS 66506, USA}

\begin{document}

\maketitle

\begin{abstract}

A theoretic framework for dynamics is obtained by transferring dynamics from state space to its dual space. As a result, the linear structure where dynamics are analytically decomposed to subcomponents and invariant subspaces decomposition based on local Koopman spectral theory are revealed. However, nonlinear dynamics are distinguished from the linear by local exponential dynamics and infinite dimension, where the latter is due to nonlinear interaction and characterized by recursively proliferated Koopman eigenspaces. The new framework provides foundations for dynamic analysis techniques such as global stability analysis (GSA) and dynamic mode decomposition (DMD) technique. Additionally, linear structure via Mercer eigenfunction decomposition derives the well-known proper-orthogonal decomposition (POD). A Hopf bifurcation process of flow past fixed cylinder is decomposed numerically by the DMD technique. The equivalence of Koopman decomposition to the GSA is verified at the primary instability stage. The Fourier modes, the least stable Floquet modes, and their high-order derived modes around the limit cycle solution are found to be the superposition of countably infinite number of Koopman modes when the flows reach periodic. The nonlinear modulation effects on the mean flow is the saturation of the superimposed monotonic Koopman modes. The nonlinear resonance phenomenon is attributed to the alignment of infinite number of Koopman spectrums. The analysis of above nonlinear dynamic process relies on the properties of continuity of Koopman spectrums and state-invariance of Koopman modes discussed in part 1. The coherent structures are found related to the state-invariant modes.

\end{abstract}

\section{Introduction}

Dynamic systems are widely studied in fields such as mathematics, physics, engineering, chemistry, biology, economics, etc. For instance, the dynamics of particles subject to external forces following Newton's law are usually described by differential equations and can be reformulated into first-order form
\begin{equation} \label{eqn:dynamics}
\dot{x} = f(x).
\end{equation}
If $f(x)$ is a linear function of $x$ (where $f(\alpha x + \beta y)=\alpha f(x) + \beta f(y)$, $\alpha, \beta \in \mathbb{R}$, $x\in\mathbb{R}^{n}$), the dynamic system is linear, otherwise nonlinear.

The categorization of linear and nonlinear dynamics is of practical importance. Linear systems are theoretically simple and well-studied. However, most systems are inherently nonlinear. Though nonlinear systems are usually approximated by linear ones through linearization, such as the linear stability analysis. Unfortunately, this only works for short term prediction.~\citet{lorenz1963deterministic} published a work on solving a nonlinear modal system numerically and revealed the contrasting fact that nonlinear dynamics could have chaotic solution for specific parameters and initial conditions. Many more works showed that nonlinear dynamics could be chaotic, unpredictable, or counter-intuitive~\citep{may1987chaos,ruelle1971nature,winfree1967biological,feigenbaum1983universal,russell1844report,strogatz2018nonlinear}.

Despite the dramatic difference between linear and nonlinear dynamic systems, efforts exist to reduce the nonlinear dynamics to some form of linear ones to make use of linear dynamic theory. For instance, by defining Koopman operator~\citep{koopman1931hamiltonian}, or its adjoint Frobenius-Perron operator~\citep{frobenius1912matrizen,perron1907theorie} on nonlinear dynamic systems, researchers instead studied dynamics of functionals whose dynamics are evolved by a linear operators~\citep{reed1972methods,lasota2013chaos}. Recently, spectral Koopman decomposition was introduced to study dynamics of ergodic systems and fluid systems~\citep{mezic2005spectral,rowley2009spectral,bagheri2013koopman}. Our work continues these efforts but wants to clarify how this `linear treatment' enhances the understanding of nonlinear dynamics. Inspired by linear dynamics and the recent development of nonlinear systems, we target two questions.
\begin{itemize}
\item Why nonlinear dynamics can be analytically decomposed to sub-dynamics?
\item Which decomposition (if exist) can better describe nonlinear dynamics?
\end{itemize}
A direct motivation for the first question is to understand various decomposition techniques, such as Fourier decomposition, proper-orthogonal decomposition (POD)~\citep{holmes1996turbulence}, and dynamic mode decomposition (DMD)~\citep{schmid2010dynamic}, on a nonlinear dynamic system where linear superposition principle does not hold. It will be clear later that the completeness of dual space of state space introduces what we called the linear structure of dynamics, resulting in the various decomposition techniques mentioned above. For the second question, invariant bases of the dynamic mapping will be adopted, which are inspired by linear dynamic systems.

The content is organized as follows. \S 2 develops a theoretical framework for linear and nonlinear dynamics by finding the solution in the dual space of the state space. This framework introduces the linear structure and invariant subspaces as two universal properties of dynamics, while the locality and infinite-dimensionality as two unique properties of nonlinear dynamics. \S 3 first shows that Koopman decomposition of a linear system is compatible with the linear dynamic theory. For a nonlinear dynamic system the hierarchy structure of Koopman decomposition provides a structural view of the dynamics. The global stability analysis is used to derive Koopman decomposition for a fluids system. The DMD and POD are revisited under the new framework in \S 4. \S 5 studies a nonlinear transition process by DMD-LS algorithm~\citep{zhang2019solving}, where the formation of a periodic solution, Fourier decomposition, Floquet solutions, nonlinear resonance phenomena are explained by Koopman decomposition.

\section{Properties of dynamics under the new framework}

As the great difference between linear and nonlinear dynamics, the two types of systems seem to develop their own theory. For linear systems, their solutions are usually obtained analytically. However, nonlinear systems lack such uniformity. The approaches used to understand these nonlinear systems may depend on the specific systems. Another discrepancy might come from the dimensionality of a system. Take the Lorenz system~\citep{lorenz1963deterministic} for an example
\begin{equation}\left\{
\begin{aligned} 
\dot{x} &= \sigma (y-x), \\
\dot{y} &= x(\rho-z)-y, \\
\dot{z} &= xy-\beta z. \\
\end{aligned} \right.
\end{equation}
$(x, \, y, \, z)^T \in \mathbb{R}^3$ is the system state. $\sigma,\, \rho,\, \beta \in \mathbb{R}$ are the parameters for the system. 
It is a system with three degrees of freedom. For a linear system (by removing above nonlinear terms), the dimension of the system is obtained by counting the freedom of the state space, that is, 3. However, the Lorenz system is infinite-dimensional because of nonlinearity. Therefore, for the linear or nonlinear system, the dimension can not consistently determined by counting their degrees of freedom in the state space. 

However, the purpose of this work is to determine whether or not there exists an uniform framework for both linear and nonlinear dynamics. If such a framework exists, it might be compatible with linear systems but differ somewhere to take into consideration the nonlinear effects. Therefore, let us first review two of the important properties in linear dynamics.

\subsection{The properties of linear dynamic systems}

\subsubsection{The superposition principle}

It is well known a linear system has the superposition principle, that is, the net response to a linear system to multiple excitations is the linear sum of each caused by the excitation individually. That is, if $x_1$, $x_2$ are two solutions, $ax_1 + bx_2$ ($a, b\in \mathbb{R}$) is also a solution, since
\begin{equation}
\frac{d}{dt}(ax_1+bx_2)= a \dot{x}_1 + b\dot{x}_2 = a f(x_1) + b f(x_2) = f(ax_1+bx_2).
\end{equation}
The principle of superposition provides great advantage for solving and understanding linear dynamics. 

\subsubsection{The invariant subspace} \label{sec:LTI}

Linear dynamics can be decomposed into invariant subspaces, which can be illustrated by a linear time-invariant (LTI) system
\begin{equation}\label{eqn:linearsystems}
\dot{\boldsymbol{x}} = A \boldsymbol{x}, \quad \boldsymbol{x}(0) = \boldsymbol{x}_0.
\end{equation}
$\boldsymbol{x} \in \mathcal{R}^n$ and $A \in \mathcal{R}^{n\times n}$. If $A$ is diagonalizable, let $A = V\Lambda V^{-1}$. The solution to the LTI system~(\ref{eqn:linearsystems}) is given
\begin{equation}
\boldsymbol{x}(t) = Ve^{\Lambda t}V^{-1}\boldsymbol{x}_0 = c_1 e^{\lambda_1 t}\boldsymbol{v}_1 + \cdots + c_n e^{\lambda_n t}\boldsymbol{v}_n.
\end{equation}
$\lambda_i$ is the eigenvalue, and $\boldsymbol{v}_i$ is the eigenvector. $\boldsymbol{c} = V^{-1}\boldsymbol{x}_0$ is for the initial condition. Linear dynamic theory says the solution is decomposed to several invariant subspaces, and each has exponential dynamics. These invariant subspaces are spanned by the eigenvectors
\begin{equation}
\mathcal{S}_i = \text{Span}\{\boldsymbol{v}_i\},
\end{equation}
since $A\boldsymbol{v}_i = \lambda_i \boldsymbol{v}_i \in \mathcal{S}_i.$ The invariance under the dynamic mapping provide excellent base for tracing dynamics.

\subsection{From state space to dual space and the linear structures of dynamics}

Viewed from linear dynamics, a prominent feature is that dynamics can be summable, that is, it can be analytically decomposed into sub-components. The capability of decomposing dynamics into subcomponents is important for understanding the dynamics. However, without the superposition principle, nonlinear systems need to justify their rigorousness in decomposition by someway else. Otherwise, decomposition such as Fourier decomposition, POD, DMD will lose their power for a nonlinear system.

On the other hand, the reader might still remember the confusion on how to count the dimension of a linear or nonlinear system at the beginning of this chapter. It is a key question to answer since the solution space which provides the correct dimension should also provide the solution itself. Therefore, the state space cannot provide the expected theoretical framework. Therefore, we should find the solution space for both linear and nonlinear systems.

The summation operation and the solution space together require a linear space of some kind, in which the linearity is defined with a summation operation. Further, considering the effort of transferring nonlinear dynamics to functional space by Koopman operator in part 1, a proper option for the linear space for the desired theoretical framework is the dual space of the system, which contains the continuous linear functionals~\citep{reed1972methods}. To give more detail, several definitions from operator theory are introduced. 

\begin{itemize}
\item A complete normed linear space is called a \emph{Banach space}. 

\item A \emph{bounded linear transformation} (or \emph{bounded operator}) from a normed linear space $<V_1, ||\,||_1>$ to a normed linear space $<V_2, ||\,||_2>$ is a function, $T$, from $V_1$ to $V_2$ which satisfies the following two conditions \\
(1) $T(\alpha v + \beta w) = \alpha T(v) + \beta T(w) $, \\
(2) For some $c \ge 0$, $||Tv||_2 \le c ||v||_1$. \\
The smallest such $c$ is called the norm of $T$
\begin{equation}
||T|| = \sup_{||v||_1 = 1} ||Tv||_2
\end{equation}

\item The set of bounded linear transformations from one Banach space $\mathbb{X}$ to another $\mathbb{Y}$ is itself a Banach space, noted by $\mathcal{L}(\mathbb{X}, \mathbb{Y})$. In the case where $\mathbb{Y}$ is the complex numbers, this space $\mathcal{L}(\mathbb{X}, \mathbb{C})$ is denoted by $\mathbb{X}^*$ and called the \emph{dual space} of $\mathbb{X}$. 
\end{itemize}
Since the elements are bounded linear transformations, the dual space $\mathbb{X}^*$ contains all the \emph{Lipschitz continuous functionals} of $\mathbb{X}$. 

It is known a metric space in which all Cauchy sequence converge is called \emph{complete}. From above definition, since the completeness of $\mathbb{X}^*$, an element $g(x)\in \mathbb{X}^*$ can find a Cauchy sequence $f_i(x)\in \mathbb{X}^*$, such that
\begin{equation} \label{eqn:cauchysequence}
g(x) = \text{lim}\sum_{i=0}^{\infty} f_i(x).
\end{equation}
Decomposition~(\ref{eqn:cauchysequence}) has an important role in dynamics analysis. The dynamics of $g(x)$, under a given dynamic system $x(t)$, is written
\begin{equation}\label{eqn:linearstructure}
g(x)\big|_{x(t)} = \left( \lim \sum_{i=0}^{\infty} f_i(x) \right) \Bigg|_{x(t)} = \lim \sum_{i=0}^{\infty} f_i(x(t)).
\end{equation}
Therefore, dynamics of $g(x)$ under a dynamic system $x(t)$, noted $g(x(t))$, are analytically decomposed into subcomponents $f_i(x(t))$, no matter the dynamic system $x(t)$ or observable $g(x)$ is linear or nonlinear. We call dynamics analytically decomposed into subcomponents via relation~(\ref{eqn:linearstructure}) to be \emph{linear structure of dynamics}.

In the case a dynamic system is uniquely determined by the state, such as the autonomous system~\ref{eqn:dynamics}, the sub-dynamics is uniquely determined by the base function and system state, that is, $f_i(x(t)) = f_i(x)$. This may be important for some numerical applications like DMD or POD, since under this situation they obtain a decomposition which is independent on the specific trajectory $x(t)$.

In part 1, a linear time variant system is considered
\begin{equation}\label{eqn:LTV}
\dot{x} = A(t)x.
\end{equation}
In the case when it is the linearization of system~\ref{eqn:dynamics} along a trajectory $x(t)$ ($x(t)$ is a solution of system~\ref{eqn:dynamics}). For such a autonomous differential equations, there is an one-to-one correspondence between state $x$ and $t$. Then $A(t) = \frac{\partial f(x)}{\partial x}\big|_{x(t)}$ is also an state dependent relation, noted $\hat{A}(x)$. Therefore, the linear system~(\ref{eqn:LTV}) is uniquely determined by the system state as well, POD or DMD can uniquely determine the corresponding linear structure. For general LTVs, the decomposition depends on time $t$ and will be more complicated.

The linear structure can be used to study the dynamics of the system $x(t)$ as well. At this time, the full-state variable $x$ is the observable, and decomposed by
\begin{equation}
x = \lim \sum_{i=0}^{\infty} c_i f_i(x).
\end{equation}
Here $c_i$ are constants, they are called modes when $x$ are vectors or field variables. $f_i(x)$ are normalized and usually called base functions. These base functions will carry unique dynamic information under a dynamic system. Two sets of dynamics-induced base functions are most familiar, they are the Koopman eigenfunctions and Mercer eigenfunctions.

\subsection{The invariant subspaces of nonlinear dynamics}

Various choices of Cauchy sequence may be available and result in different decomposition techniques. However, the choice may significantly affect dynamics analysis. Motivated by linear dynamic theory for LTI system, see \S~\ref{sec:LTI}, and linear time-variant (LTV) systems~\citep{zhou2016asymptotic,wu1974note}, it is desired to use similar linearly independent and invariant bases (if available) for nonlinear dynamics.

Unfortunately, extending the invariant subspace to the nonlinear dynamic system is not straightforward. This defines a spectral problem of some kind. No spectral theory can be defined in the state space since the system is nonlinear. An alternative is to define the spectral problem in its dual space where a linear yet infinite-dimensional map, known as Koopman operator, is defined to evolute the dynamics. Recently, the spectral decomposition of Koopman operator is introduced to dynamics analysis by~\citet{mezic2005spectral,rowley2009spectral}, as linear operator defined on linear space naturally raises spectral problem. 

However, the spectral theory of linear operator is a non-trivial extension to the finite-dimensional matrix spectral theory due to the infinite-dimensionality of the operator. It is known only bounded operator is guaranteed with bounded spectrums. Other than that, unbounded operators may or may not have spectrums~\citep{reed1972methods}. As a result, previous researchers focused on some particular dynamic system. For example, the ergodic system or periodic system define unitary operator, one type of well-propertied bounded operator~\citep{mezic2005spectral}. 

Part 1 proposed to relax the definition to the local spectrums as there were signs they depended on the state of the system for nonlinear dynamics. To be more specific, consider the dynamics of an observable $g(x)$ under the dynamics $T^{\tau}$, where $T^{\tau}$ represents the dynamic mapping $T^{\tau} : \mathbb{X}\rightarrow\mathbb{X}$ for a given time interval $\tau$. And the Koopman operator $U$ evolutes the dynamics of $g(x)$ by
\begin{equation}
U g(x) = g(T^{\tau} x)
\end{equation}
The Koopman operator is linear~\citep{reed1972methods,rowley2009spectral}. Therefore, local Koopman spectral problem is defined by
\begin{equation} \label{eqn:koopmanevolute}
\phi(T^{\tau}x) = U\phi(x) = \lambda \phi(x), \quad x\in D_{x_0}.
\end{equation}
The local eigenvalue problem is only required to hold in a open neighborhood $D_{x_0}$ of state $x_0$. 

Further, utilizing the perturbation theory of operator~\citep{reed1978methods,kato2013perturbation}, the local spectral $\lambda(x)$ and eigenfunction $\phi(x)$ can be analytically extended to the whole space under the condition that the perturbation operator is bounded. The globally extended spectral problem results in
\begin{equation}
U\phi(x) = \lambda(x)\phi(x).
\end{equation}
The eigenfunction $\phi(x)$ forms an invariant base under the dynamic mapping. For more detail, readers are referred to part 1 and references therein. The continuity requirements can be satisfied by many practical dynamic systems. With this extension, both the eigenvalue $\lambda(x)$ and the eigenfunction $\phi(x)$ will be continuous and analytical in $\mathbb{X}^*$. The resulting eigenfunctions $\phi(x)$s constitute bases for the linear structure.

\subsection{Other properties of nonlinear dynamics}

\subsubsection{The locality of Koopman spectrums}

For an LTI system, the Koopman spectral problem is globally defined, and the Koopman spectrums are constants. They are independent on the state. Therefore, they represent global exponential dynamics, for instance, the LTI systems. However, for nonlinear systems, the Koopman spectrums are no longer constants and depend on the state, therefore, locally exponential dynamics. As mentioned earlier, for those LTV systems which are linearized from nonlinear systems, the time-variant spectrums are consistent with the state-dependent Koopman spectrums.

\subsubsection{Nonlinear interaction and infinite-dimensionality of nonlinear dynamics}

\begin{figure}
\centering
\begin{subfigure}[b]{0.325\linewidth}
\includegraphics[width=1.0\textwidth]{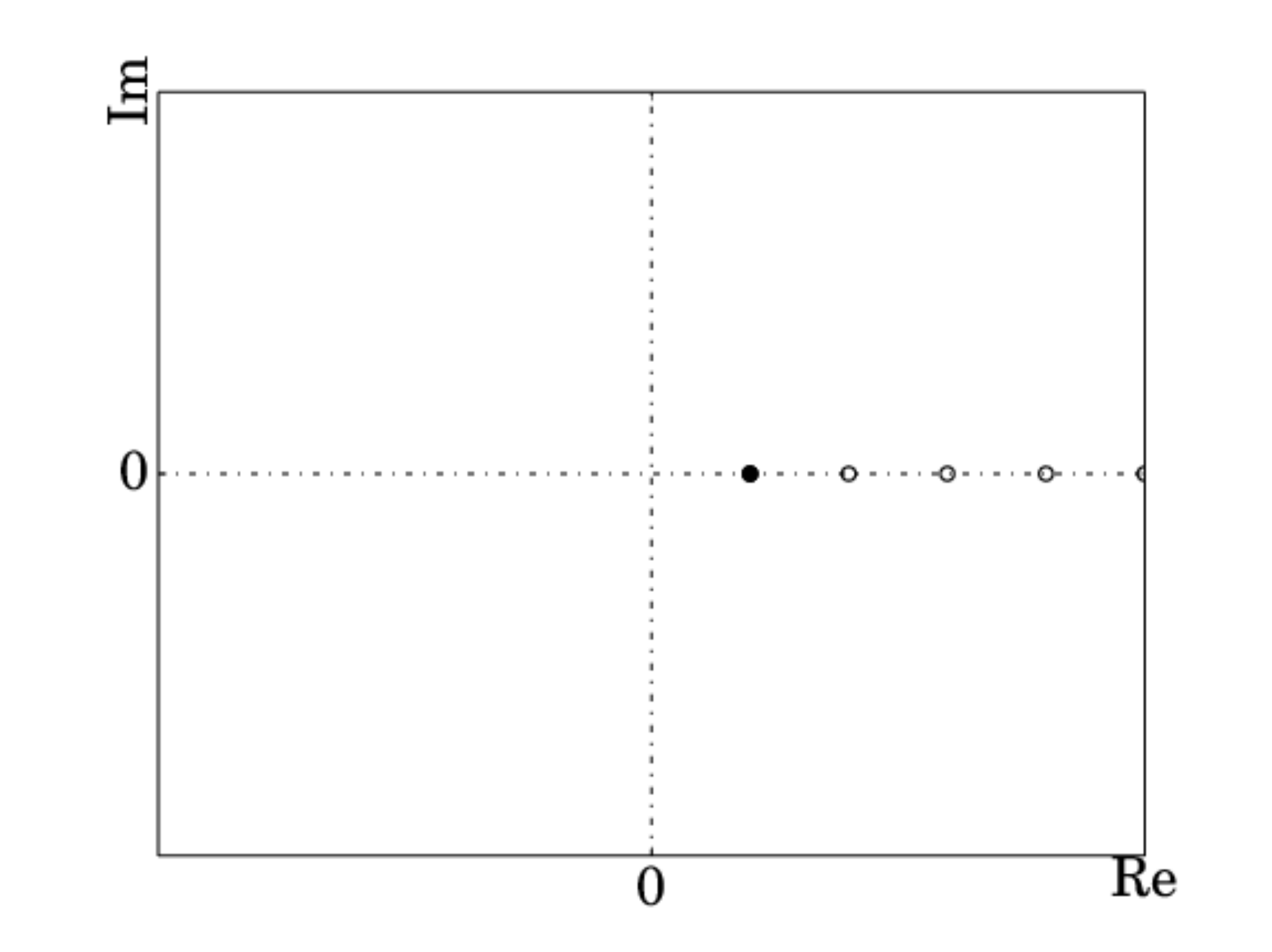}
\caption{$\lambda>0$ \label{fig:selfreal} }
\end{subfigure}
\begin{subfigure}[b]{0.325\linewidth}
\includegraphics[width=1.0\textwidth]{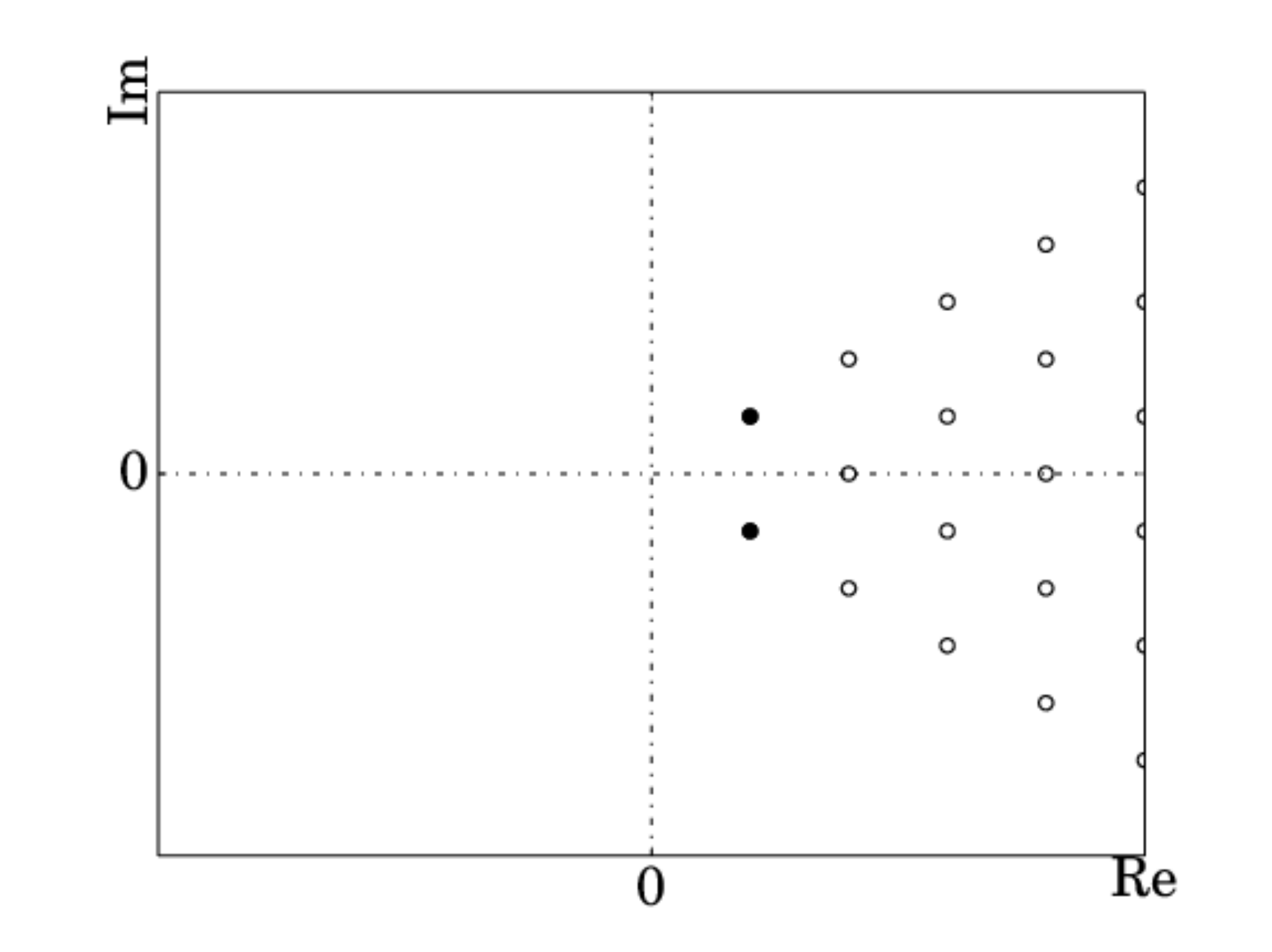}
\caption{$\text{Im}(\lambda)\neq 0$ \label{fig:selfcomplex} }
\end{subfigure}
\begin{subfigure}[b]{0.325\linewidth}
\includegraphics[width=1.0\textwidth]{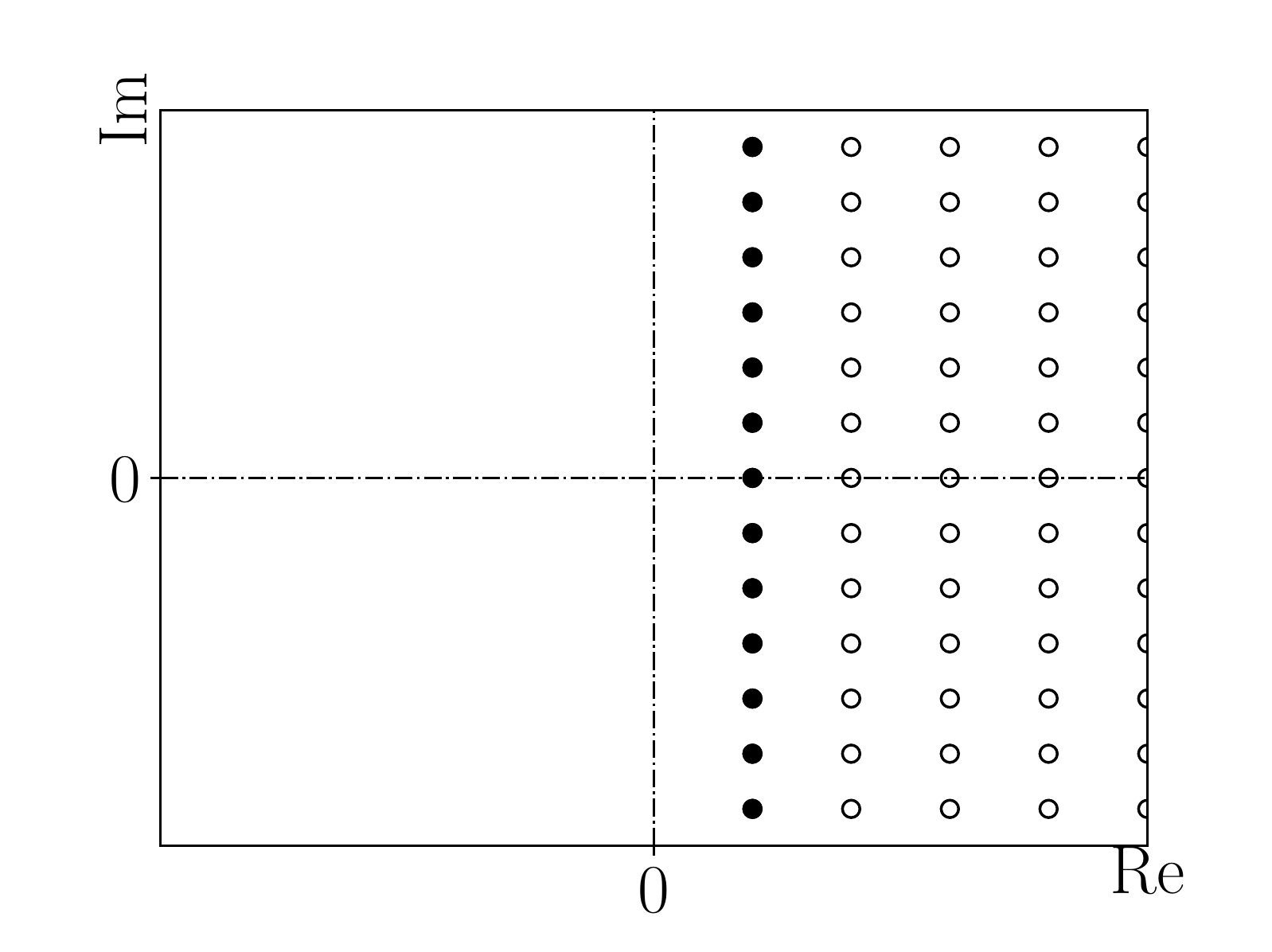}
\caption{limit cycle system \label{fig:selffloquet} }
\end{subfigure}
\caption{Self interaction of Koopman spectrums. The filled circle shows the original linear spectrums, high order Koopman spectrums are shown by hollow circle.} \label{fig:selfinteract}
\end{figure}

Another fundamental difference between linear and nonlinear dynamics is the dimensionality. It is now obvious the dimension of the system can be counted by the number of bases in the dual space to describe the dynamics, for instance, the Koopman eigenfunctions considered in this paper. For linear systems, dynamics are decoupled into invariant subspaces and no interaction between them. However, for nonlinear systems, there is nonlinear interaction between the sub-dynamics. The nonlinear interaction promotes the system to infinite-dimensional. Under Koopman decomposition, nonlinear interaction can be conveniently described by the recursive proliferation rule. For instance, some Koopman spectrum patterns generated by self-interaction are shown in figure~\ref{fig:selfinteract}. This infinite-dimensionality results in many phenomena in nonlinear dynamics and will be illustrated later.

\section{Applications of the theoretical framework to dynamic systems}

Koopman decomposition provides the dynamics of various systems. For a linear system, Koopman decomposition requires solving an eigenvalue problem. It is compatible with the linear dynamic theory. Additionally, Koopman decomposition extends flexibly to other linear or nonlinear observables. For a nonlinear system, a general solution may not be available. However, in the special case where the base dynamics is predetermined and the small perturbation on top of it exhibits disparate time scale motion, an asymptotic expansion can be used to compute the decomposition.

\subsection{Koopman decomposition for a linear dynamic system}

Let's take the LTI systems~(\ref{eqn:linearsystems}) for an example. Assuming $A$ is diagonalizable
\begin{equation}
A = V\Lambda W
\end{equation}
where $V$ and $W$ contain the right and left eigenvectors of $A$, the diagonal elements of $\Lambda$ are the spectrums, so $A V = V\Lambda$ and $WA = \Lambda W$. For a LTI system, eigenfunctions are given by the inner-product of the left eigenvectors and state variable~\citep{rowley2009spectral}.
\begin{equation}
x = V W x = V \left[\begin{array}{c} f_1(x) \\ f_2(x) \\ \vdots \\ f_n(x) \end{array} \right] = \left[\begin{array}{c} v_{11} f_1 + \cdots v_{1n} f_n \\ v_{21}f_1 + \cdots v_{2n}f_n \\ \vdots \\ v_{n1}f_1 + \cdots v_{nn}f_n \end{array} \right]
\end{equation}
where $W = \left[ w_1 \, \cdots \, w_n \right]^T$, and $f_i(x) = w_i^Tx$. Koopman decomposition of the full-state variable is obtained. The dynamics of $x(t)$ is transferred to the components $f_i(x(t))$. Though literally complex, $f_i(x(t))$ is of the simple exponential form $c_i e^{\lambda_i t}$ for LTI systems, since each eigenfunction has a global spectrum $\lambda_i$. It is then easier to trace the dynamics through these eigenfunctions $f_i(x)$. For instance, the dynamics of the system reads
\begin{equation}
x(t) = v_1 f_1(x(t)) + \cdots + v_n f_n(x(t)) = v_1 e^{\lambda_1 t} + \cdots + v_n e^{\lambda_n t}
\end{equation}
which is compatible with the linear dynamic theory. 

For some linear observables, for example, $x_1$, the first component of vector $x$, Koopman decomposition is given by 
\begin{equation}\label{eqn:x1dyn}
x_1 = v_{11} f_1 + \cdots v_{1n} f_n = v_{11} e^{\lambda_1 t} + \cdots v_{1n} e^{\lambda_n t}.
\end{equation}
Therefore, the dynamics of linear observables can be easily obtained from Koopman decomposition.

Sometimes nonlinear observables may be interested, for instance, the total kinetic energy is given by
\begin{equation}
e = \frac{1}{2}\sum_i \dot{x}_i ^2 = \sum_k \sum_l b_{lk} f_l f_k
\end{equation}
where $b_{lk} $ are the elements of $B =\frac{1}{2} \Lambda^T V^T V \Lambda$ and are constants. $f_l f_k$ is the Koopman eigenfunction proliferated by multiplying $f_l(x)$ and $f_k(x)$, see part 1. Replacing $f_l f_k$ by $f_l(x(t))f_k(x(t))$, here $ce^{(\lambda_l + \lambda_k)t}$, the dynamics of the total kinetic energy is obtained.

The above two examples show that dynamics of any linear or nonlinear observable of a LTI system can be obtained from their Koopman decomposition. It is compatible with the linear dynamic theory for the full-state observable $x$ but more flexible for other linear or nonlinear observables.

\subsection{Koopman decomposition for Navier-Stokes equations} \label{sec:GSA}

\begin{figure}
\centering
\includegraphics[width=1.0\textwidth]{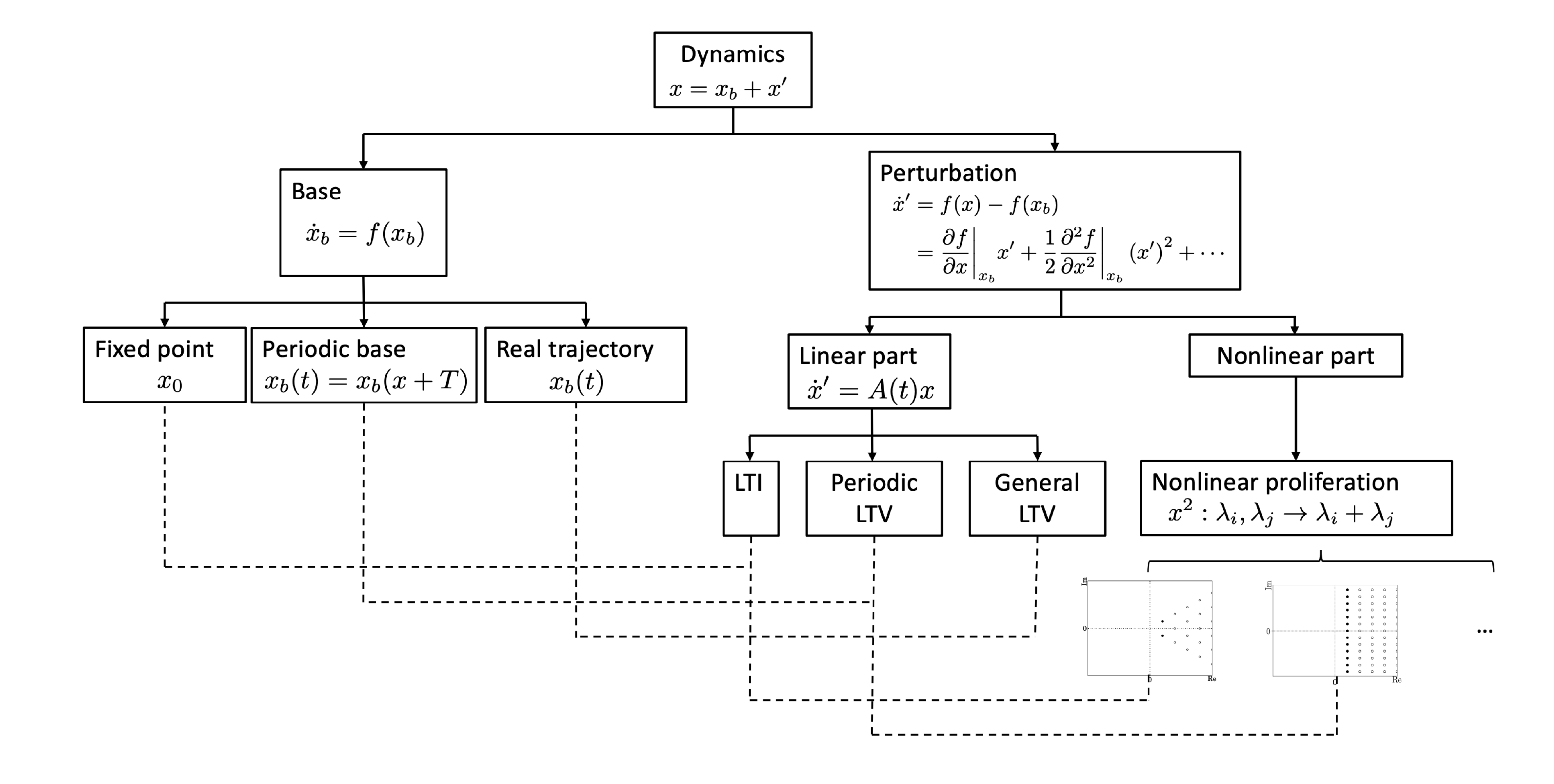}
\caption{The hierarchy of Koopman decomposition.} \label{fig:hierarchy}
\end{figure}

A general formulation of Koopman decomposition is not available for a nonlinear dynamic system. However, if augmented with other information, such as base dynamics or the information of a real trajectory of the system, the hierarchy structure of Koopman eigenspaces discussed in part 1 makes it easier to compute the Koopman decomposition. 
The hierarchy structure decomposes dynamics into base dynamics and perturbation. The base dynamics are required to be a real trajectory of the system such that it will not bring fake spectrums to the system. Typically, they are simple dynamics such as an equilibrium state, or a periodic solution whose spectrums are Fourier frequencies, or a real trajectory whose spectrums can be numerically studied by the DMD algorithm. The perturbation is described by the nonlinear perturbation equation, which is further divided into linear and nonlinear part. Koopman spectrums for the linear part are studied by the various linear systems. Further, spectrums for the nonlinear terms are obtained by recursively applying proliferation rule. Then the spectrums and the decomposition are obtained. The hierarchy is illustrated in figure~\ref{fig:hierarchy}.

In the following sections, an asymptotic expansion technique is adopted to derive the Koopman decomposition for a fluid system following the hierarchy understanding.

\subsubsection{Multi-scale assumption and asymptotic expansion}

A perturbation technique, known as multiple-scale analysis, will be used. This technique is based on the observation that the dynamics of such small perturbation usually characterized by disparate time scale motion. One is a fast scale oscillating motion. One is a much slower time scale motion describing the slow changing of magnitude. The multiple-scale analysis is accomplished by introducing a fast-scale $t$ and a slow-scale $\tau$ ($\tau=\epsilon t$) as two independent variables. In the solution process of the perturbation problem, the additional freedom, introduced by a new independent variable $\tau$, is used to remove the unwanted secular terms. The latter put constraints on the approximate solution, which are called solvability conditions~\citep{kevorkian2012multiple}.

After introducing the multiple-scale, the dynamics is expanded by the asymptotic expansion
\begin{equation}\label{eqn:asymexpansion}
\boldsymbol{q}(t) = \boldsymbol{q}_0 + \epsilon \boldsymbol{q}_1(t, \tau) + \epsilon^2 \boldsymbol{q}_2(t, \tau) + \epsilon^3 \boldsymbol{q}_3(t, \tau) + \cdots.
\end{equation}
$\boldsymbol{q}_0$ is the base flow, an assumed fixed equilibrium point in the system. $\boldsymbol{q}_i(t, \tau)$ is small perturbation with intensity of $\epsilon^i$. Substituting above asymptotic expansion into original dynamic equation, and taking into account that the slow time scale $\tau$, the following differential relation is obtained
\begin{equation}
\frac{\partial }{\partial t} \equiv \frac{\partial }{\partial t} + \epsilon \frac{\partial }{\partial \tau} .
\end{equation}
Expanding the dynamic equation and collecting terms according the different order of $\epsilon$, a set of equations are obtained, which then asymptotically solve the dynamics system.

The successful application of multiple-scale analysis relies on the implicit assumption that the time scale is disparate. Thus the application of multiple-scale analysis may fail for general cases. However, the conclusions drawn below will still be valid in the view of Koopman decomposition for complex dynamics.

\subsubsection{Fluid dynamic system}

To give a concrete example, the above asymptotic expansion is carried out on the incompressible viscous fluid system
\begin{equation}
\begin{aligned}
\frac{\partial \boldsymbol{u}}{\partial t} + \nabla \boldsymbol{u}\cdot \boldsymbol{u} & = - \nabla p + \frac{1}{Re} \nabla^2 \boldsymbol{u}, \\
\nabla \cdot \boldsymbol{u} & = 0.
\end{aligned}
\end{equation}
$\boldsymbol{u}$ is the velocity of fluids and $p$ is the pressure. $Re$ is the dimensionless Reynolds number. Let $\boldsymbol{q} = \left[ \begin{array}{c} \boldsymbol{u} \\ p \end{array} \right].$

\subsubsection{The global stability analysis}

A proper example for multiple-scale analysis in fluids is the Hopf bifurcation occurred in many flows, where the original steady flow develops periodic oscillation after an increasing or decreasing of some critical parameter. For instance,~\citet{stuart1960non} found Reynolds number controls the bifurcation of flow when he studied two-dimensional Poiseuille flow. He found the linear growth rate at $Re>Re_c$ ($Re-Re_c \ll 1$, $Re_c$ is the critical Reynolds number) is of order $\epsilon^2$ ($\epsilon^2$ is a small parameter characterizing weak nonlinear effects)
\begin{equation} \label{eqn:stuartorder}
\epsilon^2 = \frac{1}{Re_c} - \frac{1}{Re},
\end{equation}
whereas the time scale on which nonlinear interactions affect the evolution of the fundamental mode is of order (linear growth rate)$^{-1}$. The second time scale is introduced
\begin{equation}
\tau \equiv \epsilon^2 t.
\end{equation}
$\tau$ is the slow time scale, and affects the time derivative by
\begin{equation}
\frac{\partial }{\partial t} \equiv \frac{\partial }{\partial t} + \epsilon^2 \frac{\partial }{\partial \tau} 
\end{equation}

Substituting the asymptotic expansion into the incompressible Navier-Stokes equation, a series of equations at various orders of $\epsilon$ are obtained.

(i). At order $\epsilon^0$, a steady Navier-Stokes equations at $Re$ are obtained
\begin{equation}
\begin{aligned}
\nabla \boldsymbol{u}_0\cdot \boldsymbol{u}_0 &= -\nabla p_0 - \frac{1}{Re} \nabla^2 \boldsymbol{u}_0, \\
\nabla \cdot \boldsymbol{u}_0 &= 0.
\end{aligned}
\end{equation}
These equations take the original boundary condition, and have a fixed equilibrium solution, which is then chosen as the base flow for asymptotic expansion $\boldsymbol{q}_0 = [\boldsymbol{u}_0^T, p_0]^T$. Here the Koopman spectrum is $\lambda=0$, and $\boldsymbol{q}_0$ is the corresponding base Koopman mode.

(ii). At order $\epsilon^1$, an homogeneous linear systems are obtained
\begin{equation} \label{eqn:globalexp1}
\left[\begin{array}{cc} \frac{\partial }{\partial t }+ \nabla ()\cdot \boldsymbol{u}_0 + \nabla \boldsymbol{u}_0 \cdot ()-\frac{1}{Re}\nabla^2 & \nabla \\
\nabla^T & 0 \end{array} \right] \boldsymbol{q}_1 = \boldsymbol{0}
\end{equation}
with a homogeneous boundary condition. This is an eigenvalue problem. Its solution, is dominated by the most unstable modes $\boldsymbol{q}_1$
\begin{equation}
\boldsymbol{q}_1 = \left[ \begin{array}{c} \boldsymbol{u}_1 \\ p_1 \end{array} \right] = A(\tau) e^{iwt} \boldsymbol{v}_1 + c.c.
\end{equation}
$\boldsymbol{v}_1$ is the most unstable mode, and $A(\tau)$ is its magnitude varying on the slow time scale $\tau$, at initial stage $A(\tau)$ grows at $e^{\epsilon^2 t}$ or $e^{\tau}$. Therefore, $A(\tau)e^{i w t}$ provides the Koopman eigenfunction and $\boldsymbol{v}_1$ is the Koopman mode.

(iii). At order $\epsilon^2$, inhomogeneous linear equations will be obtained with homogeneous boundary condition.
\begin{equation} \label{eqn:lineardiffforced}
\left[\begin{array}{cc} \frac{\partial }{\partial t }+ \nabla ()\cdot \boldsymbol{u}_0 + \nabla \boldsymbol{u}_0 \cdot ()-\frac{1}{Re}\nabla^2 & \nabla \\
\nabla^T & 0 \end{array} \right] \boldsymbol{q}_2 = \left[ \begin{array}{l} |A|^2 \boldsymbol{F}_2^{|A|^2} + A^2 e^{2iwt} \boldsymbol{F}_2^{A^2} + \bar{A}^2 e^{-2iwt} \boldsymbol{F}_2^{\bar{A}^2} \\ 0 \end{array} \right]
\end{equation}
Left hand side contains the same linear differential operator similar to that at order $\epsilon^1$, and the right side contains the forcing terms excited by nonlinear interaction of $\boldsymbol{q}_1$ and its complex conjugate. $|A|^2$, $A^2$, $\bar{A}^2$ representing the increasing forcing magnitude. The $\boldsymbol{F}$ terms are
\begin{equation} \label{eqn:nonlinearexcitation}
\begin{aligned}
 \boldsymbol{F}_2^{|A|^2} &= -\nabla \boldsymbol{u}_1^A \cdot \boldsymbol{u}_1^{\bar{A}} - \nabla \boldsymbol{u}_1^{\bar{A}} \cdot \boldsymbol{u}_1^A \\
 \boldsymbol{F}_2^{A^2} &= -\nabla \boldsymbol{u}_1^A \cdot \nabla \boldsymbol{u}_1^A \\
 \boldsymbol{F}_2^{\bar{A}^2} &= - \nabla \boldsymbol{u}_1^{\bar{A}} \cdot \boldsymbol{u}_1^{\bar{A}} 
\end{aligned}
\end{equation}
where $\boldsymbol{u}^{A}_1$ and $\boldsymbol{u}^{\bar{A}}_1$ are the solution of equation~(\ref{eqn:globalexp1}) on $\epsilon^1$ expansion.
It is expected for the forced LTI system, the solution would have the following component 
\begin{equation*}
\boldsymbol{q}_2 = A(\tau)^{|A|^2} \boldsymbol{v}^{|A|^2} + A(\tau)^{A^2} e^{i2wt} \boldsymbol{v}^{A^2} + A(\tau)^{\bar{A}^2} e^{-i2wt} \boldsymbol{v}^{\bar{A}^2}.
\end{equation*}
$\boldsymbol{v}^{(\cdot)}$ is the symbolic spatial mode. $A(\tau)^{(\cdot)}$ is the symbolic representation for slow varying magnitude. $e^{inwt}$ is its pulsation. The modes $\boldsymbol{v}^{(\cdot)}$ are Koopman modes, the slow varying magnitude $A(\tau)$ and pulsation $e^{inwt}$ together provide the Koopman eigenfunction.

(iv). Higher-order $\epsilon^i$ expansions can be further carried out to derive the linear equations, which generate the high order Koopman modes. In those equations, left-hand side is the same linear differential operator and right-hand side contains the almost periodic but increasing forcing terms. The particular solution of these forced linear dynamic system then provides higher-order Koopman modes and Koopman eigenfunctions.

Some comments are followed:

1) Degenerated cases occur on the $\epsilon^3$ or other $\epsilon^{2i+1}$ order expansions, where secular terms, which contains $e^{iwt}$ or $e^{-iwt}$ pulsation occurs in the right-hand side as the forcing term. A compatibility condition is required to remove them, which derives the well-known Stuart-Landau equation~\citep{stuart1960non}. On $\epsilon^3$ expansion, it is
\begin{equation} \label{eqn:stuartlandau}
\frac{\partial A}{\partial \tau} = \sigma A - l|A|^2 A.
\end{equation}
It provides a good approximation of the magnitude of the critical normal mode $\boldsymbol{v}_1$.

2) The above asymptotic expansion is similar to the one used by various authors~\citep{sipp2007global,meliga2011asymptotic} to study the Hopf bifurcation of wake after blunt bodies at the critical Reynolds number but with some difference. This work expands the Navier-Stokes equation at $Re$, while the above authors expanded at $Re_c$ ($\frac{1}{Re_c} - \frac{1}{Re}\ll 1$) to study the bifurcation. The difference was characterized by a base flow modification $u^1_2$ (in Sipp's notation). As a slight difference in base flow, the eigenvalue problem is slightly different. However, as the $Re$ is so close to $Re_c$, it will be shown later, the Koopman modes numerically computed by the current authors are very similar to the modes computed by~\citet{sipp2007global} and~\citet{meliga2011asymptotic}.

3) It is found information transfers \emph{directionally}, more specifically, information transfers from low-order Koopman modes to higher-order Koopman modes. Take the expansion at $\epsilon^2$ order for an example. Equation~(\ref{eqn:lineardiffforced}) shows low order Koopman modes interacting at the right-hand side, and the higher-order modes are generated by solving the forced linear systems. This direction is also reflected by the proliferation rule, that is, $\lambda_i, \lambda_j \rightarrow \lambda_i+\lambda_j$ and only summation operation allowed in the proliferation. 

The directional transformation of information may provide a framework to describe the energy cascading. On the one hand, information transfers from low-frequency modes to high-frequency ones such as $\lambda_i \rightarrow n\lambda_i$, a typical energy cascading route. On the other hand, the information can transfer back from high frequencies to low frequencies such as $n\lambda_i, n\bar{\lambda}_i \rightarrow 2n\text{Real}(\lambda)$, resulting the so-called the backscattering phenomena~\citep{pope2001turbulent}.

4) It is known the linear differential operator has its natural frequency, under which the magnitude would be amplified. And the Koopman modes can be generated from the forced systems. Therefore, some excitations are much easier amplified than others. This resonance phenomenon will be illustrated in more detail later.

5) If the underlined base flow is time-variant, let $\boldsymbol{q}_0(t)$ the time-variant base flow. $\boldsymbol{q}_0$ is substituted by $\boldsymbol{q}_0(t)$ in the asymptotic expansion~(\ref{eqn:asymexpansion}). Then at all orders of $\epsilon^i$, a linear time-variant coefficient system is obtained, and the analysis can proceed accordingly. For example, if the asymptotic expansion is expanded around a limit cycle solution, $\boldsymbol{q}_0(t)$ is periodic, at order $\epsilon^0$, an unsteady Navier-Stokes equation is obtained and its solution will be $\boldsymbol{q}_0(t)$. At $\epsilon^1$ order, homogeneous periodic coefficient linear equations with homogeneous boundary conditions are obtained. Therefore a Floquet system is to be solved. On higher-order ($\epsilon^2$ or higher), periodic linear differential systems with a varying magnitude but almost periodic forcing terms need to be solved. Bagheri considered such a situation and reported the result in his work~\citep{bagheri2013koopman}. For solution to general LTV systems, the reader is referred to monographs such as~\citet{bittanti2009periodic,kuchment2012floquet,shmaliy2007continuous} for detail.

\section{New framework and other analysis techniques}

Linear structure is an important property of dynamics. It helps to understand the dynamics by decomposing to their sub-dynamics. Among various decomposition the framework via Koopman decomposition provides foundations to various analysis techniques for nonlinear systems. For instance, the Koopman spectral spaces are ideal for stability analysis. Therefore, the linear stability analysis or the global stability analysis can be safely performed under this framework. Section~\ref{sec:GSA} demonstrates under a particular case where dynamics of small perturbation characterized by disparate time scale motion, a global stability analysis can be performed to obtain Koopman decomposition. Other than Koopman decomposition, the following section shows that linear structure via a different set of base functions can provide other decomposition as well, for instance, the proper orthogonal decomposition.

\subsection{Understanding the DMD and POD techniques}

DMD and POD are both data-driven techniques for dynamics analysis, and often compared to each other. DMD is the data-driven tool for Koopman decomposition, or more accurately, a time-averaged approximation of the local Koopman decomposition since the sampled data last a period of time. For instance, part 1 showed DMD techniques on data collected at the asymptotic stage provides an excellent approximation of Koopman decomposition of the primary and secondary instability of flow past cylinder.

Linear structure provide the foundation for POD as well but relies on a different set of bases, the eigenfunctions of the second-order correlation function of the dynamics
\begin{equation}
R_x(t, t') = \int_{\Xi} x(\xi, t) x(\xi, t') d\xi
\end{equation}
where $t$, $t'$ is the time lapse of observation, $\xi$ is the spatial coordinate for the dynamic system. The correlation has the following properties: \emph{symmetry} as $R_x(t,t') = R_x(t',t)$, \emph{continuous} as long as $x(\xi, t)$ is continuous, and \emph{positive semi-definite}, since for any nonzero function $\psi(t)$
\begin{equation}
\int_a^b \psi^2(t) dt > 0,
\end{equation}
satisfies the following relation
\begin{equation}
\begin{aligned}
\int_a^b \int_a^b R_x(t, t') \psi(t) \psi(t') dt dt' &= \int_a^b \int_a^b \int_{\Xi} x(\xi, t) x(\xi, t') \psi(t) \psi(t') d\xi dt dt' \\
	&= \int_{\Xi} \left( \int_a^b u(\xi,t) \psi(t) dt \right)^2 d\xi \ge 0.
\end{aligned}
\end{equation}

The continuous, symmetry, positive semi-definite kernel $R_x(t,t')$ satisfies the Mercer's theorem~\citep{mercer1909xvi}, also known as Fredholm eigenvalue problem~\citep{edmunds2018spectral}. Therefore, the kernel can be expanded by series
\begin{equation}
R_x(t,t') = \sum^{\infty}_{i=1} \lambda_i e_i(t) e_i(t')
\end{equation}
where the functions $e_i(t)$ are the eigenfunctions of the following eigenvalue problem.
\begin{equation}
\int_a^b R_x(t,t') e_{xi}(t') dt' = \lambda_{xi} e_{xi}(t).
\end{equation}
The eigenfunctions $e_{xi}(t)$ then act as the bases for duration $a$ to $b$. The functional at $x$ is obtained by taking the limit $b \rightarrow a$
\begin{equation}
\phi_i(x) = \lim_{b \rightarrow a} e_{ix}(t)
\end{equation}
and use the functionals $\phi_i(x)$ for linear structure, which results in the POD decomposition. It is noticed by taking limit of above Fredholm problem, a local spectrum is defined.

It is well known that the POD gives the most efficient decomposition in the sense it captures the most energy with the same amount of modes. As $R_x(t,t')$ is symmetrical, the eigenvalues are all real numbers and therefore naturally ordered by corresponding eigenvalues (or energy). However, POD decomposition has disadvantages. Firstly, these eigenfunctions are not the eigenfunctions of the dynamics system, therefore not invariant under the dynamics mapping, therefore the sub-dynamics are coupled. Secondly, POD eigenfunctions do not support the proliferation rule. Further, Koopman spectral problem only requires the Banach space with a norm defined, while the POD requires the Hilbert space with inner-product defined. The latter may further limit its application. Therefore, Koopman decomposition may be more suitable for dynamic analysis. 

\begin{figure}
\centering
\includegraphics[width=1.0\linewidth]{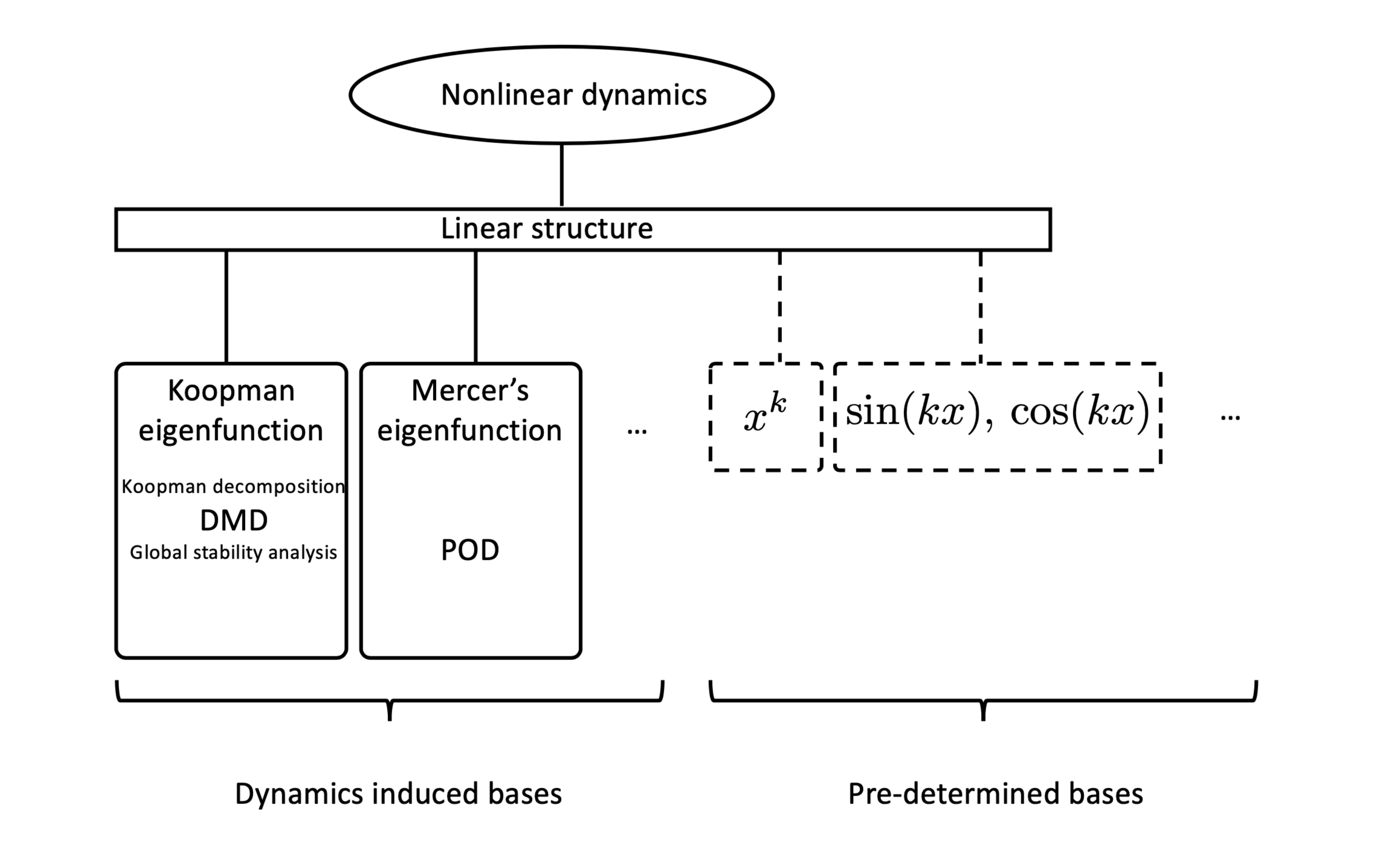}
\caption{The linear structure and bases}\label{fig:linearstructurebases}
\end{figure}

Both the Koopman eigenfunctions and eigenfunctions of Mercer's theorem are dynamics induced bases. Besides them, there exists fixed bases as well, see figure~\ref{fig:linearstructurebases}, such as polynomials $1, x, x^2, \cdots$ or the trigonometric series. The polynomials are extensively used to derive the proliferation rule. Though these bases are readily available, they may subject similar disadvantages of Mercer eigenfunctions.

\section{Numerical examples for Koopman decomposition}

In part 1, Koopman decomposition of the fluid flows is computed by DMD algorithm at the asymptotic stages. In the following section, a Hopf bifurcation process is studied to show how the new framework can be used to study a nonlinear transition process.



Hopf bifurcation is a common phenomenon in nonlinear dynamics. It describes a critical point where a system loses stability and transits to a periodic solution. It is often caused by a change of control parameters of the system, during which a pair of stable complex conjugate characteristic eigenvalues move to the unstable region. For example, the instability may be influenced by fluid viscosity, which is usually presented by the dimensionless Reynolds number, named after Osborne Reynolds.
\begin{equation}
Re = \frac{\text{inertial force}}{\text{viscous force}} = \frac{\rho U L}{\mu}.
\end{equation}
For the fluid past cylinder problem, the reference velocity $U$ is the incoming velocity, and $L$ is the diameter of the cylinder. Experiments show the bifurcation occurs around $Re\approx 50$~\citep{tritton1959experiments,roshko1954development} when the wake breaks symmetry and the original two steady counter-rotating vortices become unstable and shed alternatively from the cylinder, resulting in the well-known K\'arm\'an vortex.

\subsection{Koopman decomposition for primary instability}
\label{sec:dmd}

\begin{figure}
\centering
\begin{subfigure}[b]{0.49\linewidth}
\includegraphics[width=1.0\textwidth]{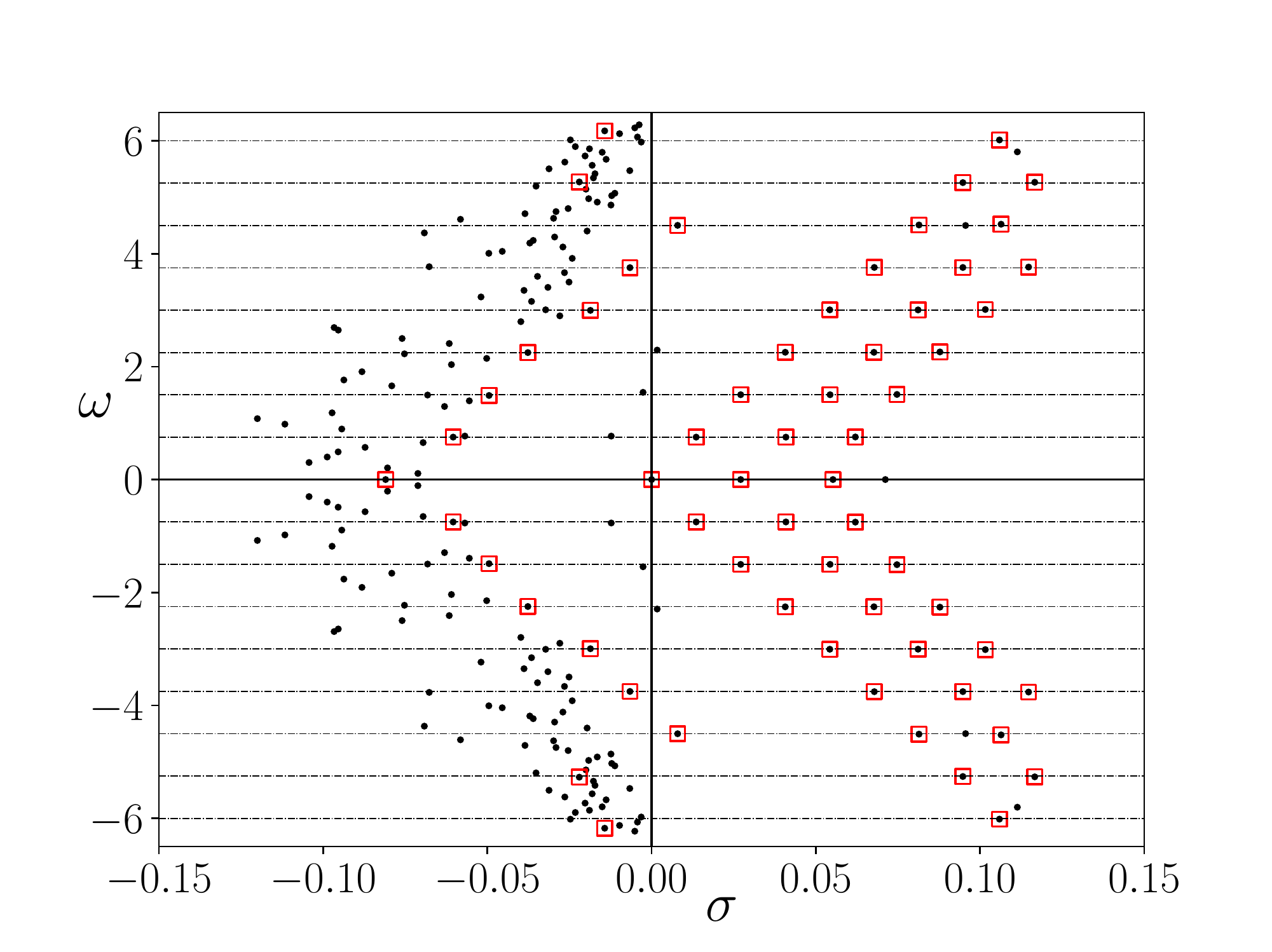}
\caption{initial stage \label{fig:spectruminit}}
\end{subfigure}
\begin{subfigure}[b]{0.49\linewidth}
\includegraphics[width=1.0\textwidth]{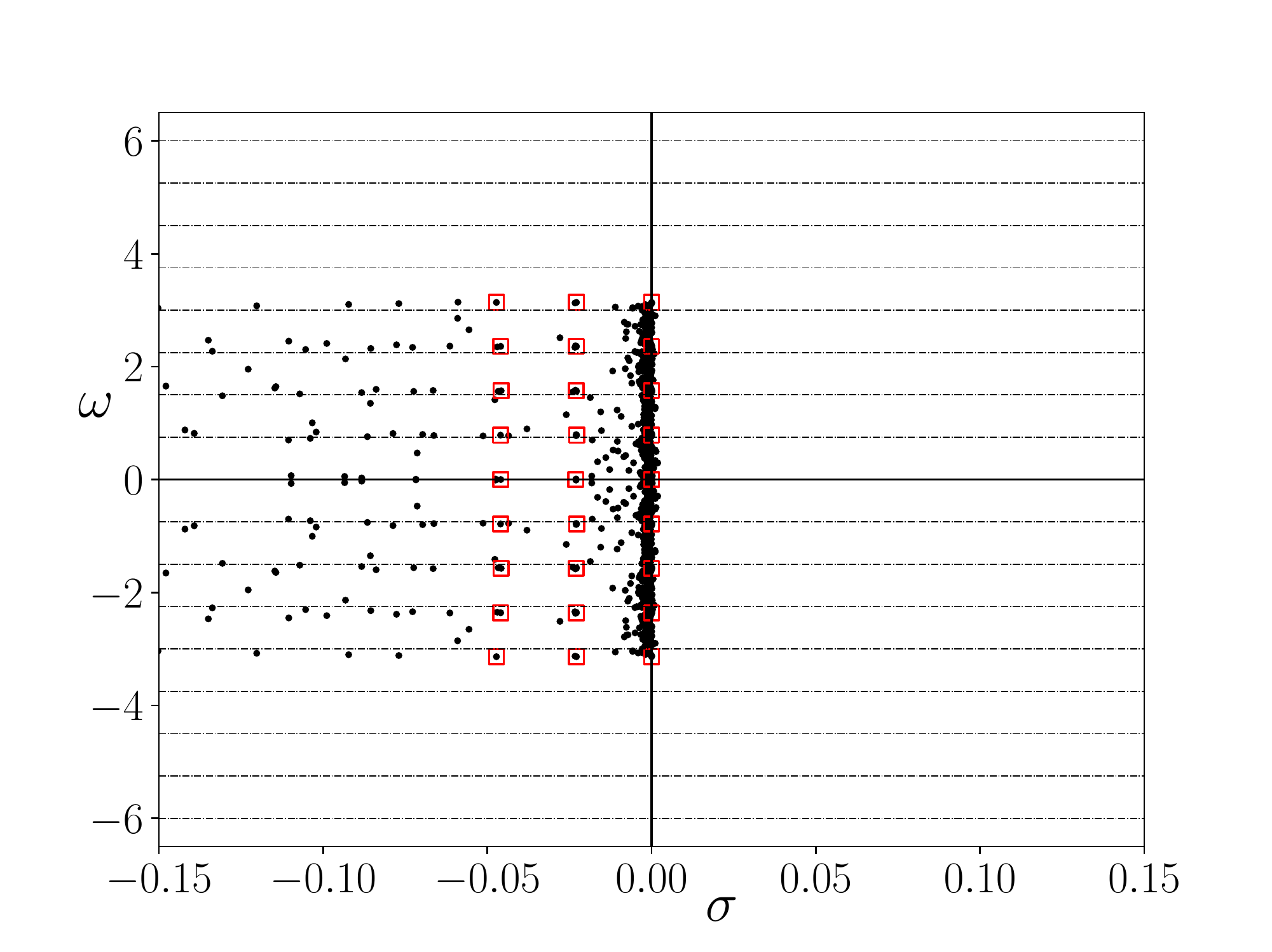}
\caption{finial stage \label{fig:spectrumfinal}}
\end{subfigure}
\caption{Koopman spectrums of primary instability of flow past fixed cylinder at $Re=50$. ($\bullet$) Black dots show all DMD modes. ({\color{red}{$\Box$}}) Red square indicates the most 'significant' modes. (a) Spectrums around the unstable equilibrium state. (b) Spectrums around the stable limit cycle solution.} \label{fig:primaryspectrum}
\end{figure}

As stated earlier, at the primary instability stage the normal mode grows exponentially at a small magnitude. The growth saturates as the perturbation grows and finally reaches periodic. Thus the two-dimensional primary instability is divided into two phases. The initial phase describes the initial growth of perturbation around the unstable equilibrium state. The final phase described the saturation of perturbation around the stable limit cycle solution. Koopman spectrums at each phase are studied separately. More detail about the data collected can be found in part 1. The Koopman spectrums computed for both stages are shown in part 1. For convenience, they are also listed in figure~\ref{fig:primaryspectrum}.

\subsubsection{Koopman modes at initial stage}

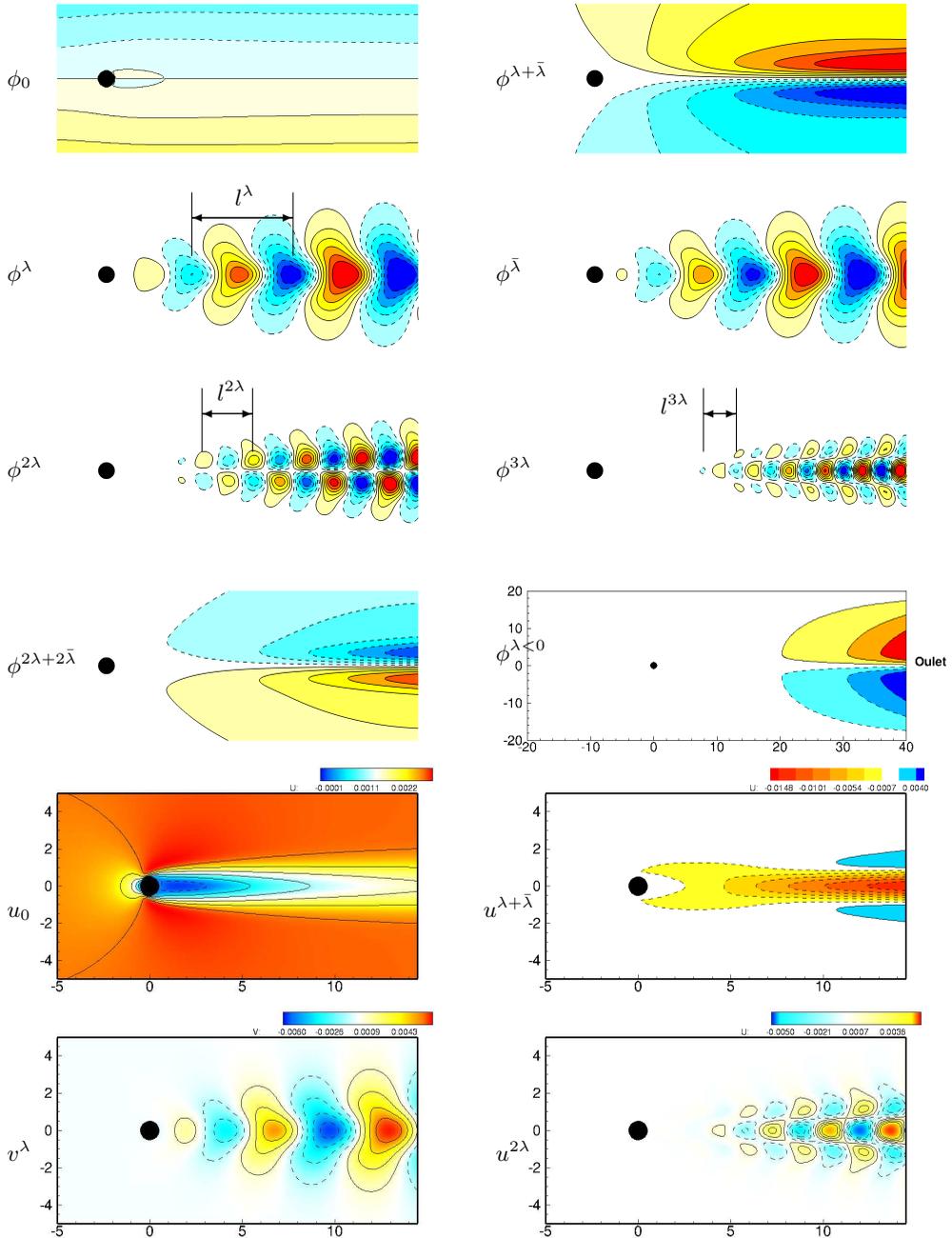
\begin{figure}
\centering
\begin{subfigure}[t]{0.495\linewidth}
\begin{overpic}[width=1.0\linewidth]{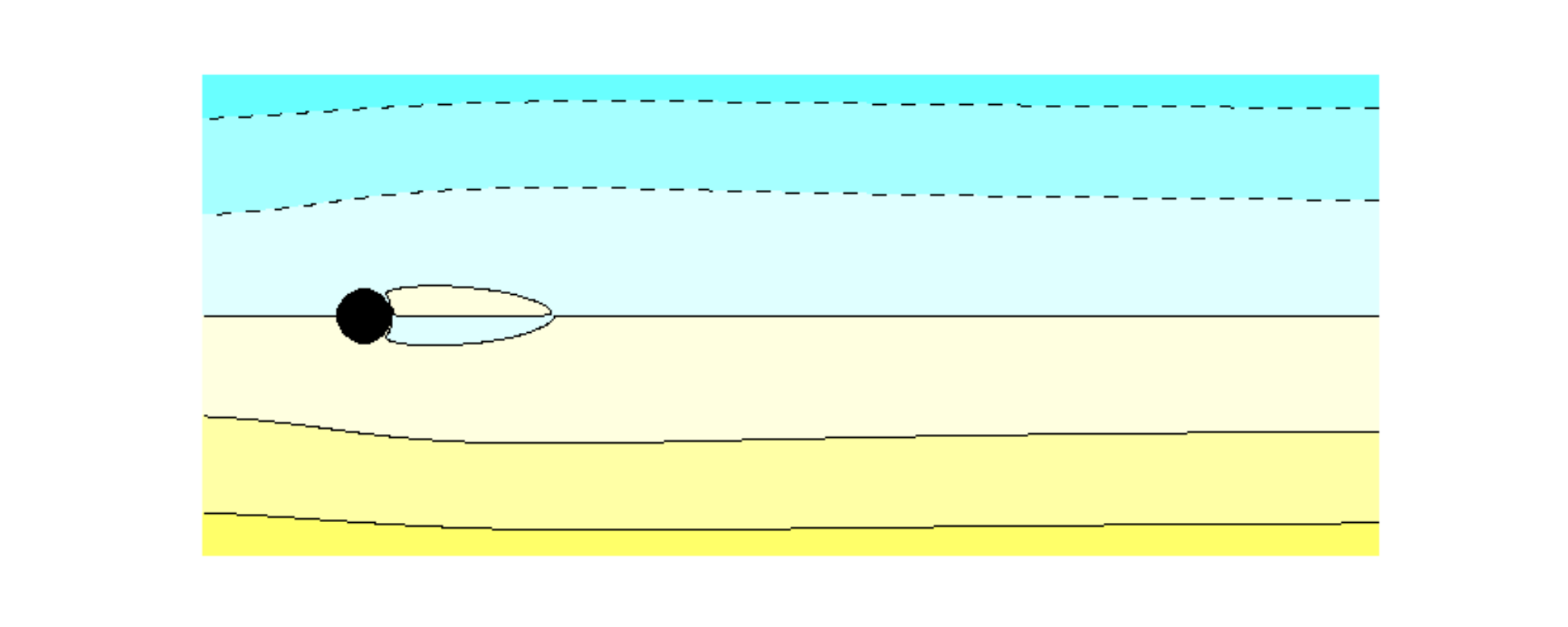}
\put(5,35){$\phi_0$}
\end{overpic}
\end{subfigure}
\begin{subfigure}[t]{0.495\linewidth}
\begin{overpic}[width=1.0\textwidth]{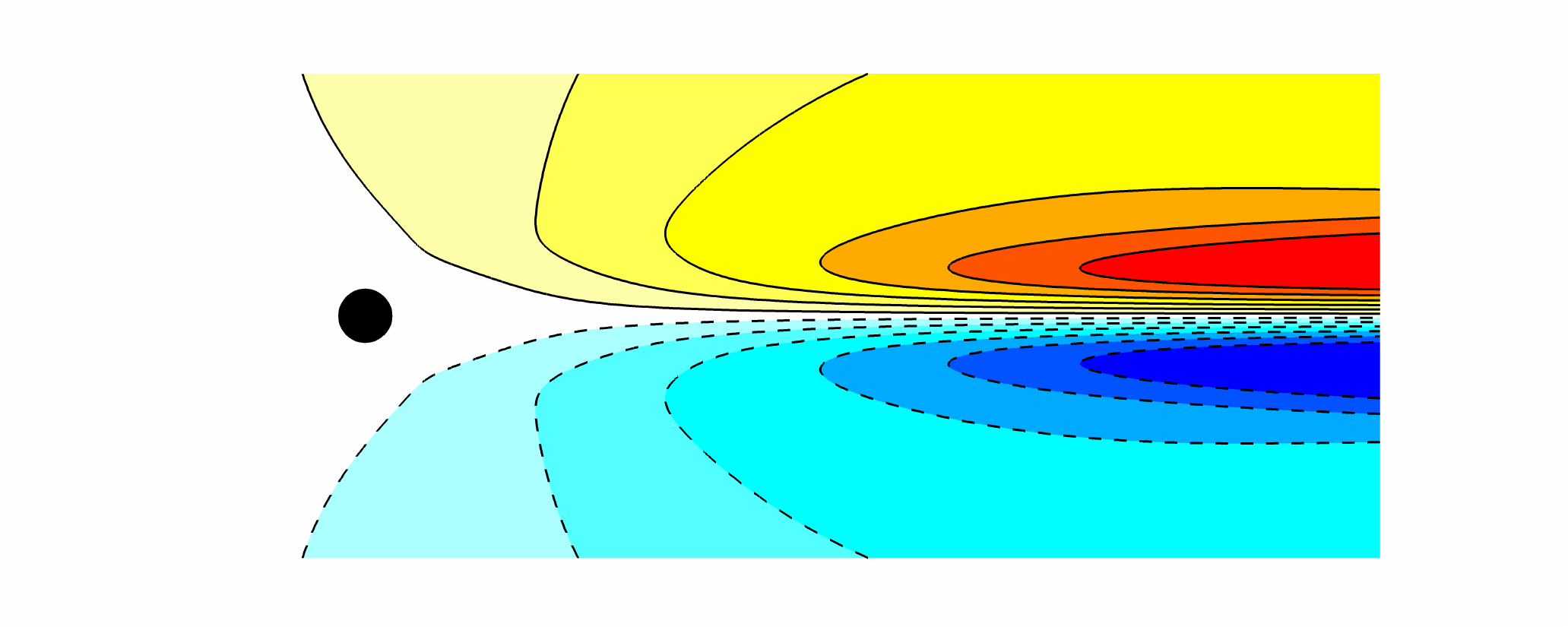}
\put(5,35){$\phi^{\lambda + \bar{\lambda}}$}
\end{overpic}
\end{subfigure}\\
\begin{subfigure}[t]{0.495\linewidth}
\begin{overpic}[width=1.0\linewidth]{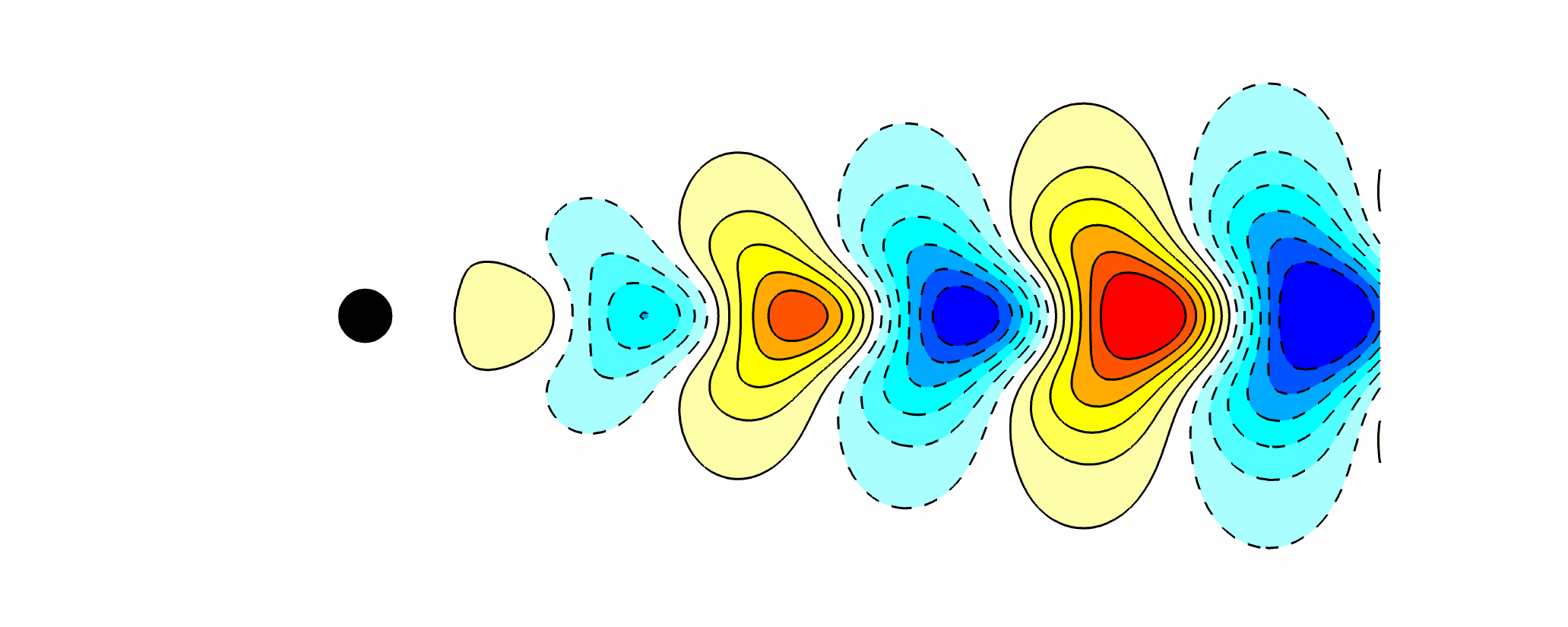}
\put(5,35){$\phi^{\lambda}$}
\put(78,45){\line(0,1){25}}
\put(118,45){\line(0,1){25}}
\put(78,60){\vector(1,0){40}}
\put(118,60){\vector(-1,0){40}}
\put(95,65){$l^{\lambda}$}
\end{overpic}
\end{subfigure}
\begin{subfigure}[t]{0.495\linewidth}
\begin{overpic}[width=1.0\textwidth]{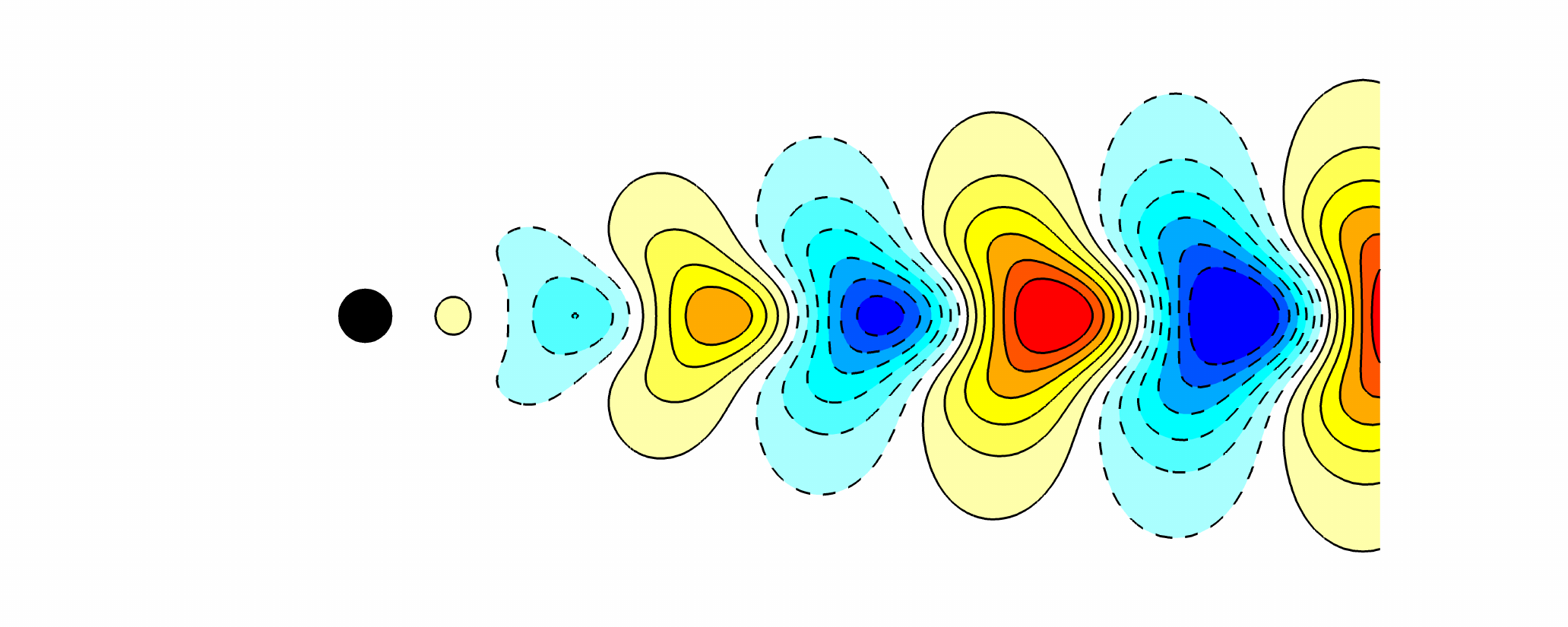}
\put(5,35) {$\phi^{\bar{\lambda}}$}
\end{overpic}
\end{subfigure}\\
\begin{subfigure}[t]{0.495\linewidth}
\begin{overpic}[width=1.0\linewidth]{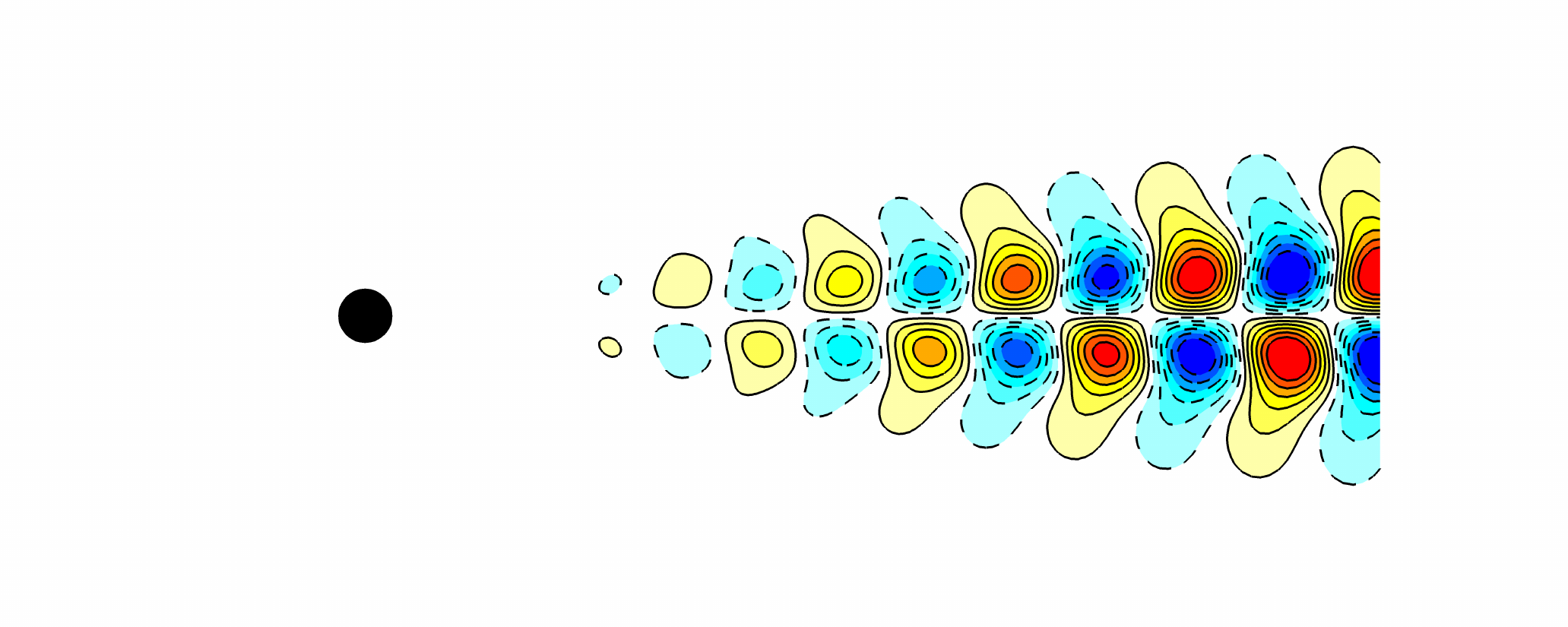}
\put(5,35){$\phi^{2\lambda}$}
\put(82,45){\line(0,1){25}}
\put(102,45){\line(0,1){25}}
\put(82,60){\vector(1,0){20}}
\put(102,60){\vector(-1,0){20}}
\put(88,65){$l^{2\lambda}$}
\end{overpic}
\end{subfigure}
\begin{subfigure}[t]{0.495\linewidth}
\begin{overpic}[width=1.0\textwidth]{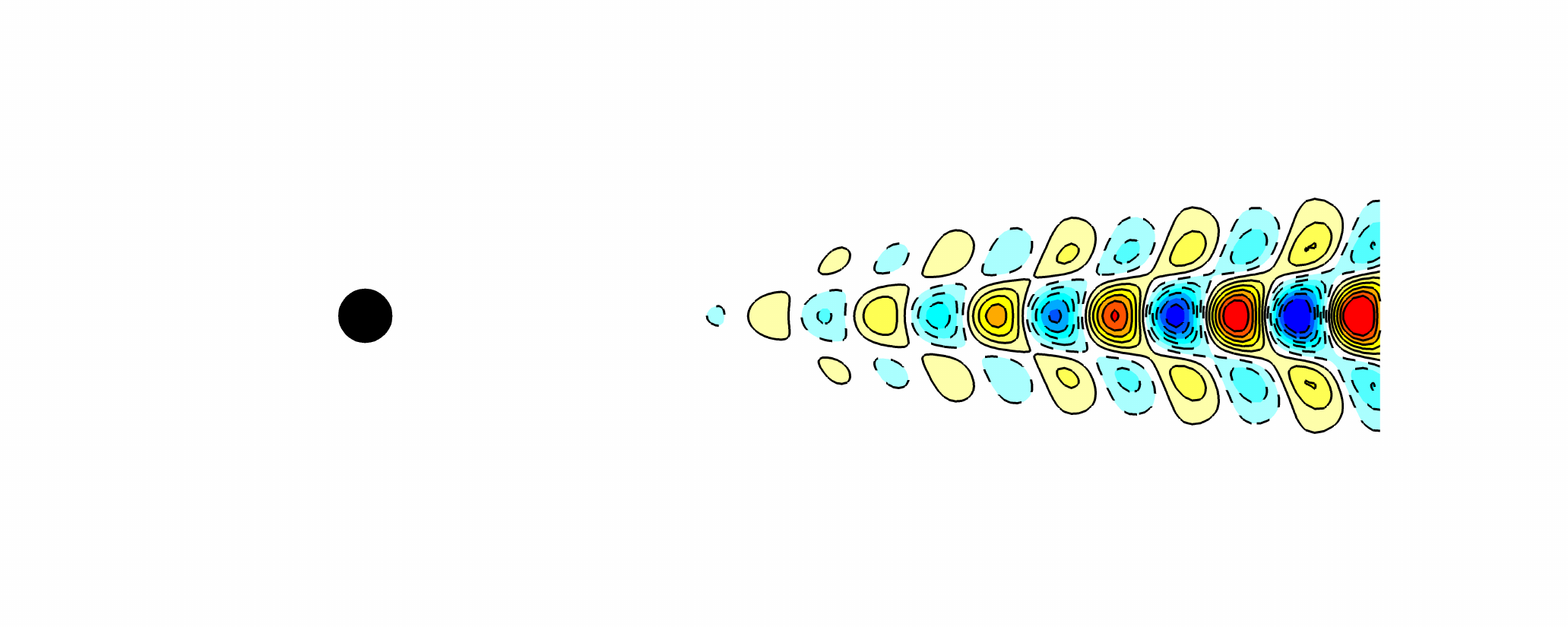}
\put(5,35){$\phi^{3\lambda}$}
\put(87,45){\line(0,1){25}}
\put(100,45){\line(0,1){25}}
\put(87,60){\vector(1,0){13}}
\put(100,60){\vector(-1,0){13}}
\put(70,60){$l^{3\lambda}$}
\end{overpic}
\end{subfigure}\\
\begin{subfigure}[t]{0.495\linewidth}
\begin{overpic}[width=1.0\linewidth]{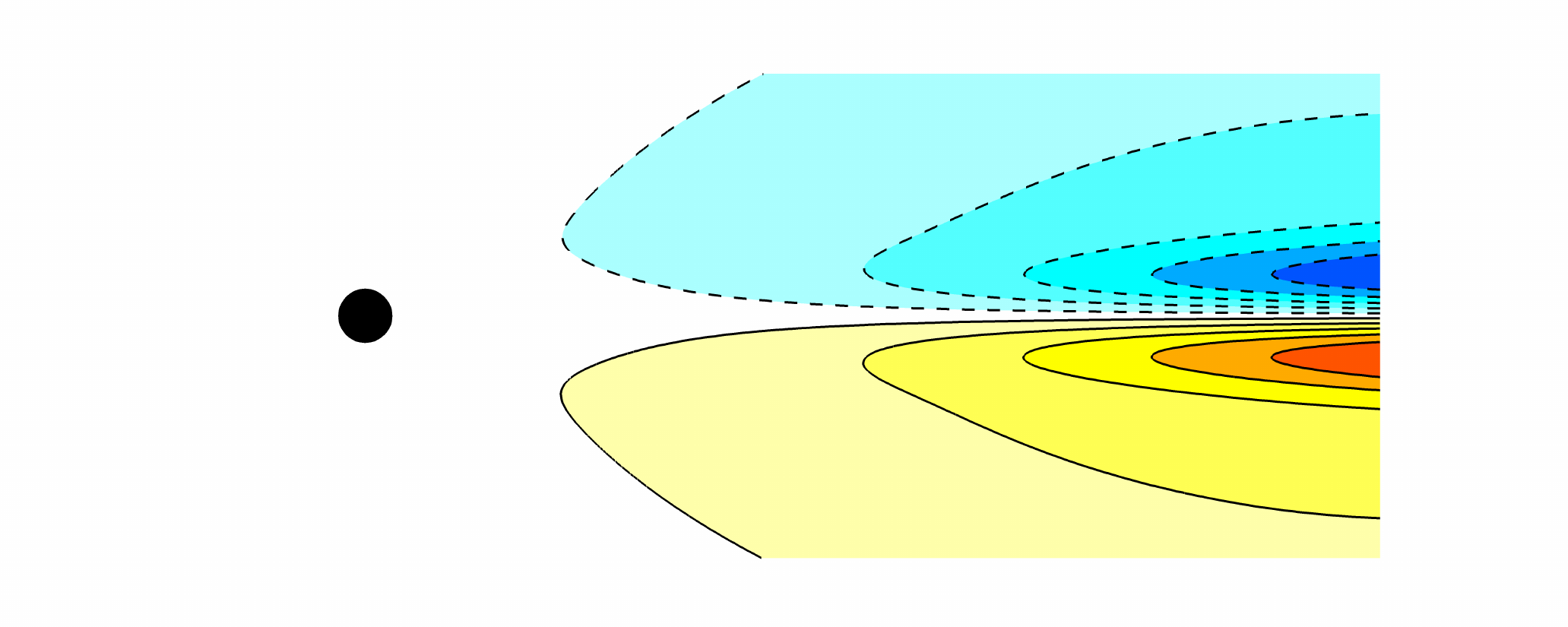}
\put(5,35){$\phi^{2\lambda+2\bar{\lambda} }$}
\end{overpic}
\end{subfigure}
\begin{subfigure}[t]{0.495\linewidth}
\begin{overpic}[width=1.0\textwidth]{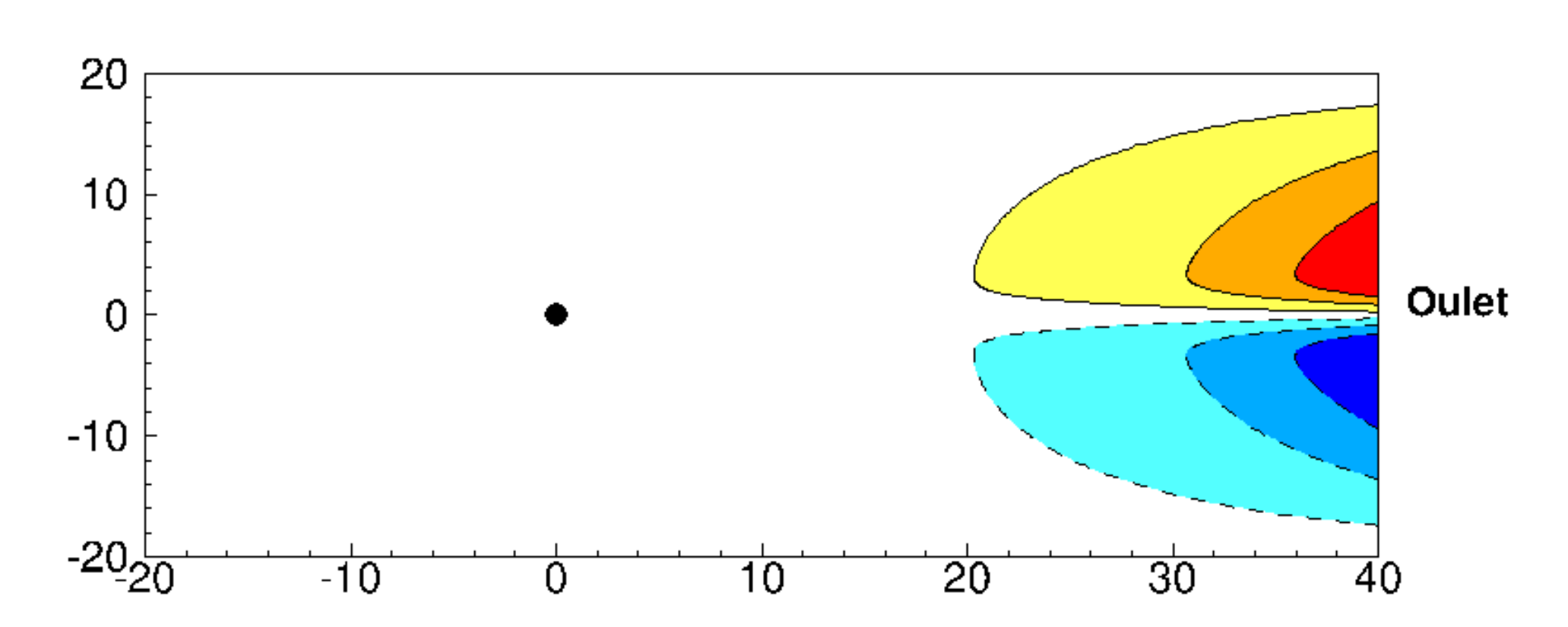}
\put(5,40){$\phi^{\lambda < 0}$}
\end{overpic}
\end{subfigure}
\begin{subfigure}[t]{0.495\linewidth}
\begin{overpic}[width=1.0\linewidth]{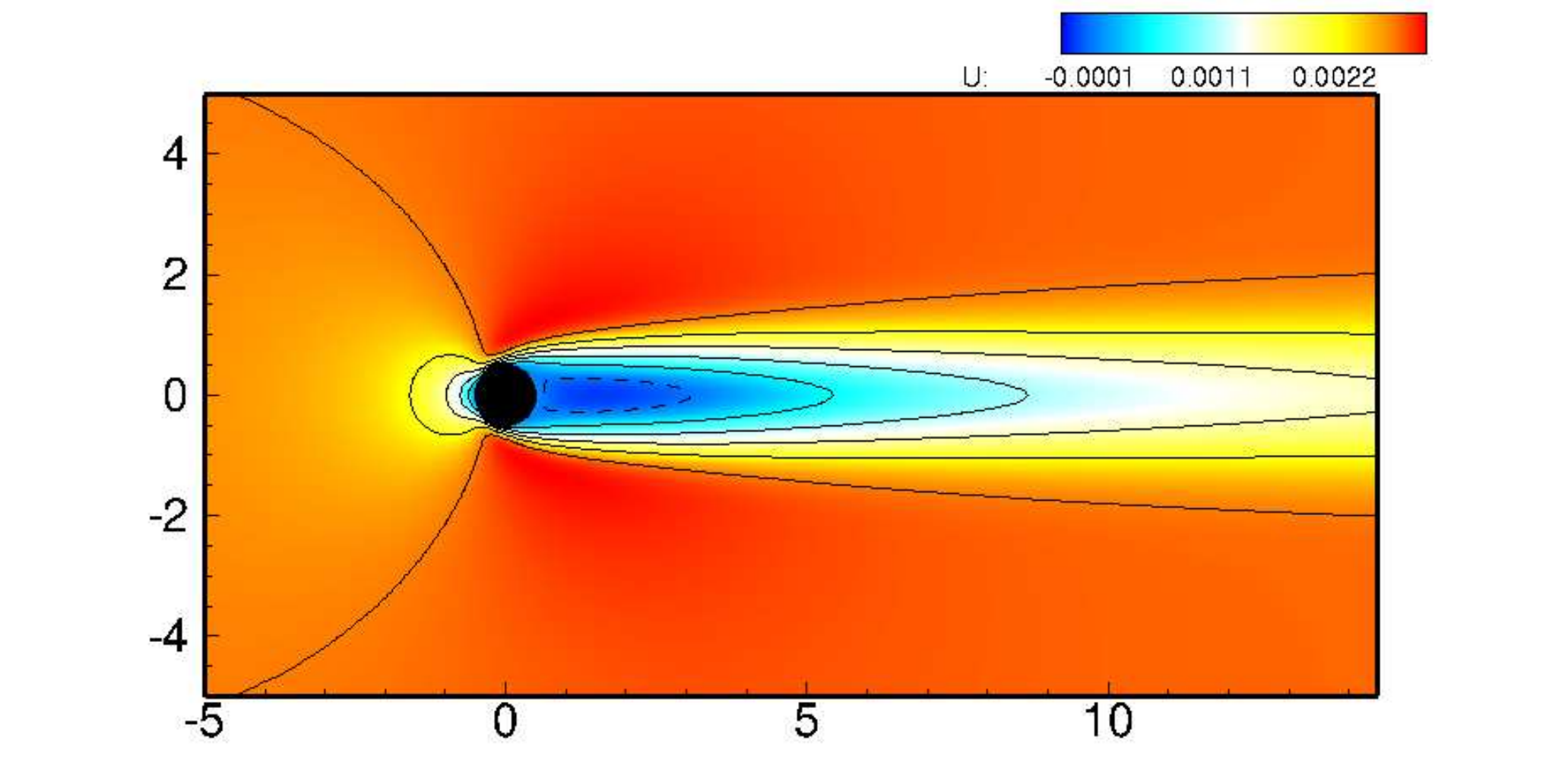}
\put(5,35){$u_0$}
\end{overpic}
\end{subfigure}
\begin{subfigure}[t]{0.495\linewidth}
\begin{overpic}[width=1.0\textwidth]{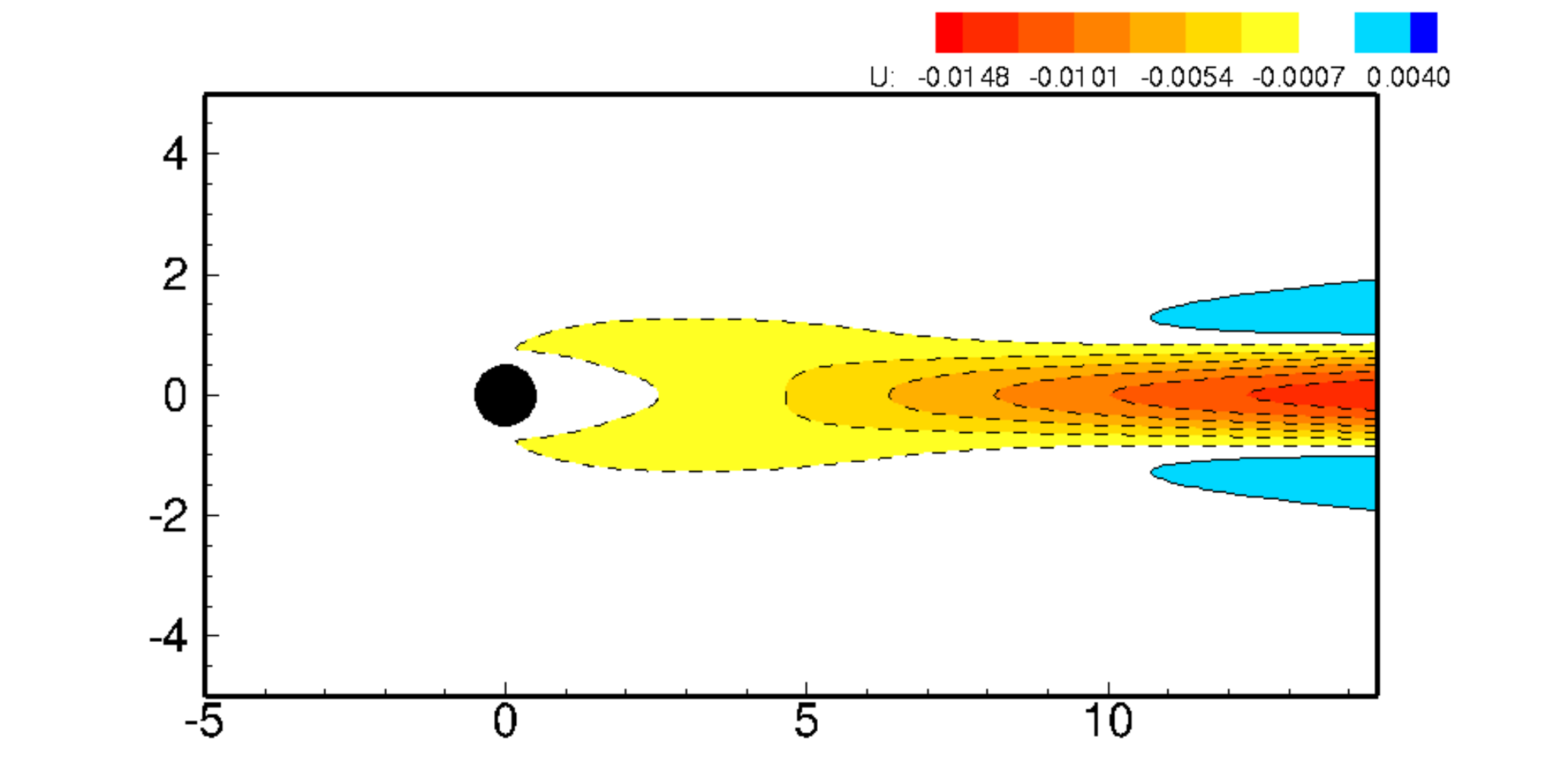}
\put(0,35){$u^{\lambda+\bar{\lambda}}$}
\end{overpic}
\end{subfigure}\\
\begin{subfigure}[t]{0.495\linewidth}
\begin{overpic}[width=1.0\linewidth]{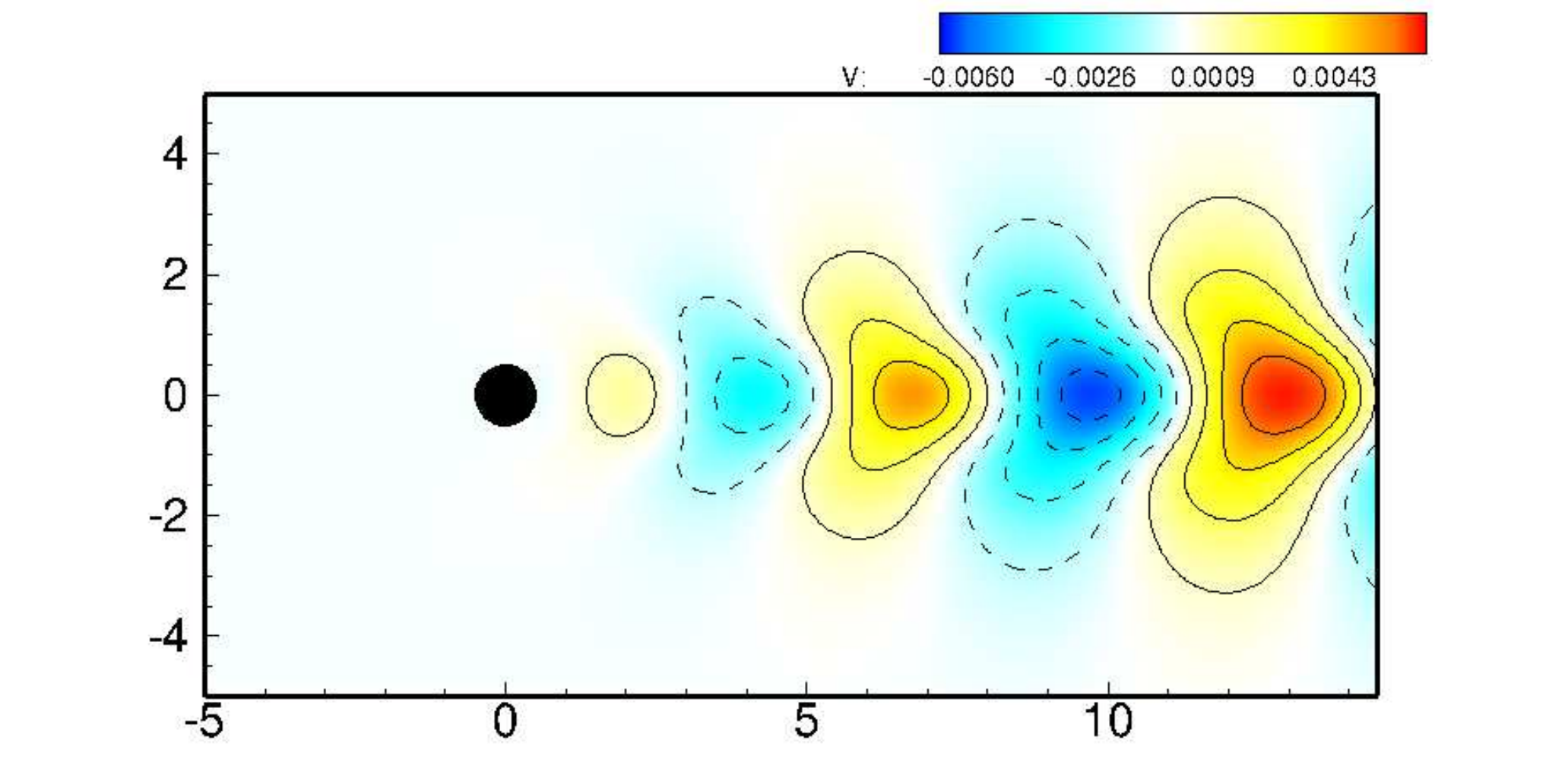}
\put(5,35){$v^{\lambda}$}
\end{overpic}
\end{subfigure}
\begin{subfigure}[t]{0.495\linewidth}
\begin{overpic}[width=1.0\textwidth]{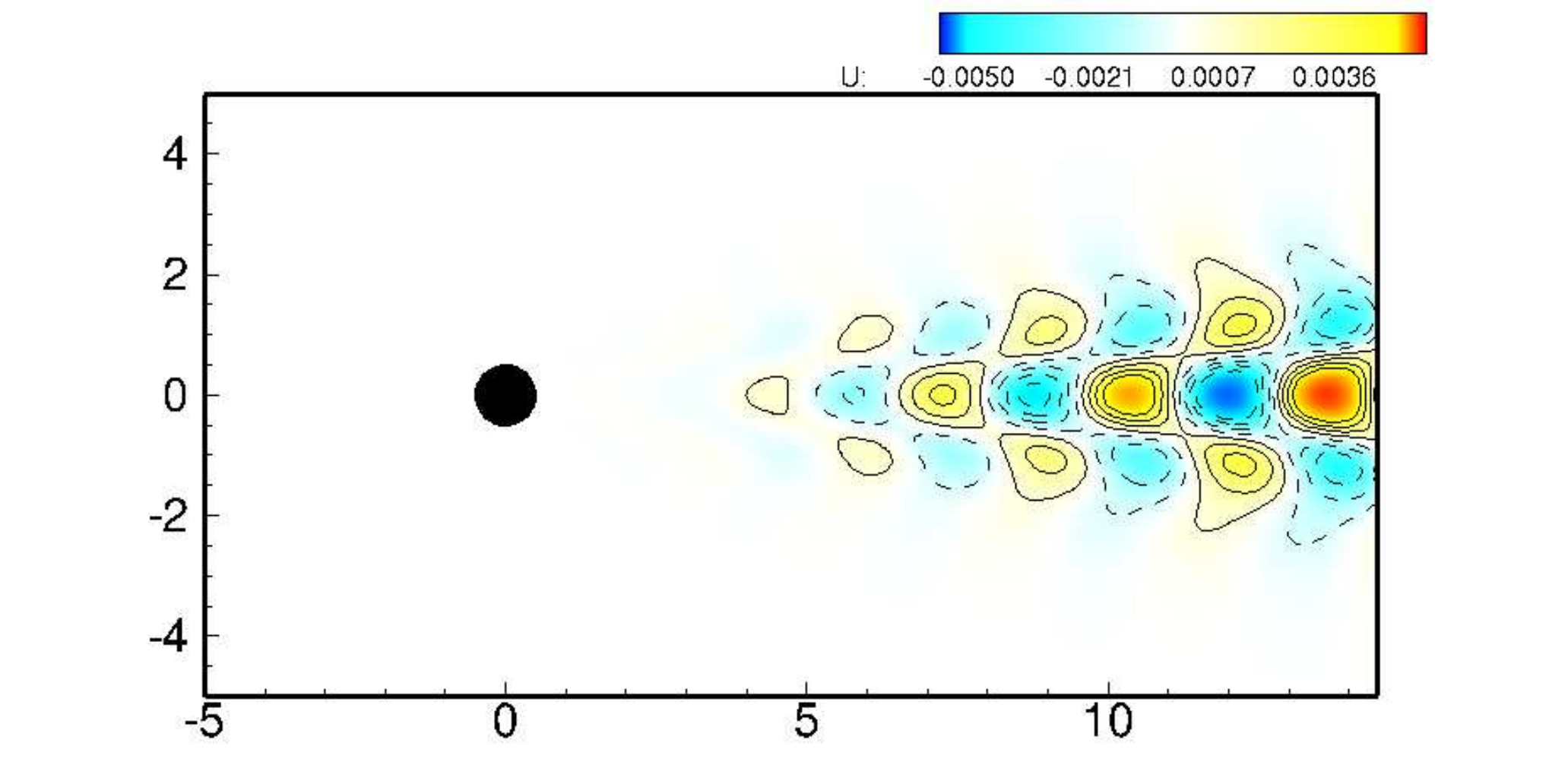}
\put(5,35){$u^{2\lambda}$}
\end{overpic}
\end{subfigure}
\caption{DMD modes captured at the initial stage. First 4 rows show the stream function of the DMD modes with labels on the left. The streamwise velocity of the $u_0$, $u^{\lambda+\bar{\lambda}}$ and $u^{2\lambda}$ and cross flow velocity $v^{\lambda}$ are also shown in the last two rows for comparison with~\citet{sipp2007global}.  The wavenumber of $\phi^{\lambda}$, $\phi^{2\lambda}$, $\phi^{3\lambda}$ are 0.16, 0.31, 0.45, which have an ratio $0.16:0.31:0.45\approx 1:2:3$, and the corresponding wave length is labelled on the figure. } \label{fig:2dinitialmodes}
\end{figure}

Some Koopman modes are shown in figure~\ref{fig:2dinitialmodes}, the mean flow $\boldsymbol{u}_0$ captures the base flow, the equilibrium solution at $Re$. This is the unperturbed state, symmetry with respect to centerline. The critical normal modes $\boldsymbol{u}^{\lambda}$ and $\boldsymbol{u}^{\bar{\lambda}}$ are a complex conjugate pair, so the stream function $\phi^{\lambda}$ and $\phi^{\bar{\lambda}}$ of the real and imaginary part are shown. These normal modes are the same with the ones computed by~\citet{jackson1987finite}, who numerically solved the eigenvalue problem of the homogeneous linearized Navier-Stokes equation by Arnoldi method at $Re=50$. They are also similar to the global modes computed by~\citet{sipp2007global} by solving the homogeneous linearized equation at $Re_c=46.6$. By definition our mean flow is different from Sipp et. al. in the sense our mean flow is the combination of their mean flow and Reynolds modification.
\begin{equation*}
\boldsymbol{u}_0 = \boldsymbol{u}_0^{Sipp} + \epsilon \boldsymbol{u}^{Sipp}_{F_2^1}.
\end{equation*}
$\epsilon \boldsymbol{u}^{Sipp}_{F_2^1}$ is the Reynolds modification as the increase of $Re_c$ to $Re$. In practice, as $Re$ is close to $Re_c$, the mean mode, the normal modes $\boldsymbol{u}^{\lambda}$, $\boldsymbol{u}^{\bar{\lambda}}$ and the high order derived modes $\boldsymbol{u}^{2\lambda}$, $\boldsymbol{u}^{3\lambda}$ computed here ($Re=50$) are all close to the one obtained by~\citet{sipp2007global} and~\citet{meliga2011asymptotic} at $Re=46.6$ or 47. Mode $\boldsymbol{u}^{\lambda + \bar{\lambda}}$ and $\boldsymbol{u}^{2\lambda + 2\bar{\lambda}}$ are the monotonic growing modes (spectrums with pure positive real part). These are symmetry modes which is generated by the interaction of $\boldsymbol{u}^{\lambda}$, $\boldsymbol{u}^{\bar{\lambda}}$ or $\boldsymbol{u}^{2\lambda}$, $\boldsymbol{u}^{2\bar{\lambda}}$. These monotonic modes' effect is to `modulate' the base flow, to short the recirculation zone after the cylinder until the base flow reaches the final equilibrium state.

An interesting phenomenon is that these Koopman modes wavenumber increases proportionally with eigenvalues. For instance, the wavenumber of the critical mode $\phi^{\lambda}$ and its high-order derived modes $\phi^{2\lambda}$, $\phi^{3\lambda}$, based on the average distance of those `bubbles' are estimated to be 0.16, 0.31, 0.45 (per diameter of the cylinder) in the streamwise direction. These wavenumbers approximately satisfy the ratio of $0.16:0.31:0.45\approx 1:2:3$, that is, proportional to the ratio of eigenvalues. This wavenumber-eigenvalue relation also approximately holds at the cross-flow direction. This proportional relation is preserved in the final periodic solution, see modes in Figure~\ref{fig:2dfinalmodes}. It is noticed that the wavenumber-eigenvalue relation exists for the secondary instability flow after an infinite-long fixed cylinder at $Re=200$ (see Figure~10 in part 1), where the critical Floquet modes have a spanwise wavenumber of 0.25, while its second-order higher ones have 0.5.

This phenomenon may be explained by the global stability analysis (GSA) in \S~\ref{sec:GSA}, where the high-order modes are the response to an excitation by a linear system
\begin{equation}
\mathcal{T}q_k = \mathcal{B}(q_i, q_j).
\end{equation}
Here $\mathcal{T}$ is an linear operator for the dynamic system (see equation~\ref{eqn:globalexp1}) and $\mathcal{B}$ is an bilinear operator on $q_i$ and $q_j$ for the excitation (see equation~\ref{eqn:nonlinearexcitation}). The interaction of $q_i$ and $q_j$ raised the order of spectrum as well as wavenumber, therefore, the response $q_k$ has corresponding summed wavenumber and spectrum.

Besides the unstable normal modes and their high order derived modes, a decaying mode $\boldsymbol{\Phi}^{\lambda < 0}$ ($\lambda=-0.0810$) is also presented. From the figure, this mode is only significant at the outlet. This monotonic decaying mode is believed caused by the incompatibility of the initial uniform flow field and the convective outlet boundary condition ($\frac{\partial \boldsymbol{u}}{\partial t} + U_{in} \frac{\partial \boldsymbol{u}}{\partial x} = 0$) used in the simulation.
$U_{in}$ is the incoming flow velocity. This mode disappeared as the incompatibility was reduced as the simulation continuous. Though this is an artificial mode caused by the initial flow, its cross interaction with the unstable Koopman modes is predicted by proliferation rule and captured by DMD algorithm.

\subsubsection{Koopman modes at the finial stage}

\begin{figure}
\centering
\begin{subfigure}[b]{0.325\linewidth}
\includegraphics[width=1.0\textwidth]{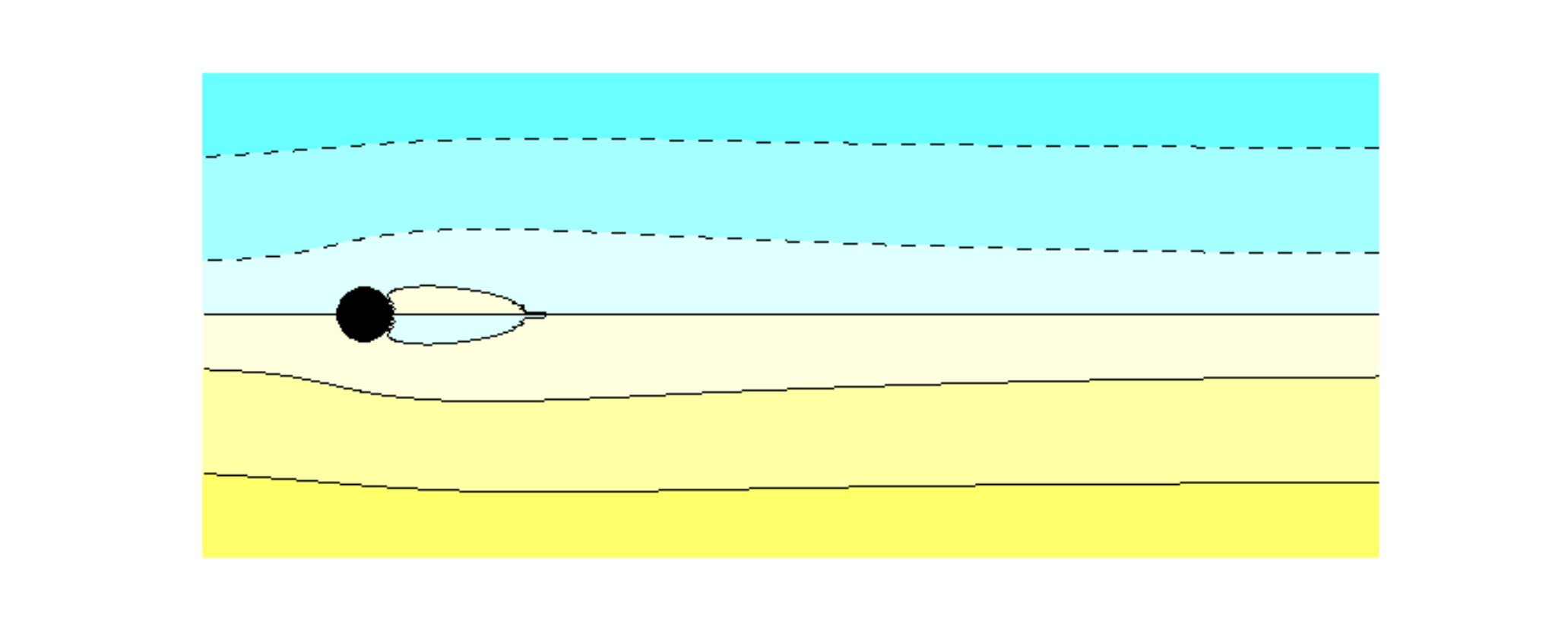}
\includegraphics[width=1.0\textwidth]{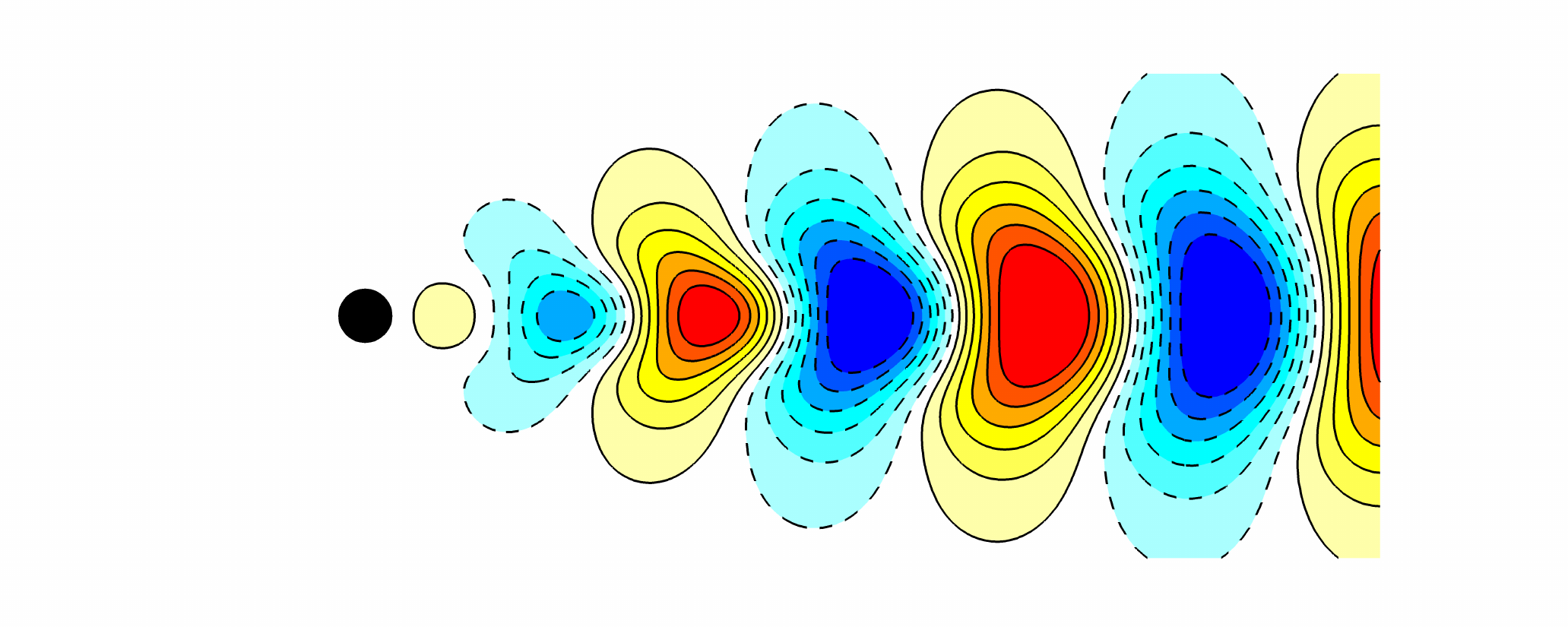}
\includegraphics[width=1.0\textwidth]{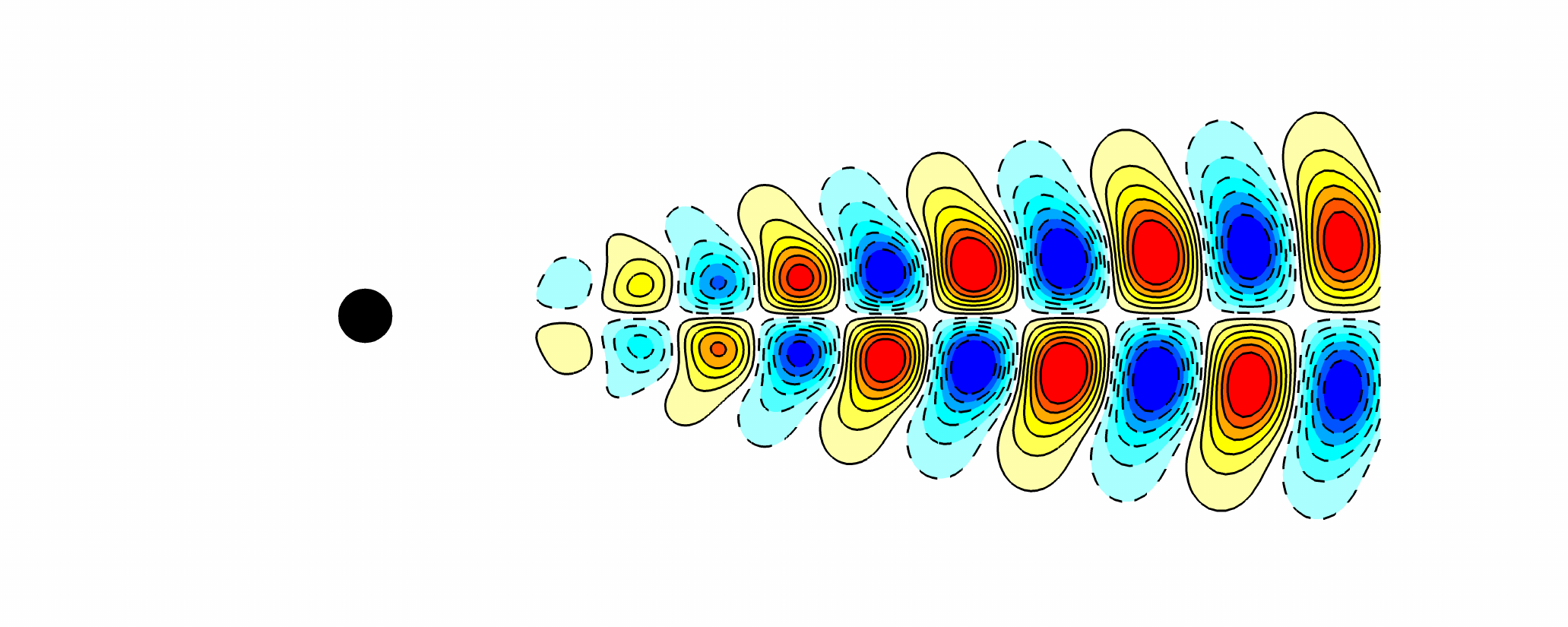}
\includegraphics[width=1.0\textwidth]{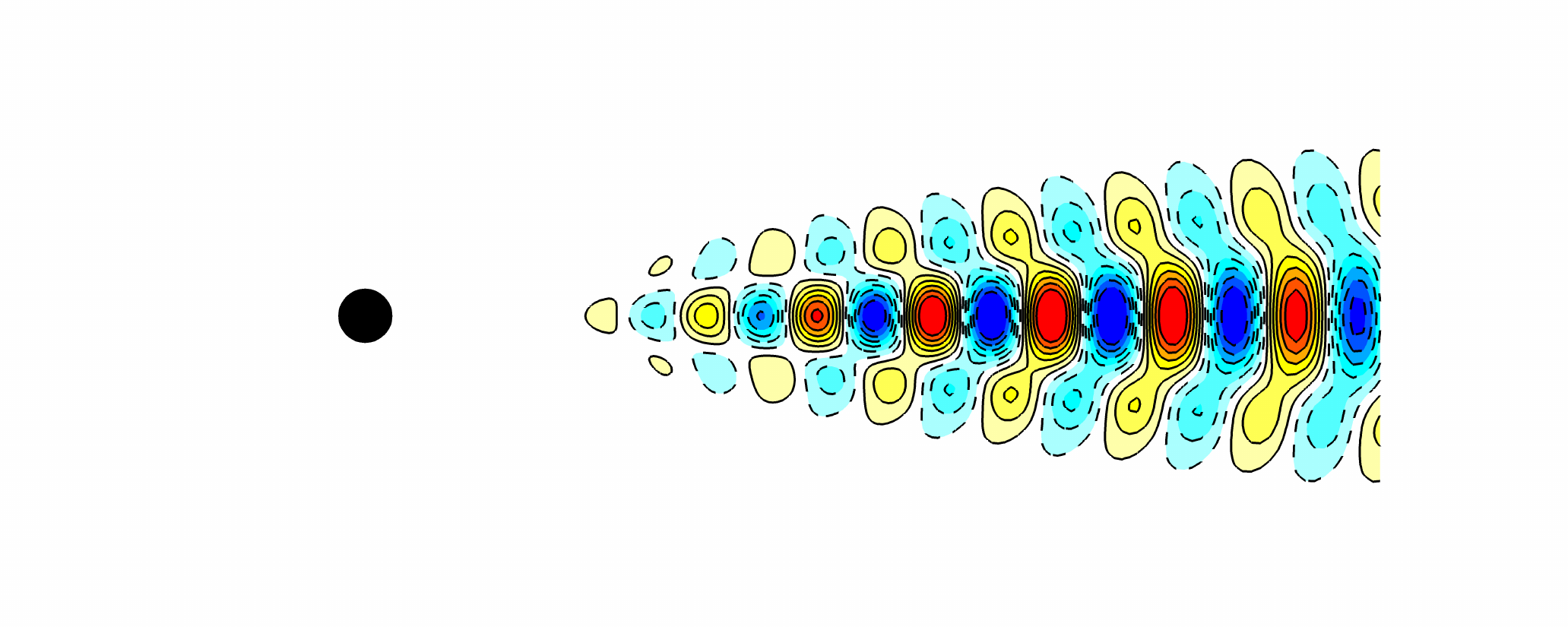}
\caption{$\sigma=0$} \label{fig:2dfinalmodesa}
\end{subfigure}
\begin{subfigure}[b]{0.325\linewidth}
\includegraphics[width=1.0\textwidth]{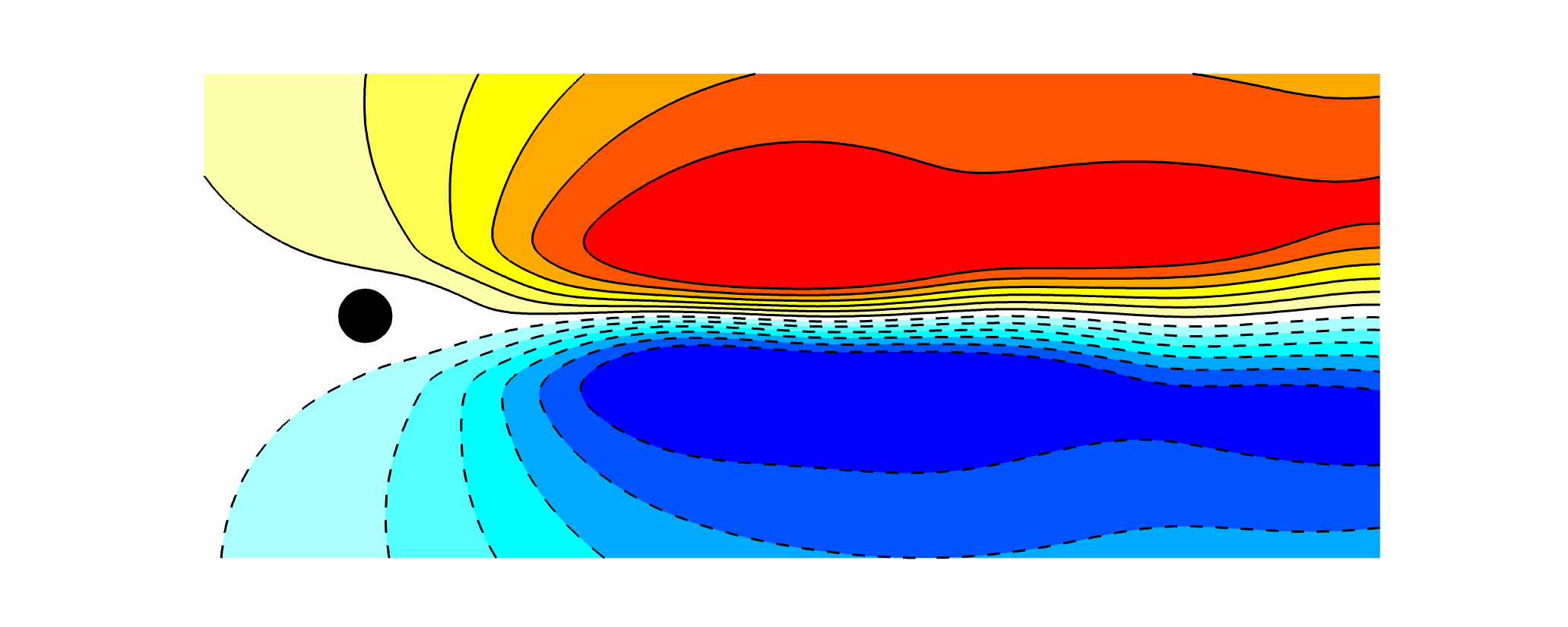}
\includegraphics[width=1.0\textwidth]{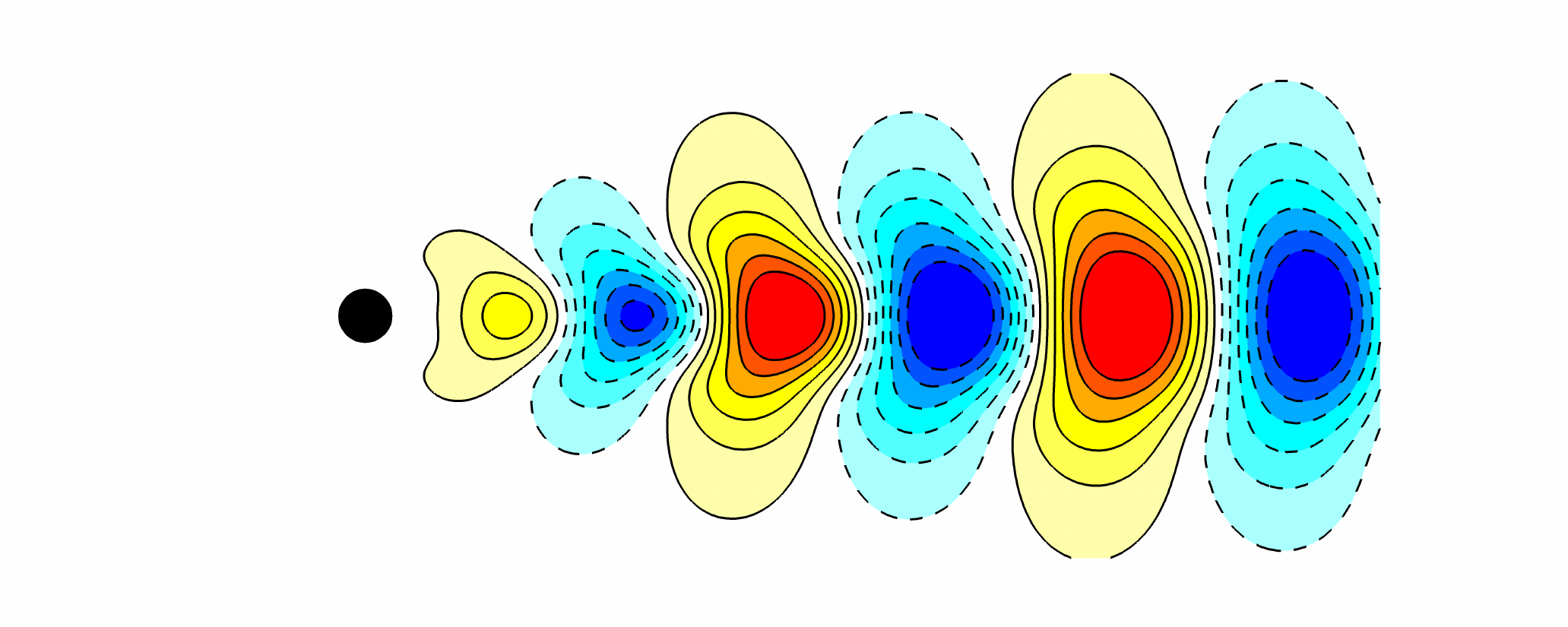}
\includegraphics[width=1.0\textwidth]{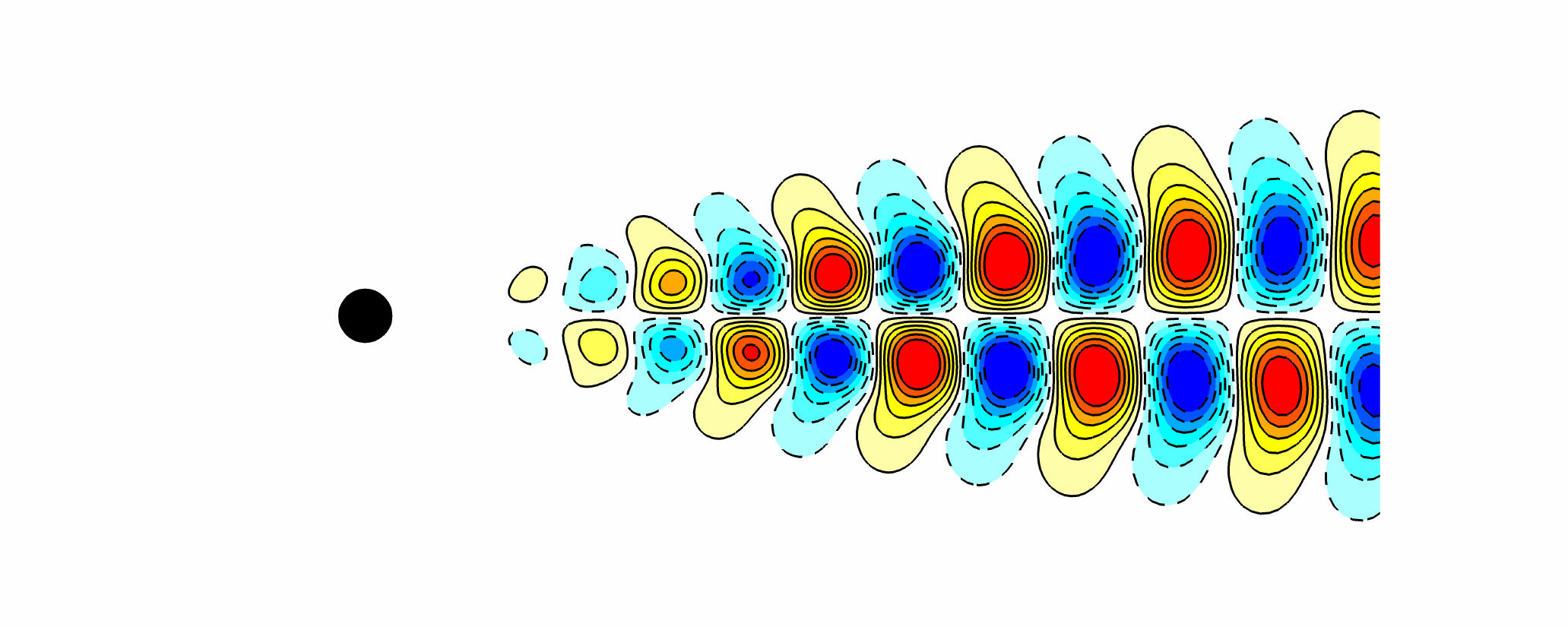}
\includegraphics[width=1.0\textwidth]{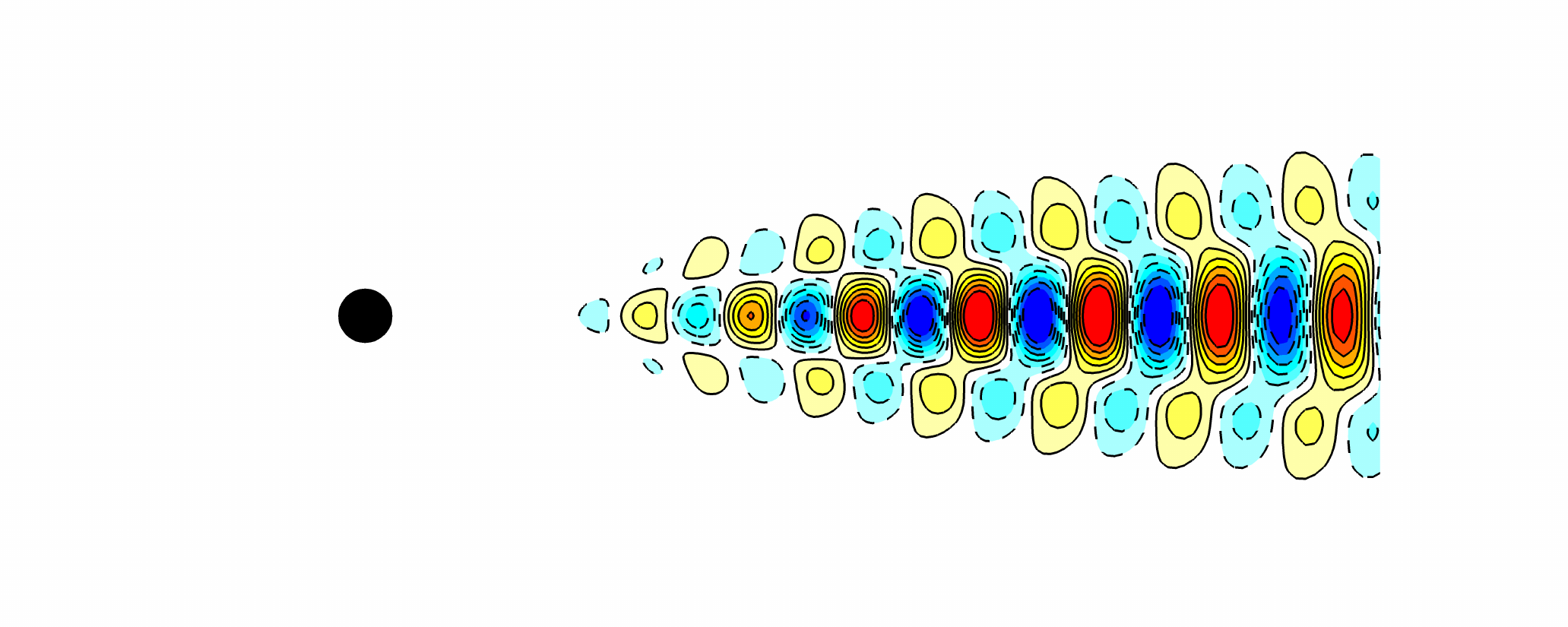}
\caption{$\sigma=-0.023$}
\end{subfigure}
\begin{subfigure}[b]{0.325\linewidth}
\includegraphics[width=1.0\textwidth]{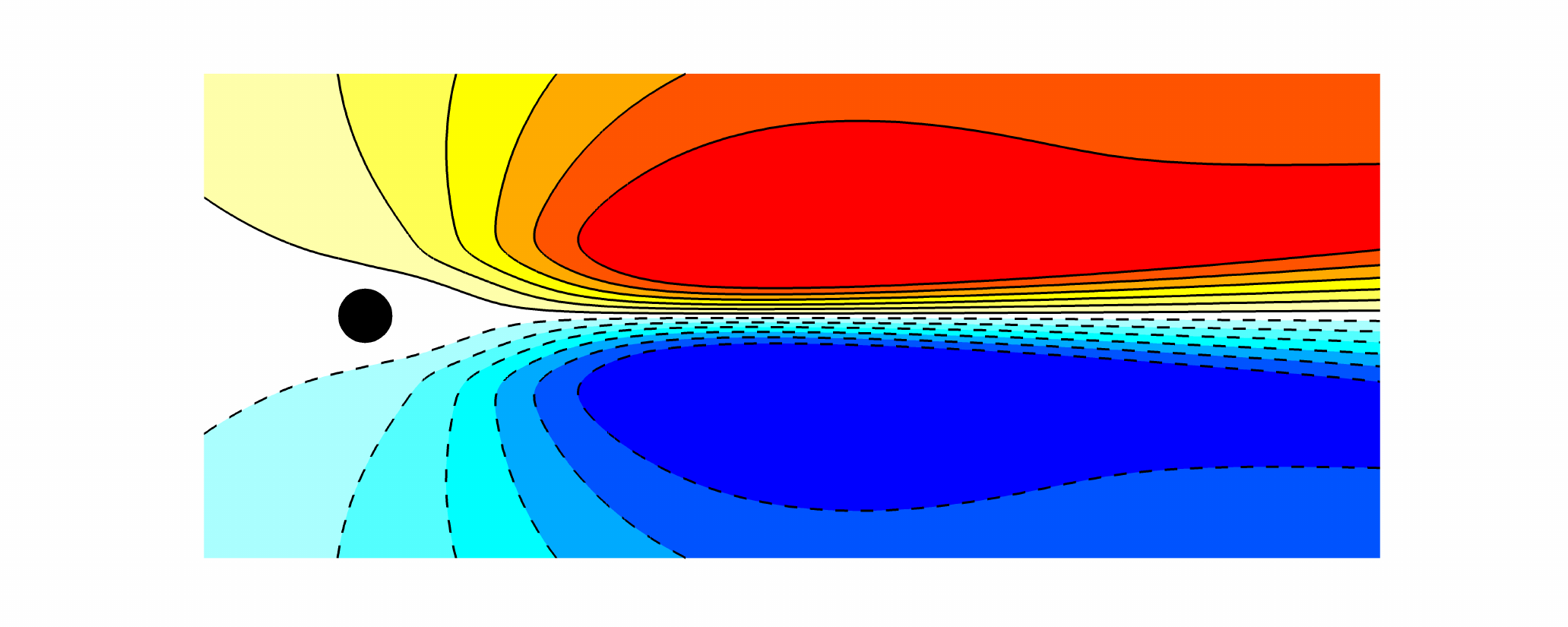}
\includegraphics[width=1.0\textwidth]{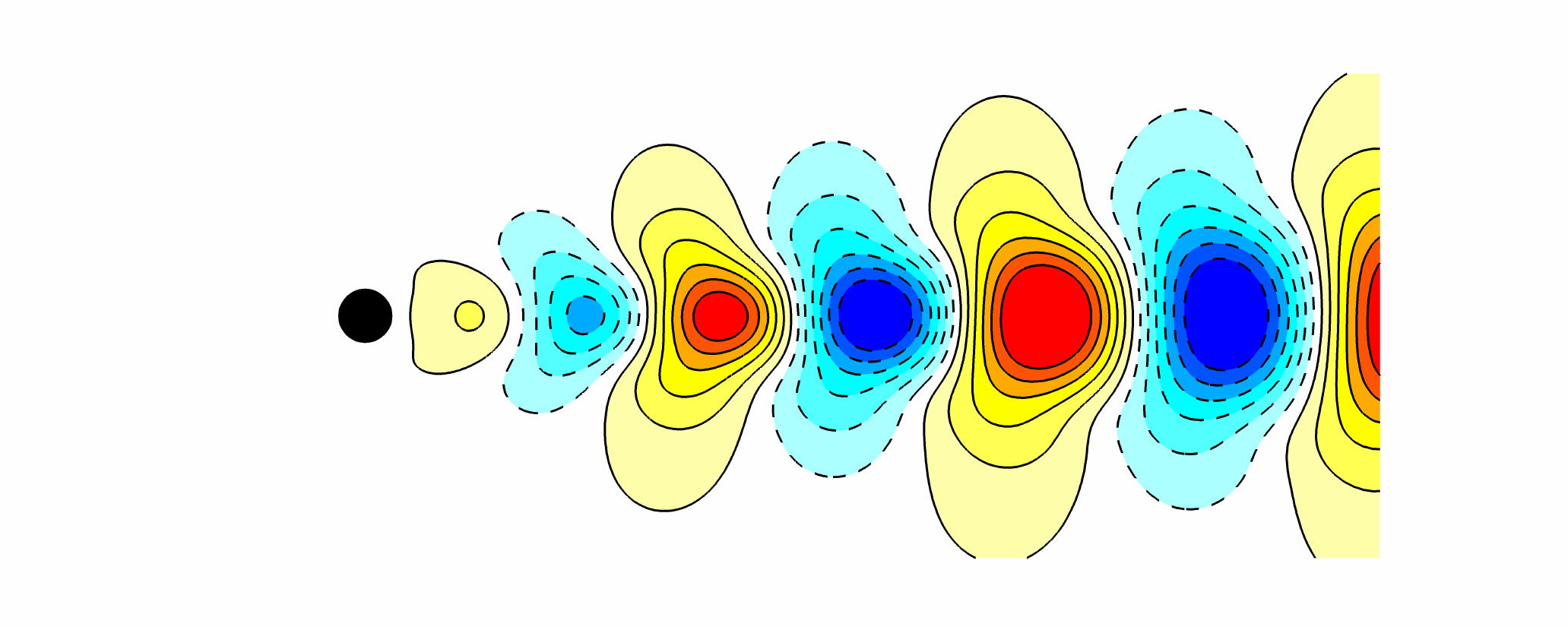}
\includegraphics[width=1.0\textwidth]{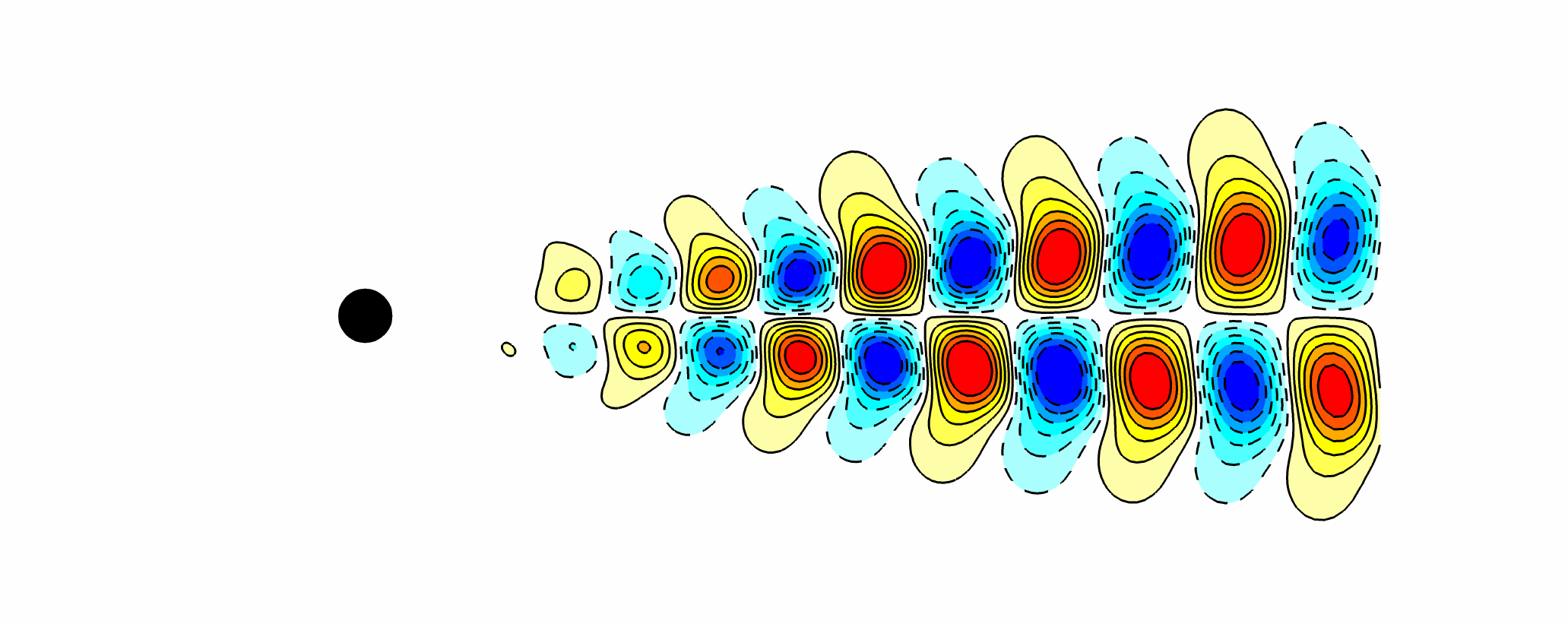}
\includegraphics[width=1.0\textwidth]{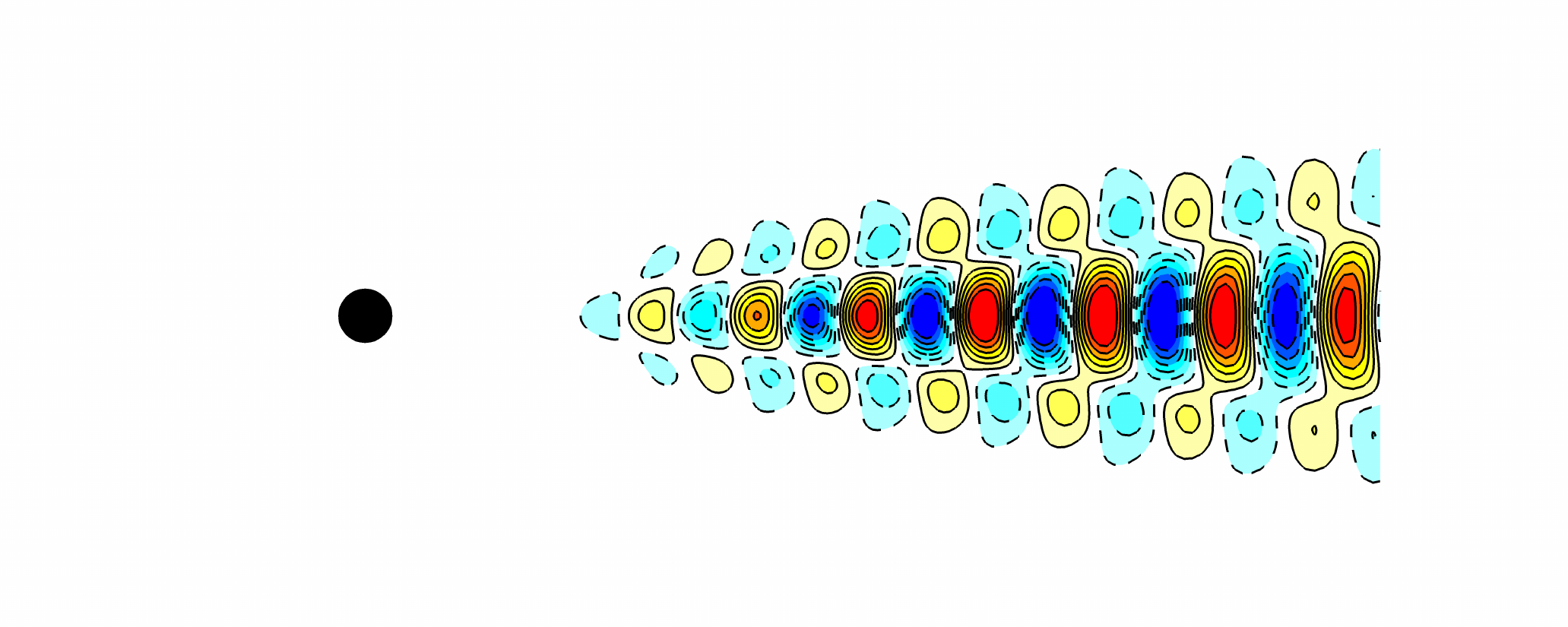}
\caption{$\sigma=-0.046$}
\end{subfigure}
\caption{DMD modes capture the limit cycle solution at finial stage. (a) The first column modes capture the periodic wake after cylinder. (b) The second column modes capture the least stable Floquet mode. (c) The third column modes capture the high order derived modes from the least stable Floquet mode.}\label{fig:2dfinalmodes}
\end{figure}

DMD modes of the asymptotic periodic flow are shown in figure~\ref{fig:2dfinalmodes}. Fourier modes of the periodic base flow, the Least stable Floquet modes, and the high-order derived mode of the least stable Floquet mode are shown in the first, second, and third columns, respectively. Note that wavenumber-frequency relation also holds among these modes.

\subsection{Dynamics of nonlinear transition via Koopman decomposition} 
\label{sec:wholeprimary}


After showing the Koopman modes at two asymptotic stages, we now focus on the goal of analyzing a nonlinear transition process. For that purpose, let us recall the three characteristics of Koopman decomposition from part 1. Firstly, the \emph{proliferation rule} recursively proliferates the spectrums, eigenfunctions, and modes to infinite dimension because of nonlinear interaction. The spectrums are orderly organized, for example, some possible distributions are shown in figure~\ref{fig:selfinteract}. Secondly, from operator perturbation theory, a system everywhere continuous differentiable has \emph{continuous Koopman spectrums}. The continuity extends the local Koopman decomposition to the global domain. As a result, the Koopman modes are \emph{invariant}, which is the third important property of Koopman decomposition. The three properties of Koopman decomposition are the major tools to study the dynamics of complicated nonlinear systems.

\subsubsection{Fourier expansion for periodic dynamics}
\label{chap:formationfloquetmode}

For a LTI system, periodic solution results from a center equilibrium point, who has spectrums $\pm j \omega$ and other spectrums are stable, here $j = \sqrt{-1}$. However, for nonlinear dynamics, the periodic solution is usually not the result of a center point. Therefore, it might be interesting to study a nonlinear periodic process. One of the examples is the mentioned Hopf bifurcation process of flow past cylinder. 

\begin{figure}
\centering
\begin{subfigure}[b]{0.325\linewidth}
\begin{overpic}[width=1.0\textwidth]{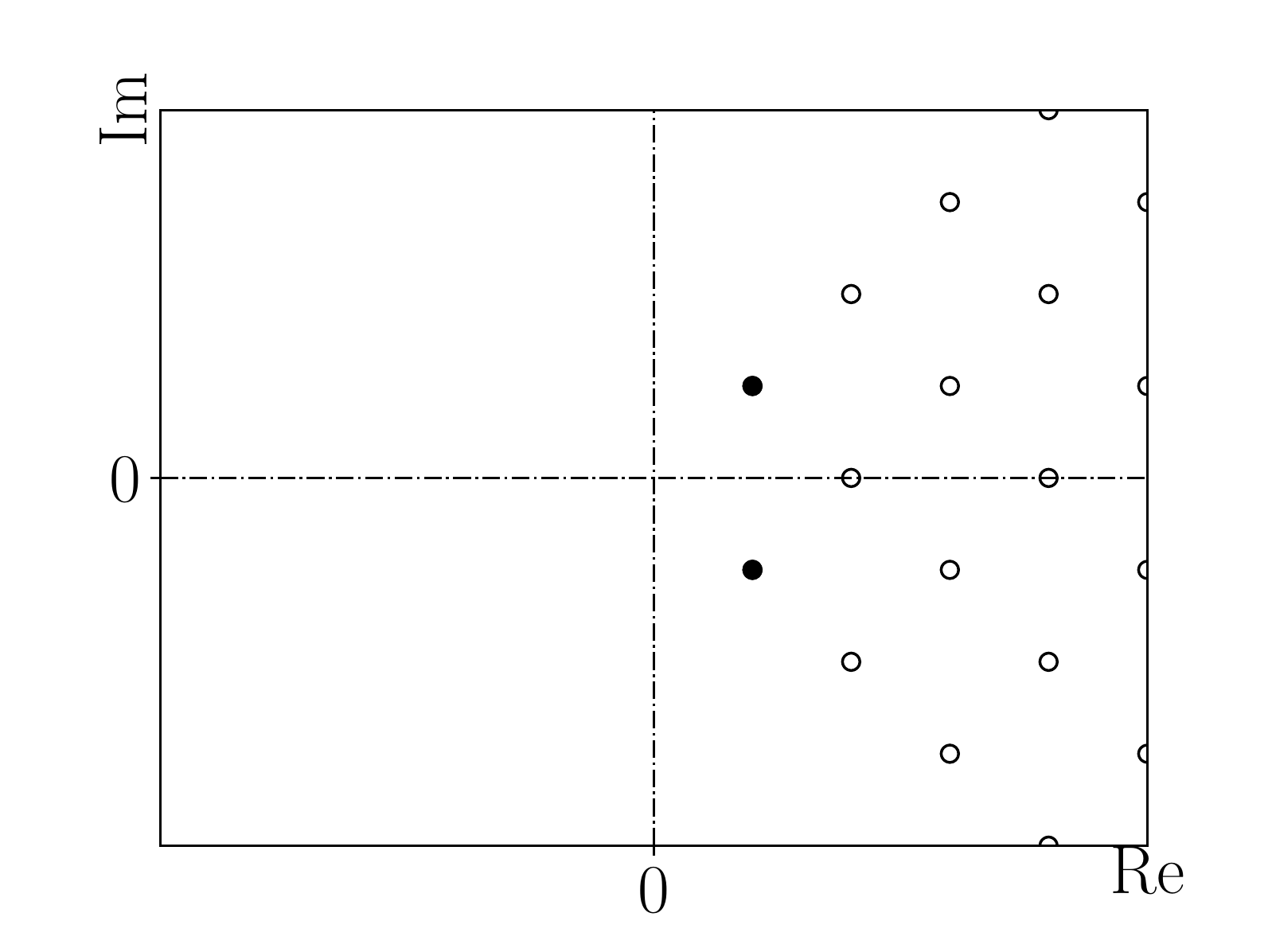} 
\put(72,56){\color{red}\vector(-1,0){10}}
\put(72,37){\color{red}\vector(-1,0){10}}
\put(70,3){$\sigma$}
\put(5,55){$\omega$}
\end{overpic}
\caption{$t_{init}$}
\end{subfigure}
\begin{subfigure}[b]{0.325\linewidth}
\begin{overpic}[width=1.0\textwidth]{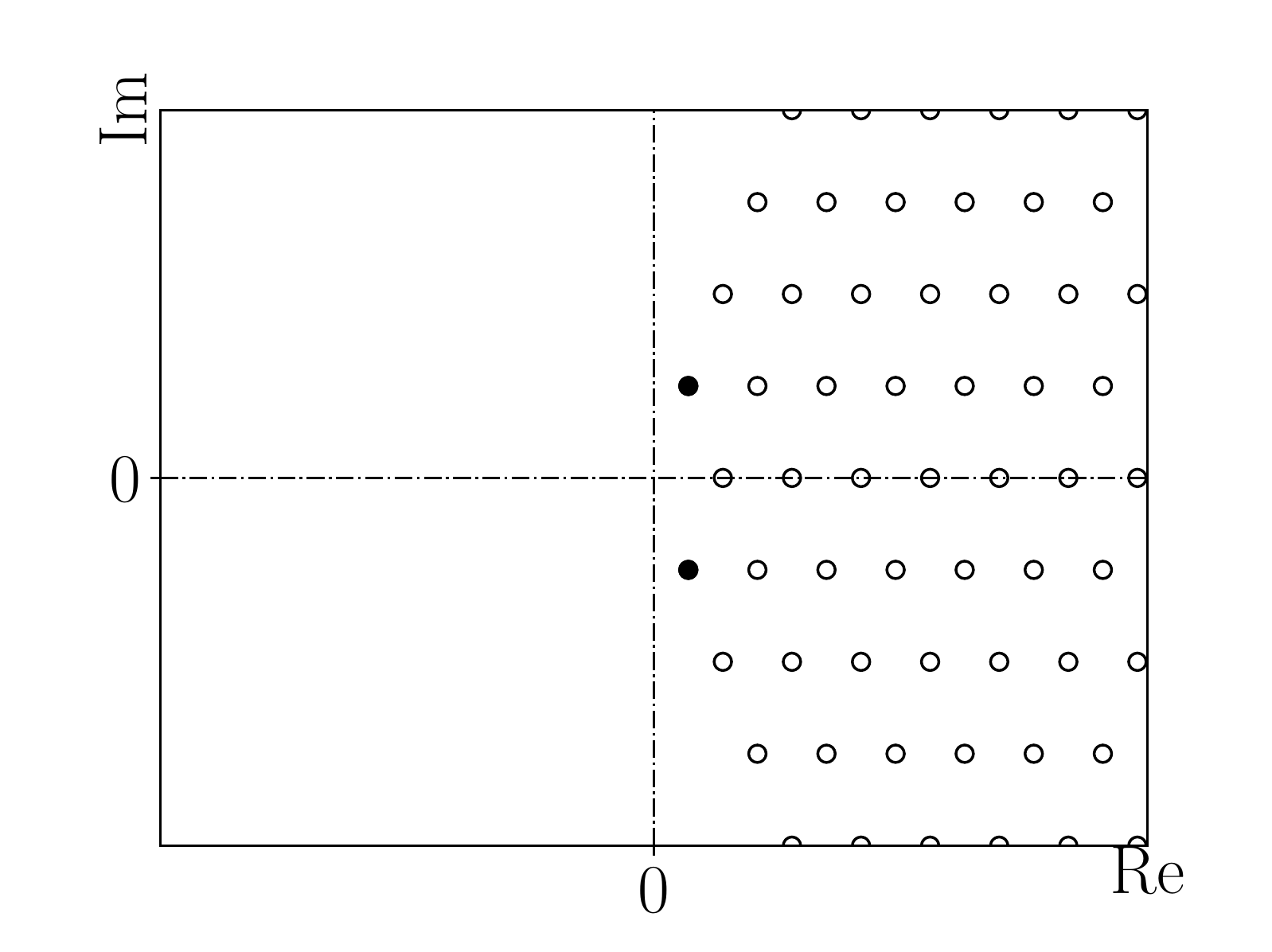} 
\put(66,56){\color{red}\vector(-1,0){3}}
\put(66,37){\color{red}\vector(-1,0){3}}
\put(70,3){$\sigma$}
\put(5,55){$\omega$}
\end{overpic}
\caption{$t_{mid}$}
\end{subfigure}
\begin{subfigure}[b]{0.325\linewidth}
\begin{overpic}[width=1.0\textwidth]{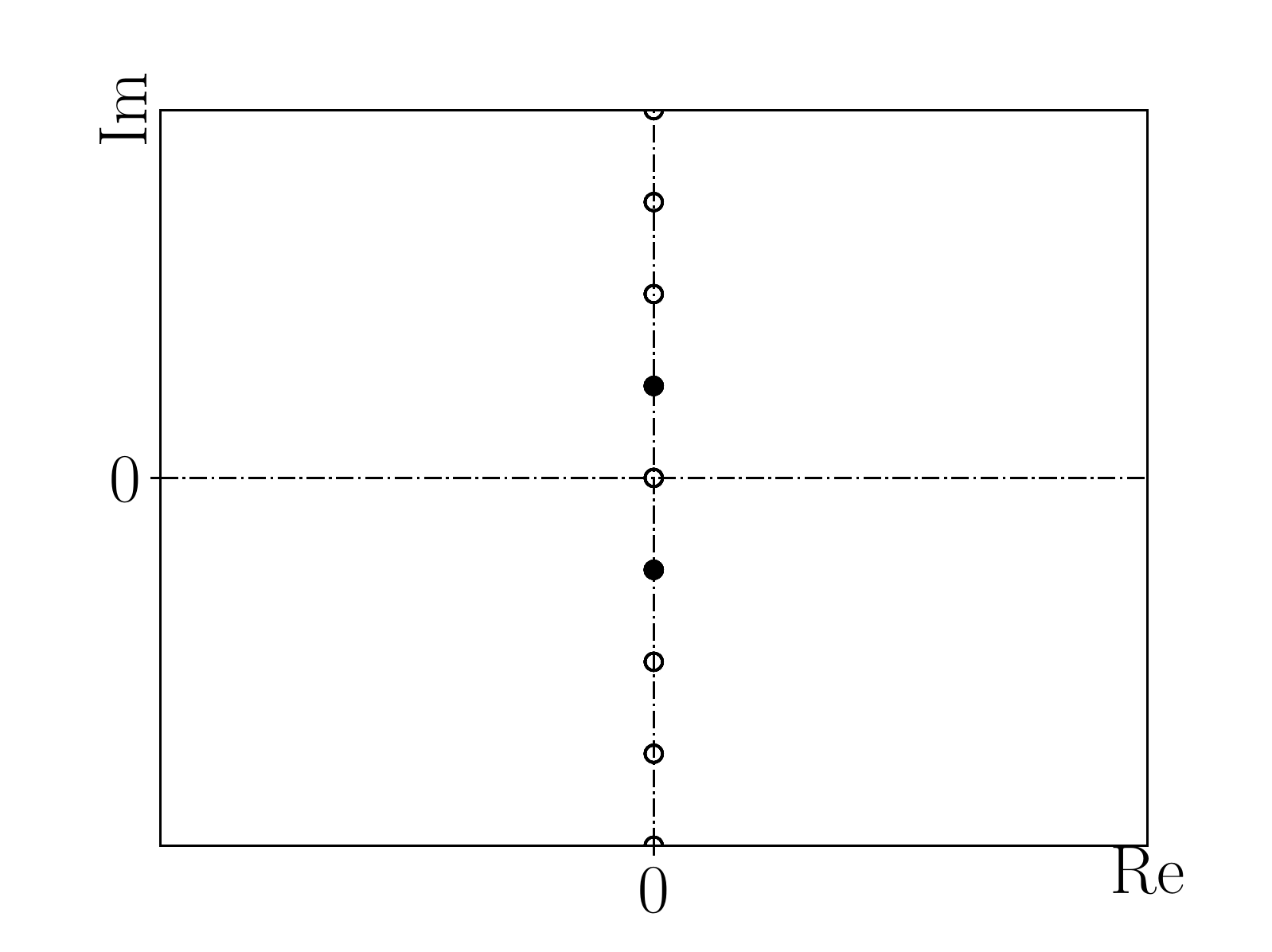} 
\put(70,3){$\sigma$}
\put(5,55){$\omega$}
\end{overpic}
\caption{$t_{final}$}
\end{subfigure}
\caption{Koopman spectrums move to imaginary axis when growth rate decreases to 0 (saturation). (a) (b) (c) the Koopman spectrums at three time instance. (c) shows infinite number of Koopman spectrums fall on the imaginary axis, producing periodic solution.} \label{fig:saturation}
\end{figure}
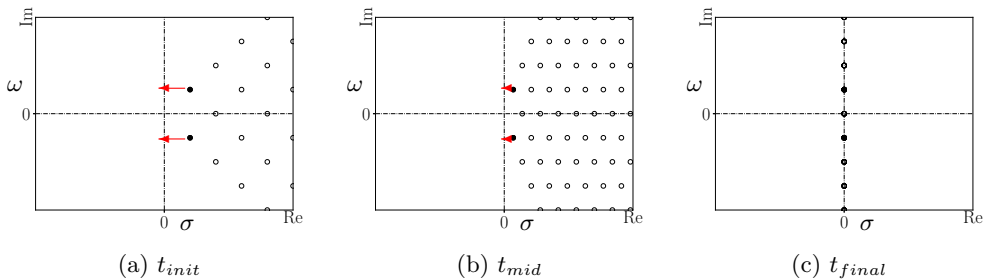

Figure~\ref{fig:spectruminit} shows the critical Koopman spectrums $\lambda=0.0136\pm 0.751j$ and their derived spectrums are unstable at the initial stage. The system then slides away from the unstable state until it reaches some kind of stable trajectory. Here it is the stable periodic solution, whose critical Koopman spectrums have zero growth rate and all other spectrums are stable, see figure~\ref{fig:spectrumfinal}. During the nonlinear transition process, the movement of Koopman spectrums is sketched in figure~\ref{fig:saturation}. As the growth rates of the critical Koopman modes decrease, all these Koopman spectrums move to the imaginary axis simultaneously, keeping the triad-chain formation. As they asymptotically fall on the imaginary axis, periodic solution is obtained.

It is well known that periodic dynamics can be expanded by Fourier series. It can be explained by Koopman decomposition using the Hopf bifurcation case. As all critical Koopman spectrums fall on the imaginary axis as shown by figure~\ref{fig:saturation}, there are countably infinite number of Koopman spectrums fall on each $\pm jn\omega$, resulting in enhanced dynamics with pulsation $e^{\pm jn\omega t}$. Note these are Fourier frequencies as well, therefore, Fourier expansion is obtained. It is interesting to notice that these high-frequencies are multiples of the base frequency $\omega$. The fact that Fourier frequencies are integer multiple of base frequency also conforms to the proliferation rule.

\begin{figure}
\centering
\includegraphics[width=0.9\textwidth, trim={100 498 200 65}, clip]{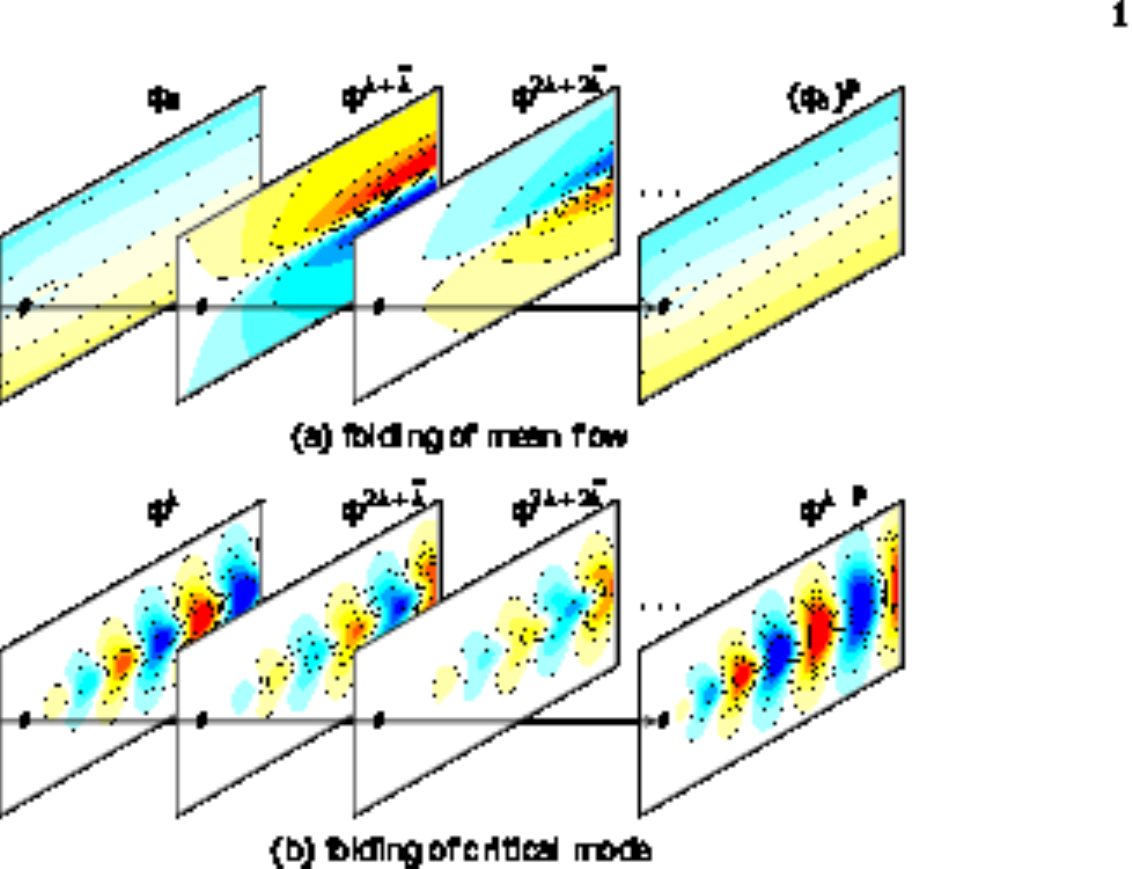}
\caption{Folding of Koopman modes as the flow reaches saturation.} \label{fig:modulation}
\end{figure}

Fourier modes are therefore generated by the superposition of Koopman modes. As flow reaches periodic, the growth rates of critical Koopman modes decrease to 0. Modes with spectrums $m\sigma + j n \omega$, $m=1, \cdots,$ can not be distinguished since they have the same dynamics $e^{j n \omega t}$ ($\sigma\rightarrow 0$). As a result, these infinite-dimensional Koopman modes with the same frequency superimpose on top of each other creating a `new' mode, or the so-called Fourier mode. This folding process is illustrated in figure~\ref{fig:modulation}.

There was guess that saturation of Hopf bifurcation was as the result of the `nonlinear modulation' of mean flow~\citep{landau1959course,sipp2007global}. From current study, the magic `nonlinear modulation' effect is as simple as superposition of Koopman modes
\begin{equation} \label{eqn:formationmean}
\left(\phi_0\right)^p = \phi_0 + \epsilon \phi^{\lambda + \bar{\lambda}} + \epsilon^2 \phi^{2\lambda+2\bar{\lambda}} + \cdots + \epsilon^n \phi^{n\lambda+n\bar{\lambda}} + \cdots.
\end{equation}
where $\left(\phi_0\right)^p$ is the mean flow at the final periodic stage. $\phi_0$ is the initial base flow, $\phi^{\lambda + \bar{\lambda}}$, $\phi^{2\lambda + 2\bar{\lambda}},\,\cdots$ are monotonic modes (with real spectrums), which are generated by the critical pair Koopman modes (with $\sigma \pm j\omega$) and high-order ones (with $2\sigma \pm j2\omega$). The $\epsilon$ terms are borrowed from GSA to represent the magnitude of these modes. The superposition of the base flow is illustrated in figure~\ref{fig:modulation}(a).

Similarly the modulation on the critical modes $\phi^{\lambda}$ is also the result of linear superposition of Koopman modes, which is illustrated by figure~\ref{fig:modulation}(b). By superimposing all the modes with frequency $\omega$, the final $ \left(\phi^{\lambda}\right)^p$ mode is obtained
\begin{equation}
 \left(\phi^{\lambda}\right)^p = \phi^{\lambda} + \epsilon \phi^{2\lambda + \bar{\lambda}} + \epsilon^2 \phi^{3\lambda+2\bar{\lambda}} + \cdots + \epsilon^n \phi^{(n+1)\lambda+n\bar{\lambda}} + \cdots.
\end{equation}

For the current Hopf bifurcation case showed above, as $\epsilon$ is on the order
\begin{equation}
\epsilon = \sqrt{\frac{Re-Re_c}{Re Re_c}} \ll \frac{1}{Re_c},
\end{equation}
see equation~\ref{eqn:stuartorder}. Koopman modes $\phi_0, \cdots, \phi^{n\lambda}$ play a dominant role in the final Fourier modes. Therefore Koopman modes $\phi^{\lambda}$, $\phi^{2\lambda}$, $\phi^{3\lambda}$ presented in figure~\ref{fig:2dinitialmodes} are similar to the corresponding Fourier modes showed in figure~\ref{fig:2dfinalmodesa} but with slight difference.

For convenience, we will call all modes in figure~\ref{fig:saturation} the \emph{critical Koopman modes}. The reason is twofold. Firstly, they are derived from the critical Koopman mode (the unstable normal mode~\citep{jackson1987finite}). Secondly, they dominate the flow and develop the final periodic flow.

\subsubsection{Floquet solution around limit cycle dynamics}

The solution of a periodic LTV system
\begin{equation}
\dot{x} = A(t) x, \quad A(t+T) = A(t)
\end{equation}
is given by Floquet theory~\citep{coddington1955theory}. Here $x\in \mathbb{R}^n, A(t) \in \mathbb{R}^{n\times n}, t, T \in \mathbb{R}$ and $T>0$. The solution is given by
\begin{equation}
x(t) = \sum_i e^{\mu_i t} f_i(t)
\end{equation}
where $\mu_i$ is a constant complex number called Floquet exponent, and $f_i(t)$ is a periodic function. Further expanding periodic $f_i(t)$ by Fourier series, time-invariant Koopman spectrums $\mu_i \pm jn\omega$ ($\omega = \frac{2\pi}{T}$ and $j = \sqrt{-1}$) and modes are obtained, see Appendix A in part 1. If there is any $\text{Real}(\mu_i)> 0$, the periodic solution is unstable. This theory is a valuable tool to study stability of the limit cycle solution of nonlinear dynamics via linearization. Among the above Floquet exponents, there is one with $\mu_i = 0$ providing the periodic base flow, whose formation is already studied in the previous section. The rest Floquet modes are generated similarly but with difference.

\begin{figure}
\begin{subfigure}[b]{1.0\linewidth}
\includegraphics[width=0.325\textwidth]{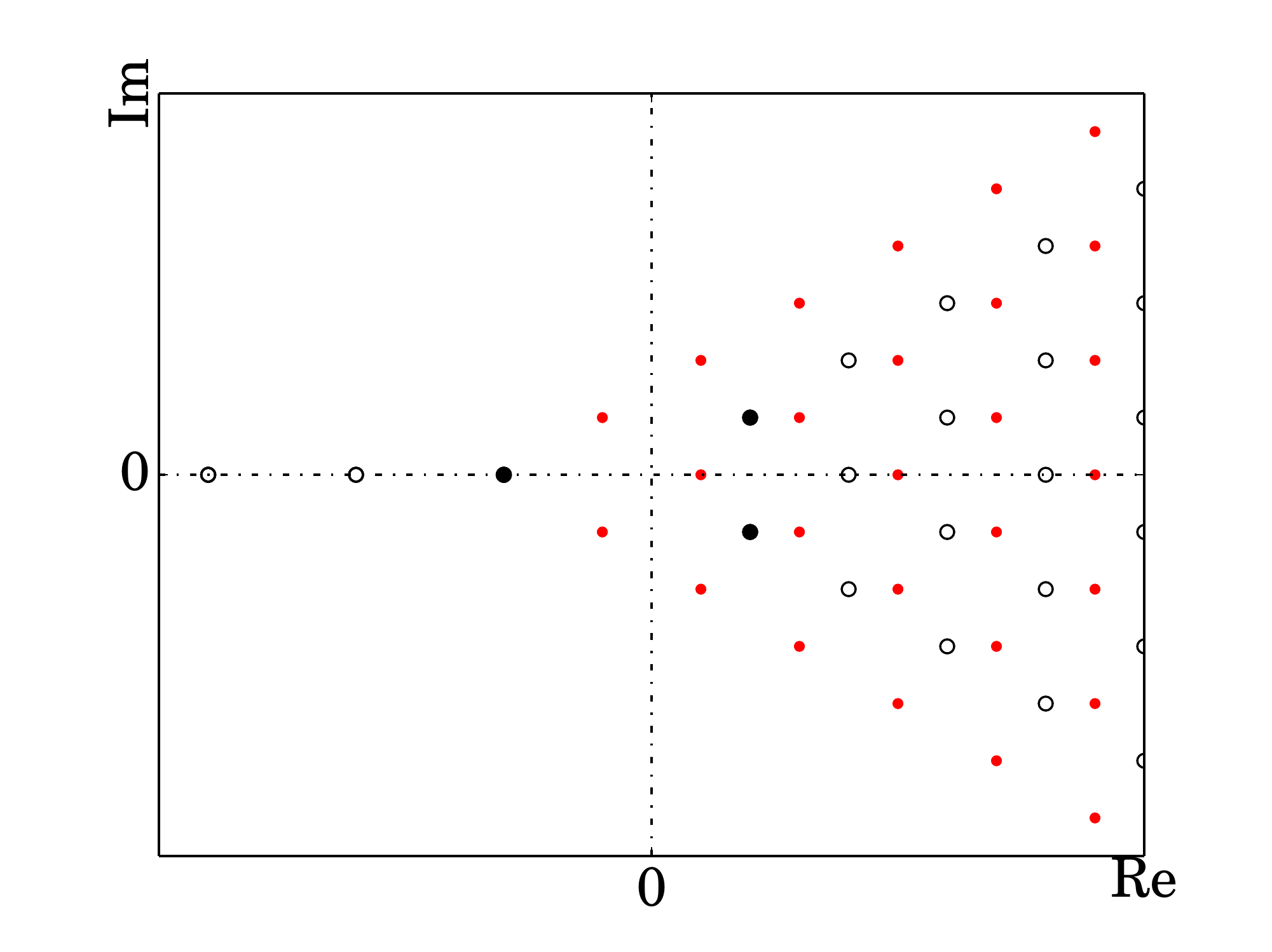}
\includegraphics[width=0.325\textwidth]{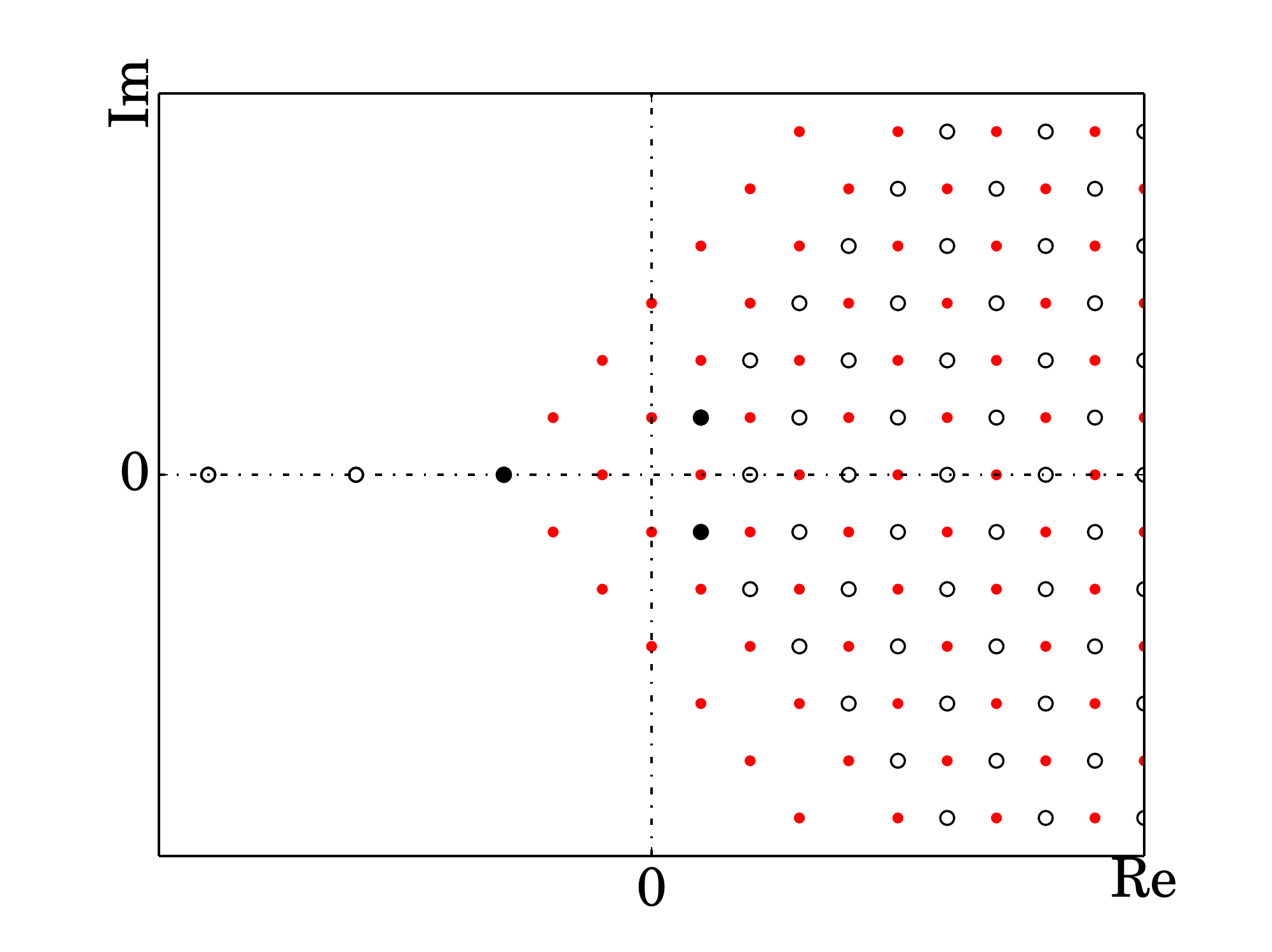}
\includegraphics[width=0.325\textwidth]{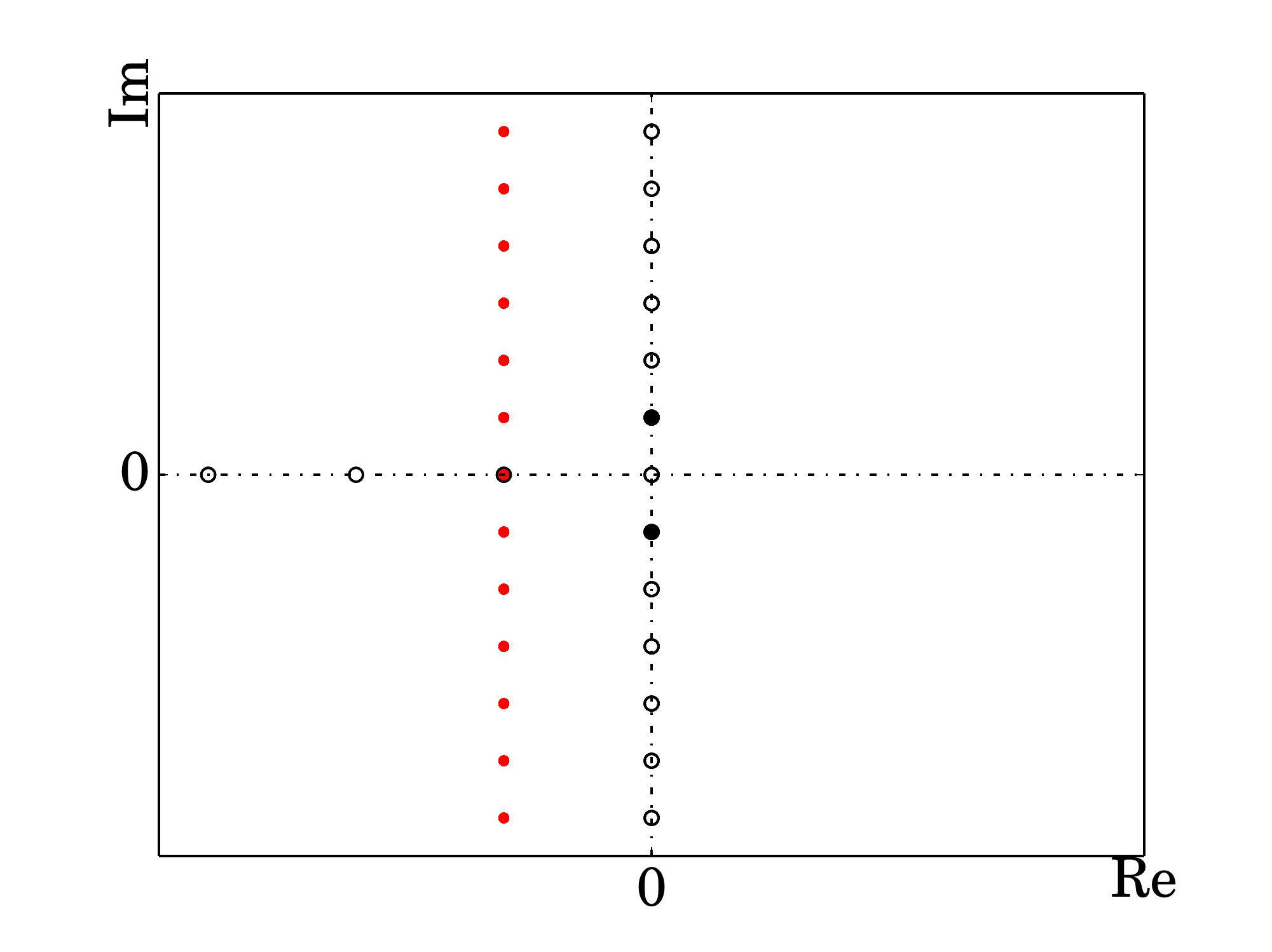}
\caption{case 1, generation of Floquet modes.}
\end{subfigure}
\begin{subfigure}[b]{1.0\linewidth}
\includegraphics[width=0.325\textwidth]{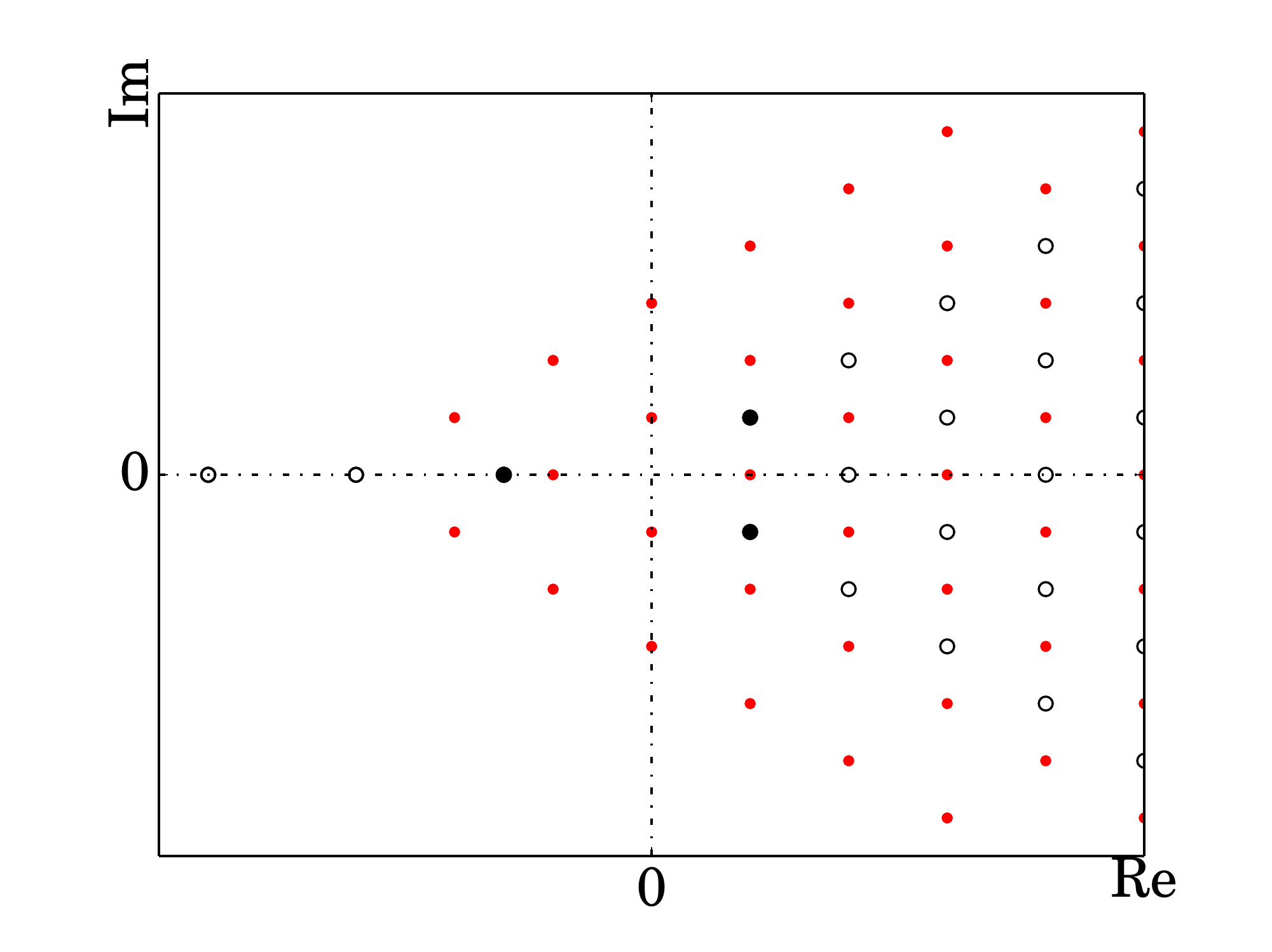}
\includegraphics[width=0.325\textwidth]{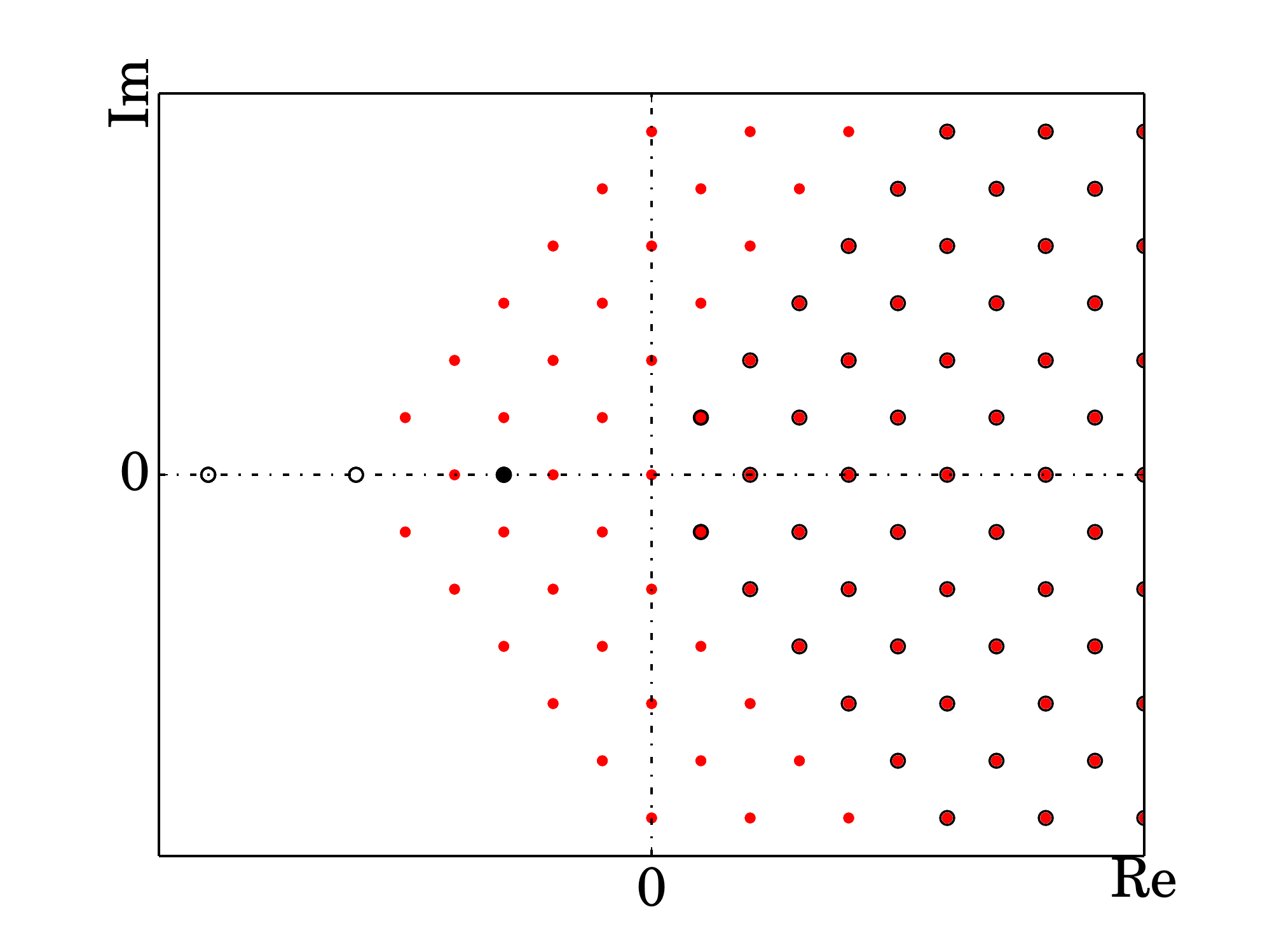}
\includegraphics[width=0.325\textwidth]{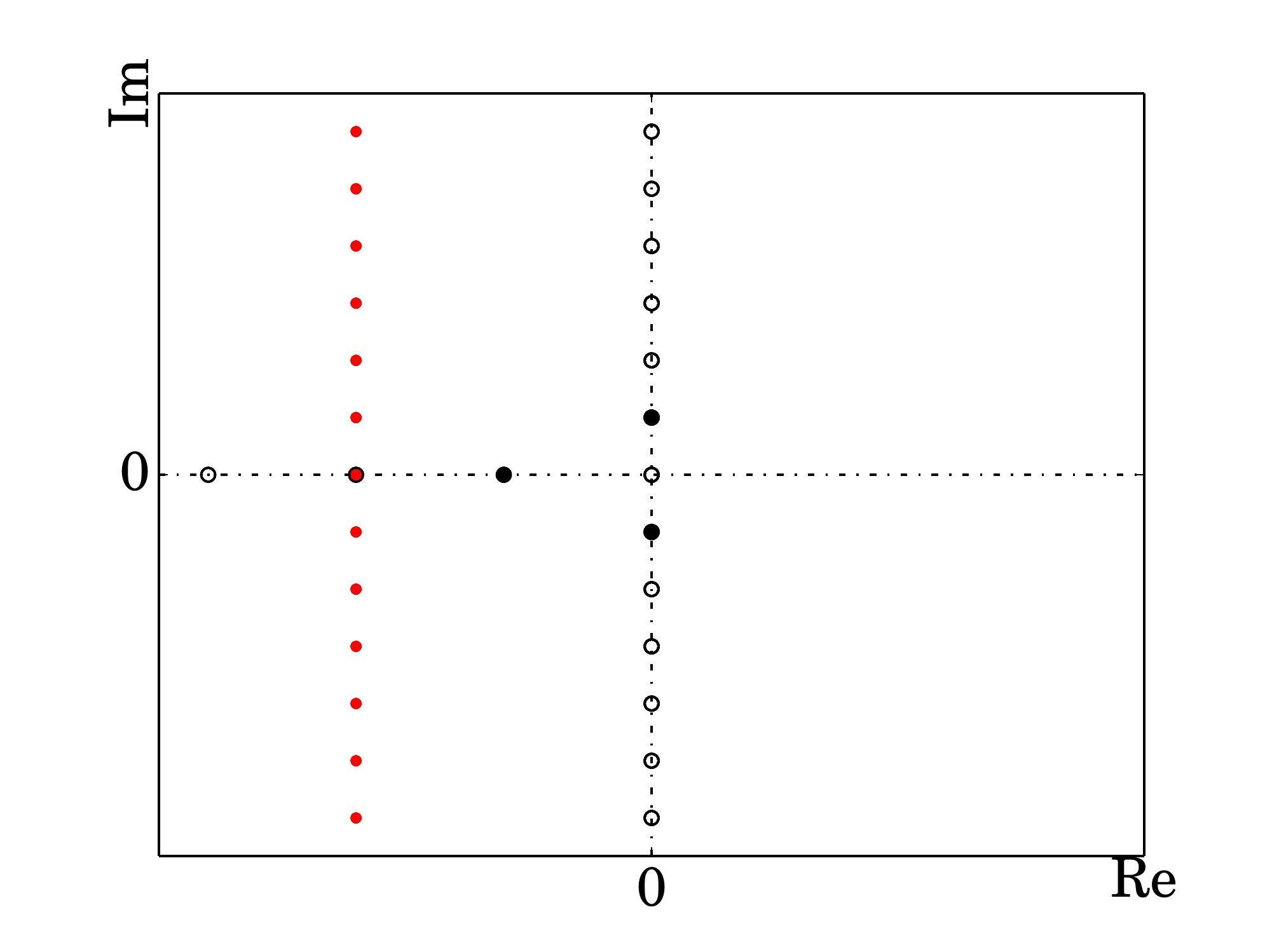}
\caption{case 2, generation of high-order derived Floquet modes.}
\end{subfigure}
\begin{subfigure}[b]{1.0\linewidth}
\includegraphics[width=0.325\textwidth]{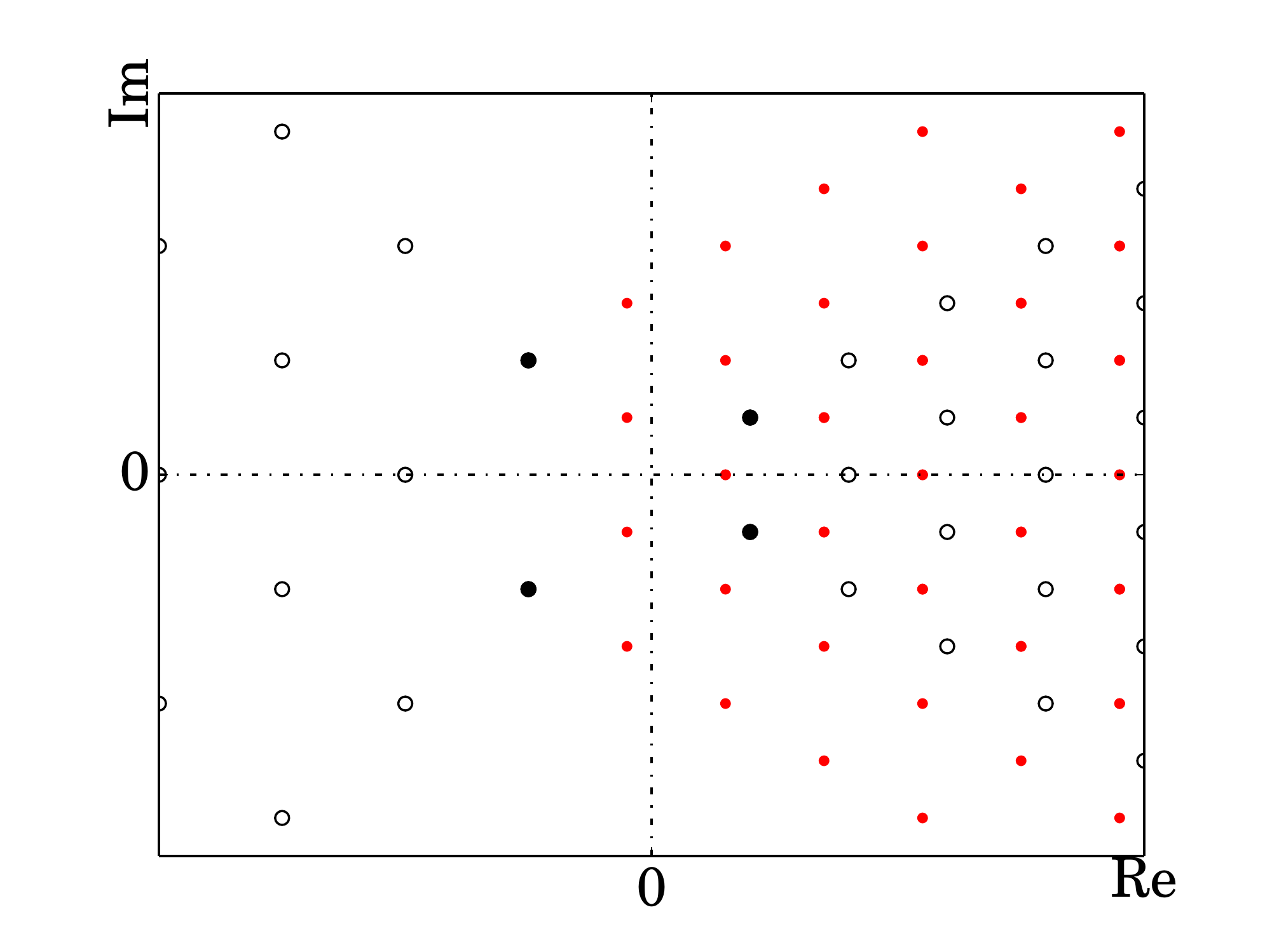}
\includegraphics[width=0.325\textwidth]{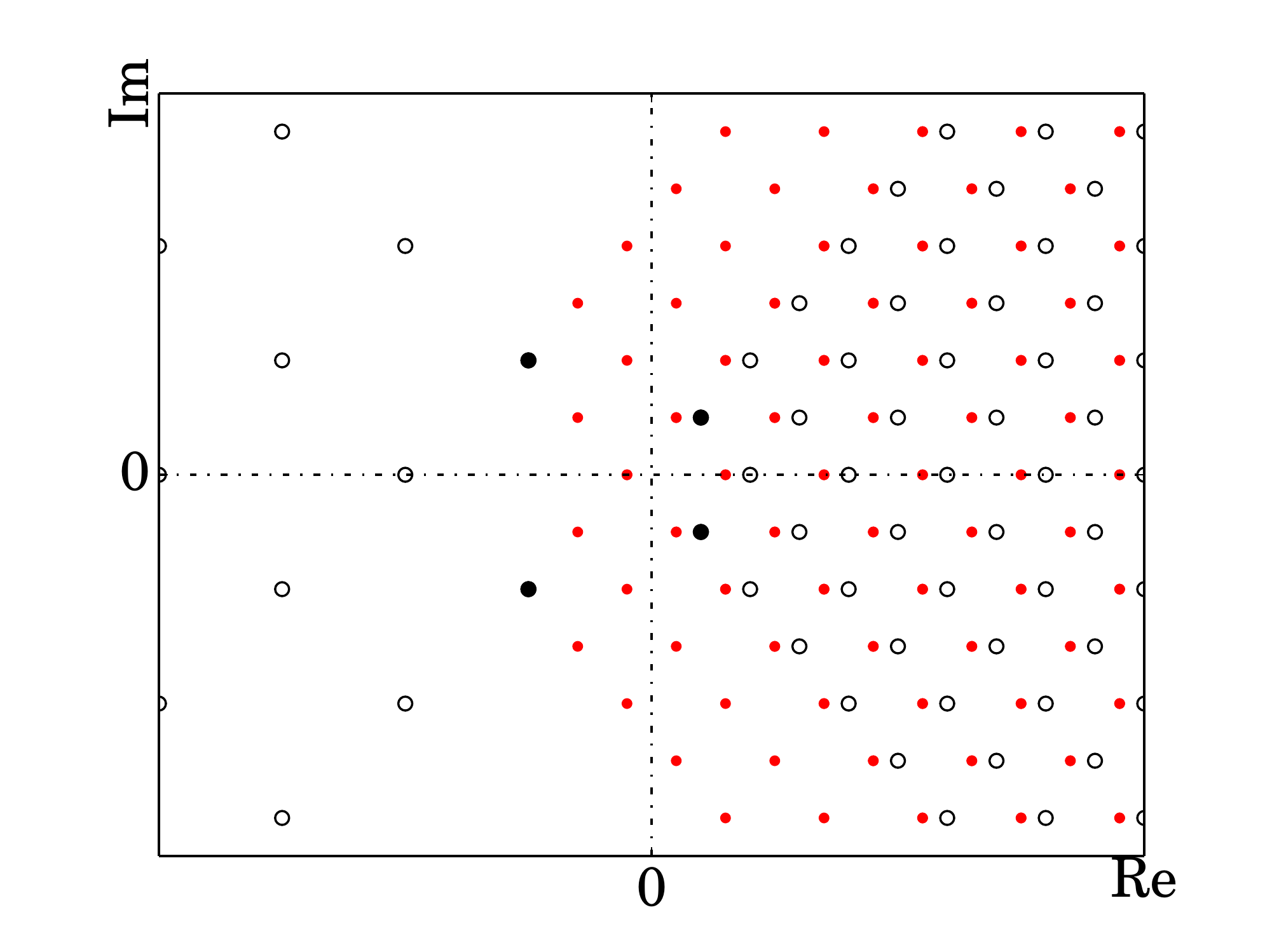}
\includegraphics[width=0.325\textwidth]{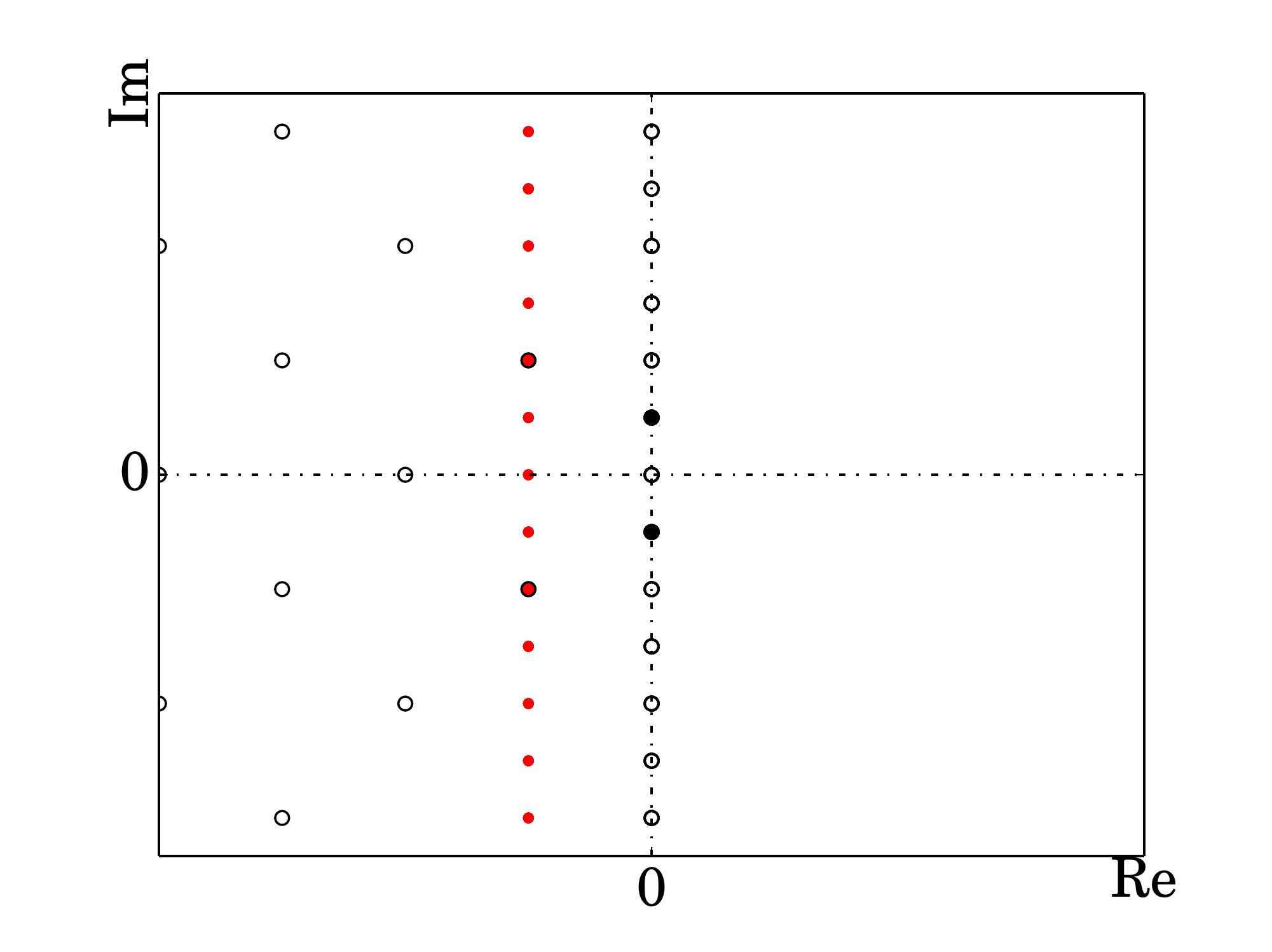}
\caption{case 3, generation of Floquet modes.}
\end{subfigure}
\begin{subfigure}[b]{1.0\linewidth}
\includegraphics[width=0.325\textwidth]{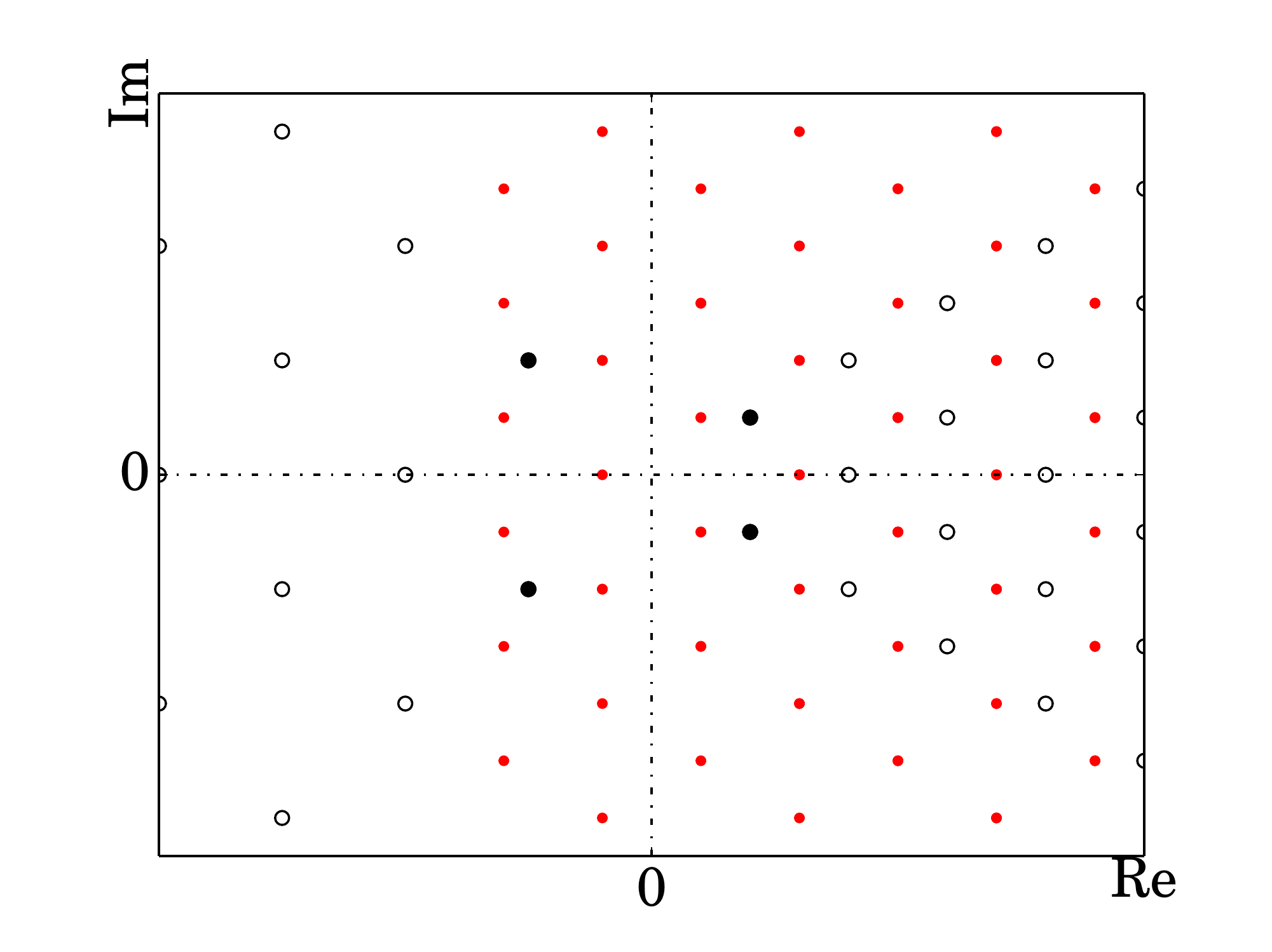}
\includegraphics[width=0.325\textwidth]{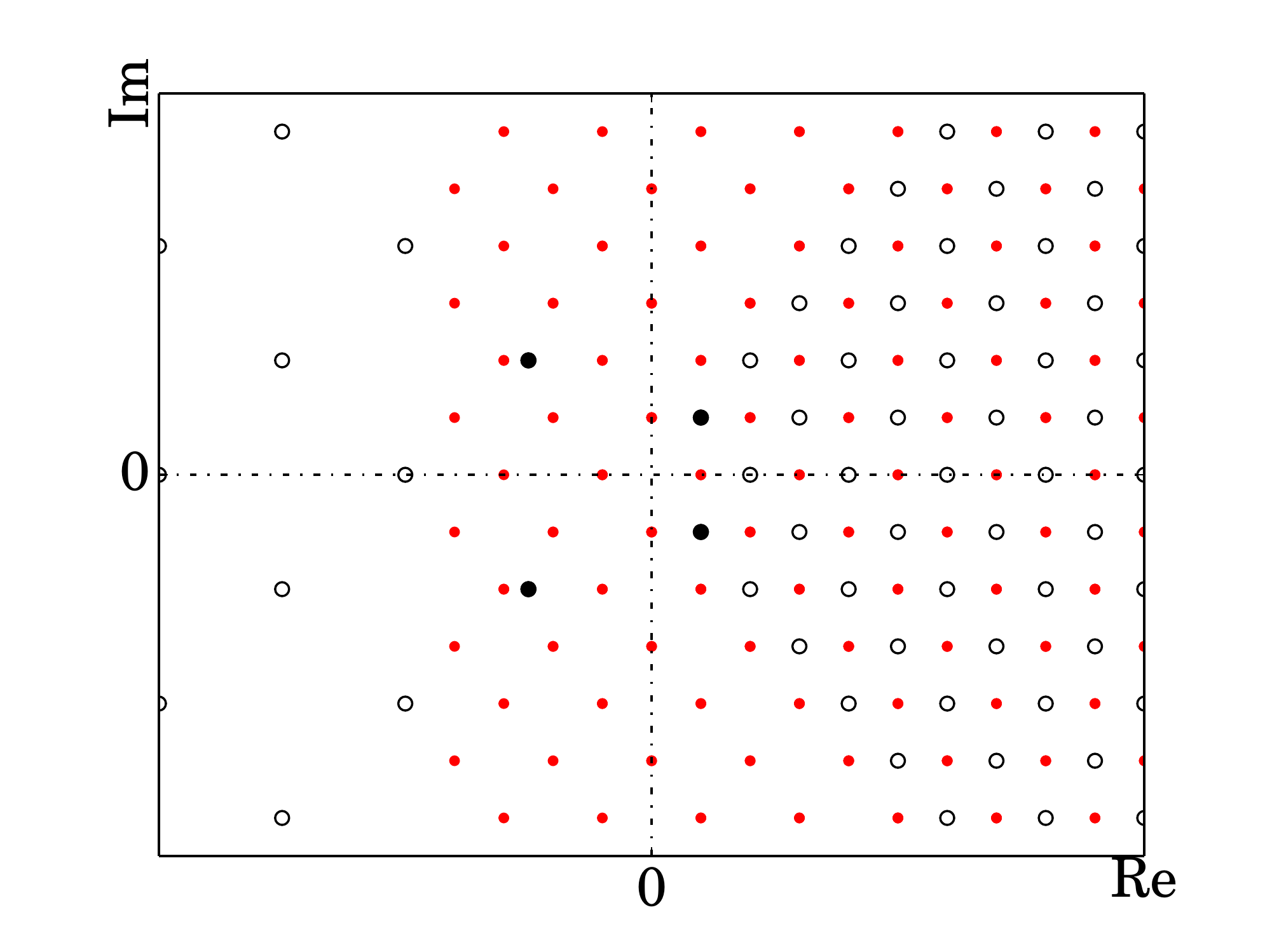}
\includegraphics[width=0.325\textwidth]{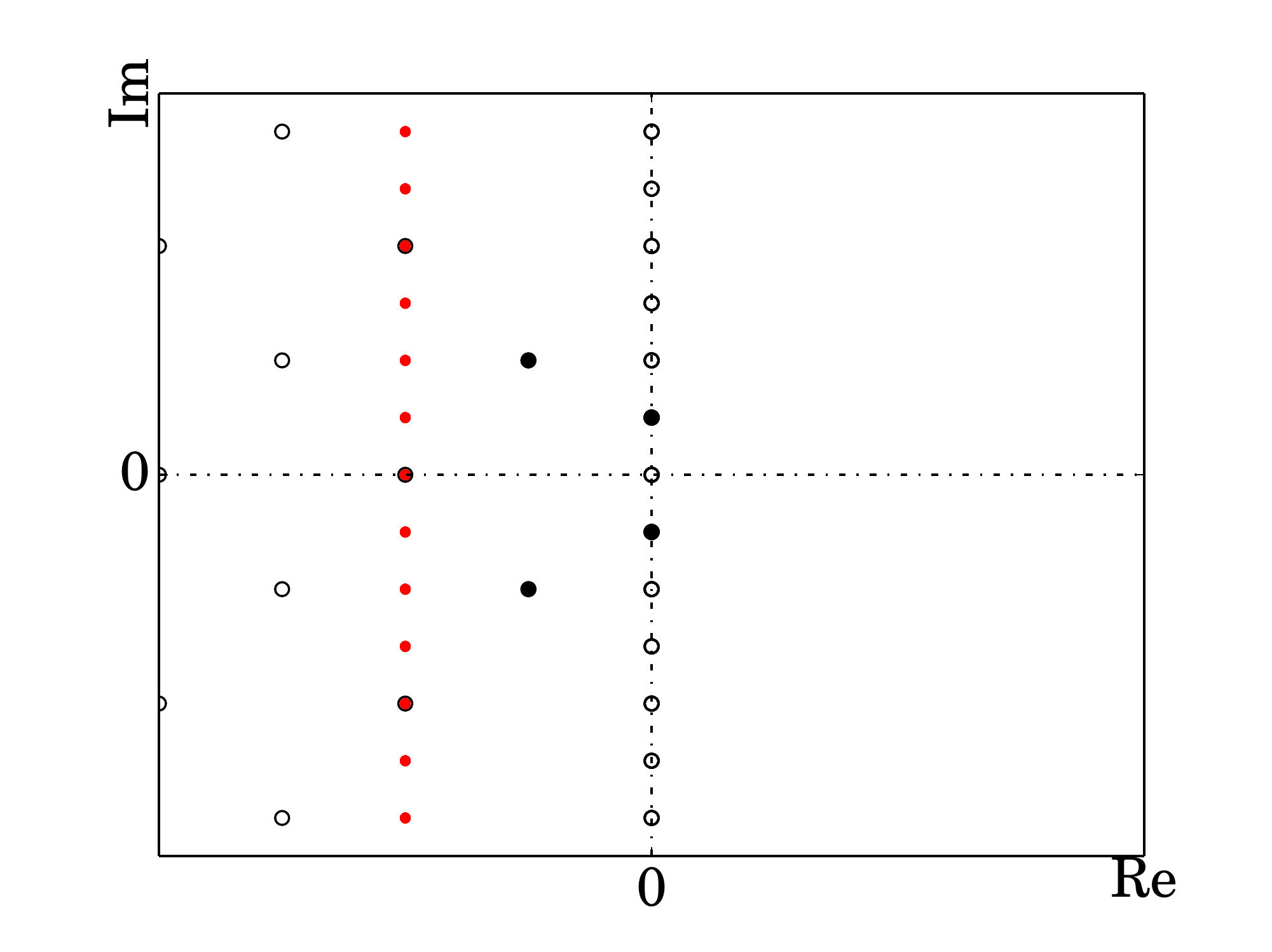}
\caption{case 4, generation of high-order derived Floquet modes.}
\end{subfigure}
\caption{4 Cases to show folding of Koopman modes to form the Floquet modes as the growth rate decrease to 0.}\label{fig:floquetformation}
\end{figure}

The Floquet modes are the cross interaction of the critical Koopman modes shown in figure~\ref{fig:saturation} with some other Koopman mode/modes. This can be illustrated by figure~\ref{fig:floquetformation}. Four cases generating Floquet modes are studied. Case 1 and case 3 show a real and a pair of complex conjugate mode/modes cross interaction with the critical Koopman modes. Case 2 and 4 studied the high-order derived modes cross interaction with the critical Koopman modes. These cases are summarized in Table~\ref{tab:floquetcases}.
\begin{table}
\centering
\caption{Cases to generate Floquet modes.} \label{tab:floquetcases}
\begin{tabular}{cccc}
$\lambda_1$ & $\lambda_2$ & Floquet modes & high-order derived modes \\
$0.2\pm 0.15j$ & $-0.3\pm 0j$  & case 1  & case 2     \\
$0.2\pm 0.15j$ & $-0.25\pm 0.3j$ & case 3  & case 4      
\end{tabular}
\end{table}

For detail, case 1 show the interaction between critical Koopman modes generated by $\lambda_1 = 0.2\pm 0.15j$ and spectrum $\lambda_2=-0.3$. Red dots (small solid dots) are the cross interaction spectrums between $-0.3$ and the critical Koopman modes (big hollow dots). As the growth rate of $\lambda_1$ decreases from 0.2 to 0, the spectrums of critical Koopman modes fall on the imaginary axis. At the same time, the cross spectrums fall on the line $\sigma=-0.3$, resulting in the Floquet spectrums. The infinite-dimensional Koopman modes at the same frequency are superimposed on top of each other, creating the least stable Floquet mode.

It is known that nonlinearity will proliferate the linear Floquet modes to high-order ones. This is illustrated by case 2 in figure~\ref{fig:floquetformation}. The red dots show the interaction of $-0.6$ (-0.3+-0.3) with the critical Koopman modes generated by $\lambda_1$. As the growth rate of $\lambda_1$ decreases to 0, all cross spectrums fall on the line $\sigma=-0.6$, generating high-order Koopman modes of the Floquet mode in case 1.

Similarly, the cross interaction of a complex conjugate pair with the critical Koopman modes generating the Floquet modes and their high-order derived modes are discussed in case 3 and 4, illustrated in figure~\ref{fig:floquetformation}(c) and (d).

\subsubsection{The resonance phenomena}

Another inspiring phenomenon is the frequencies of Floquet modes picked by the residue criterion~\citep{zhang2019solving} are multiples of the base frequency, see figure~\ref{fig:spectrumfinal}. This phenomenon is not answered by the Floquet theory, where $\mu_i$ can be any complex number.

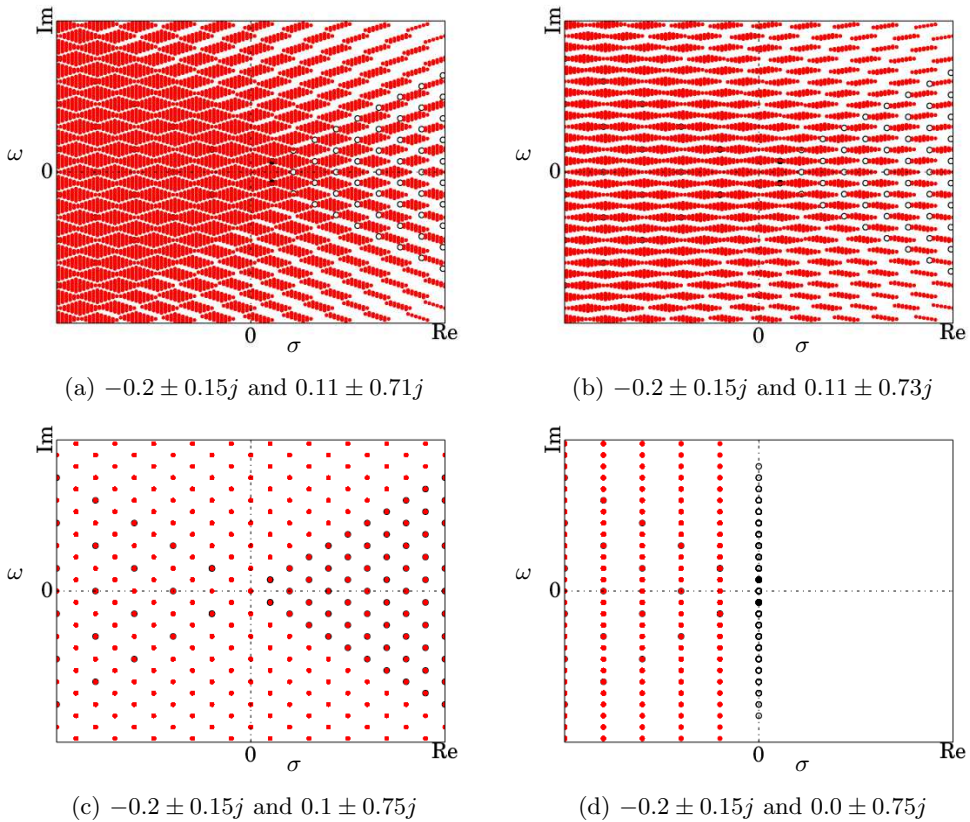
\begin{figure}
\centering
\begin{subfigure}[b]{0.49\linewidth}
\begin{overpic}[width=1.0\textwidth]{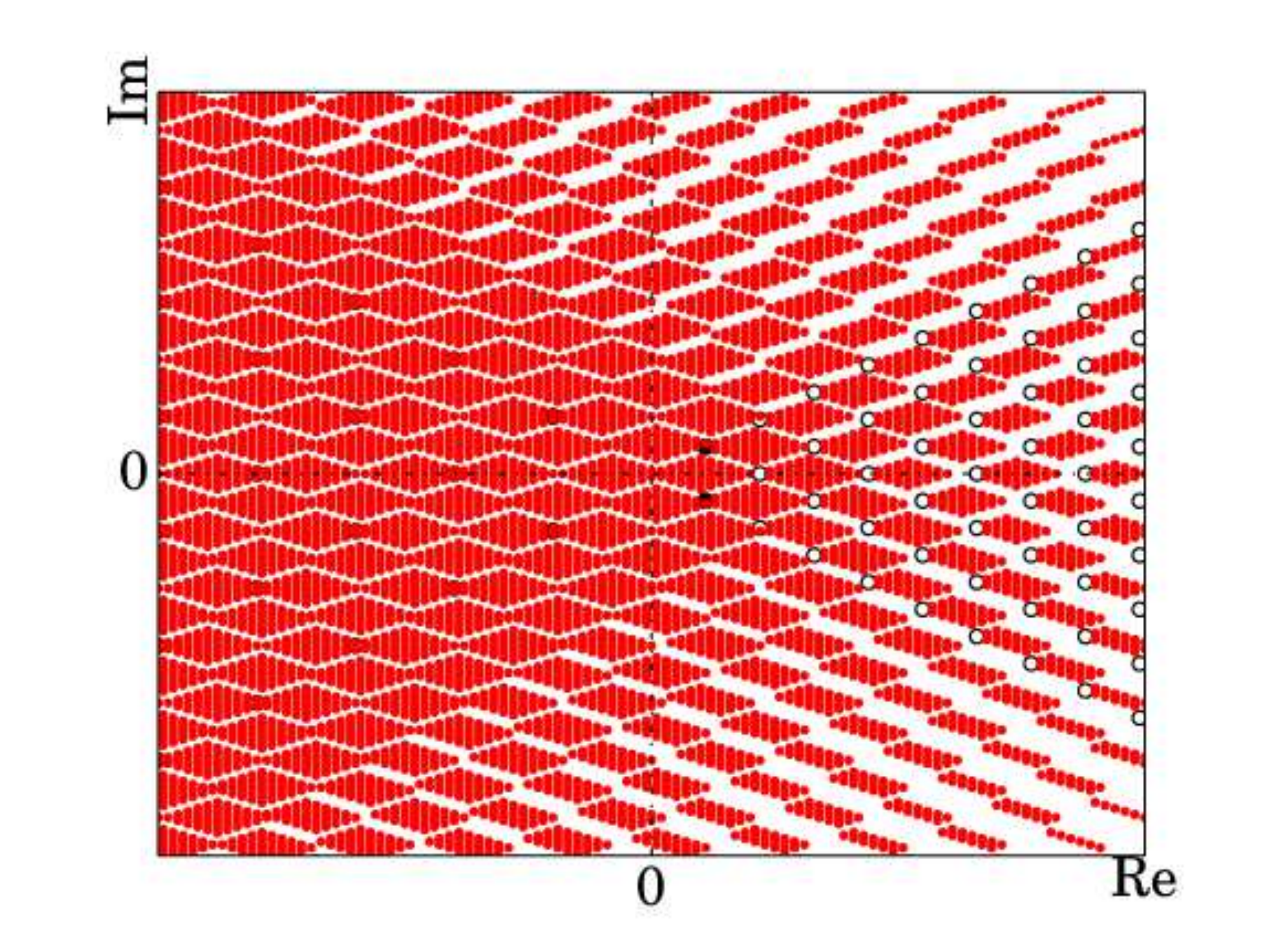} 
\put(110,3){$\sigma$}
\put(5,75){$\omega$}
\end{overpic}
\caption{$-0.2\pm 0.15j$ and $0.11\pm 0.71j$} \label{fig:resonancea}
\end{subfigure}
\begin{subfigure}[b]{0.49\linewidth}
\begin{overpic}[width=1.0\textwidth]{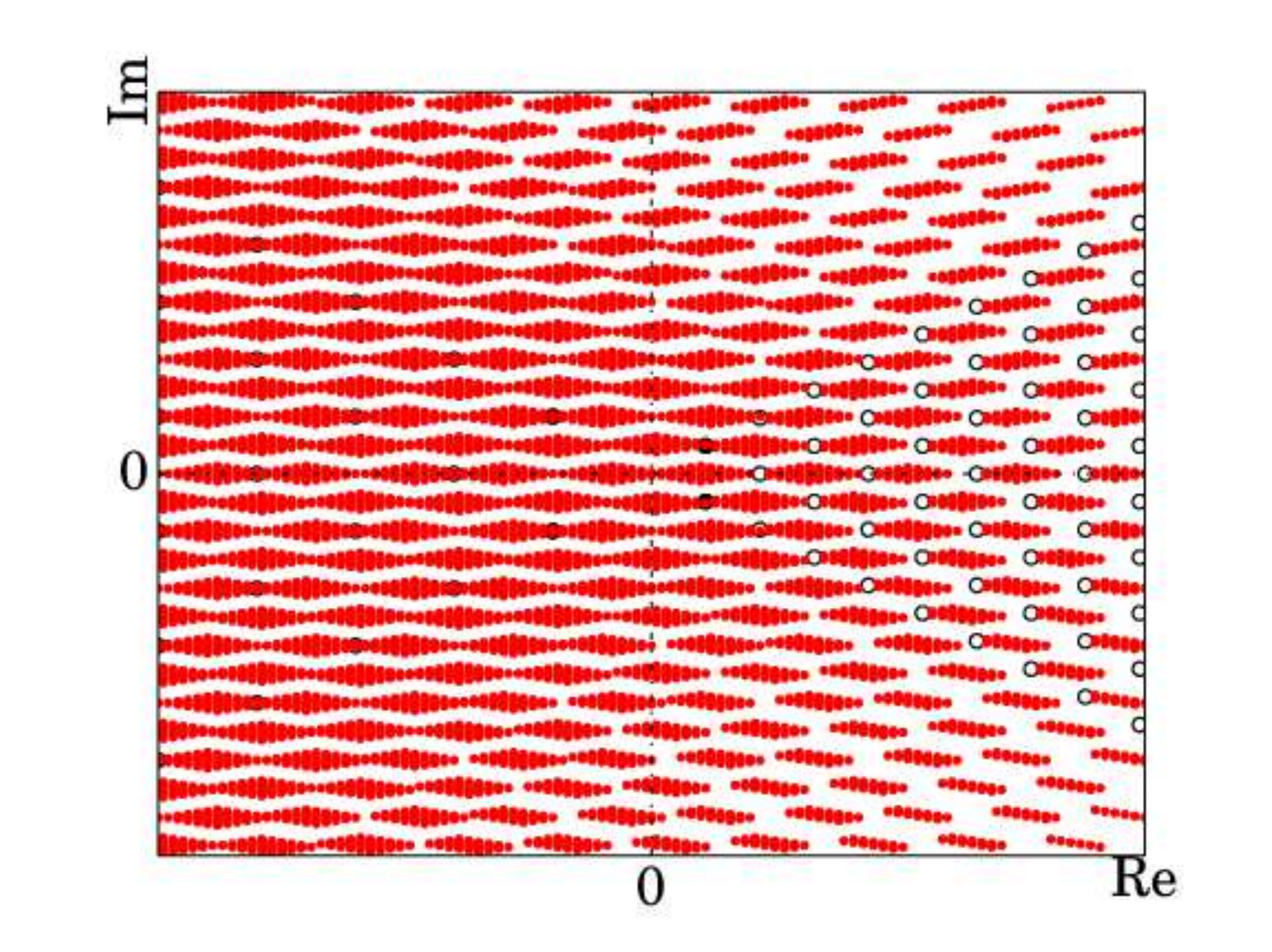} 
\put(110,3){$\sigma$}
\put(5,75){$\omega$}
\end{overpic}
\caption{$-0.2 \pm 0.15j$ and $0.11\pm 0.73j$} \label{fig:resonanceb}
\end{subfigure} \\
\begin{subfigure}[b]{0.49\linewidth}
\begin{overpic}[width=1.0\textwidth]{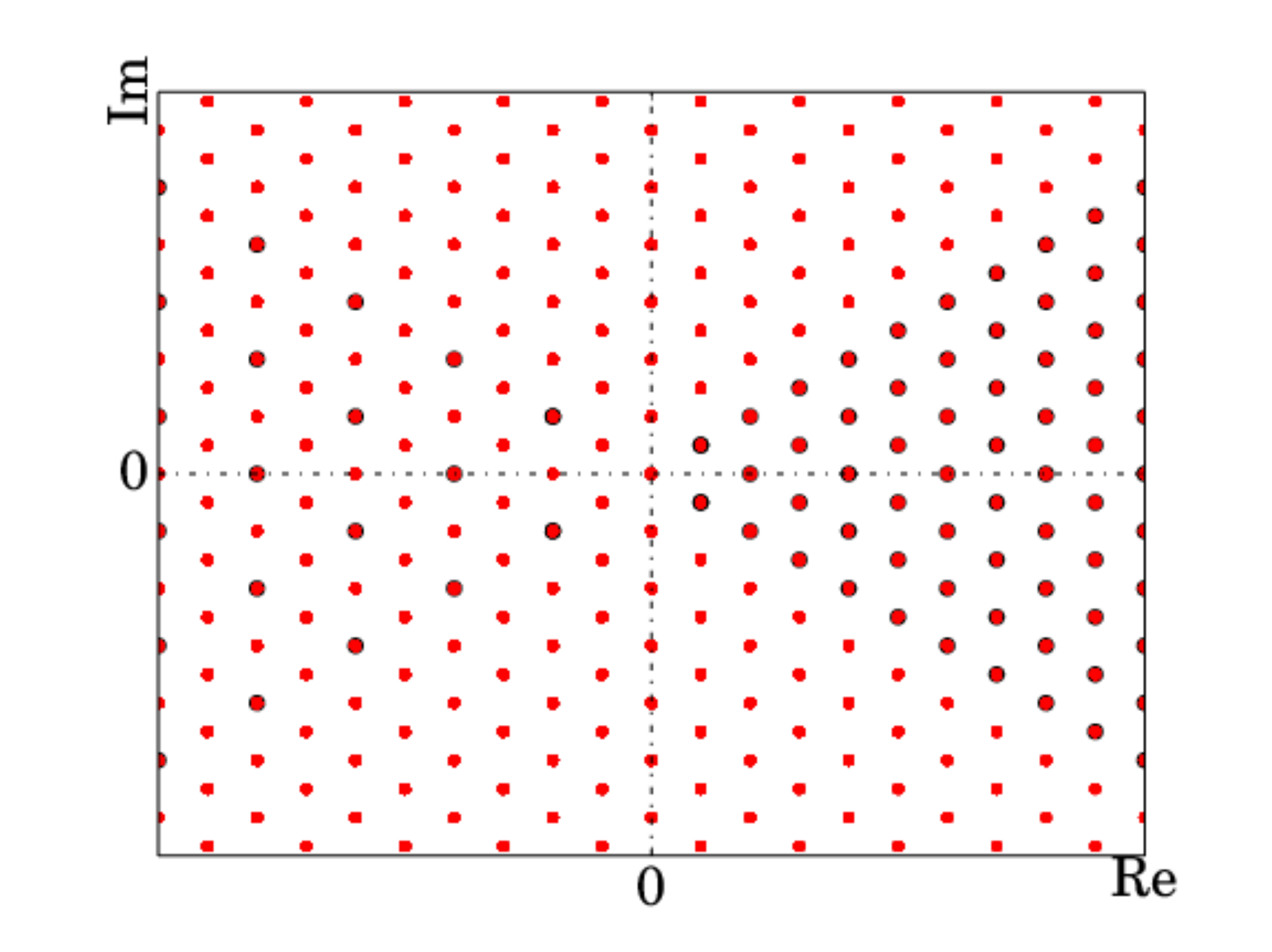} 
\put(110,3){$\sigma$}
\put(5,75){$\omega$}
\end{overpic}
\caption{$-0.2 \pm 0.15j$ and $0.1\pm 0.75j$} \label{fig:resonancec}
\end{subfigure}
\begin{subfigure}[b]{0.49\linewidth}
\begin{overpic}[width=1.0\textwidth]{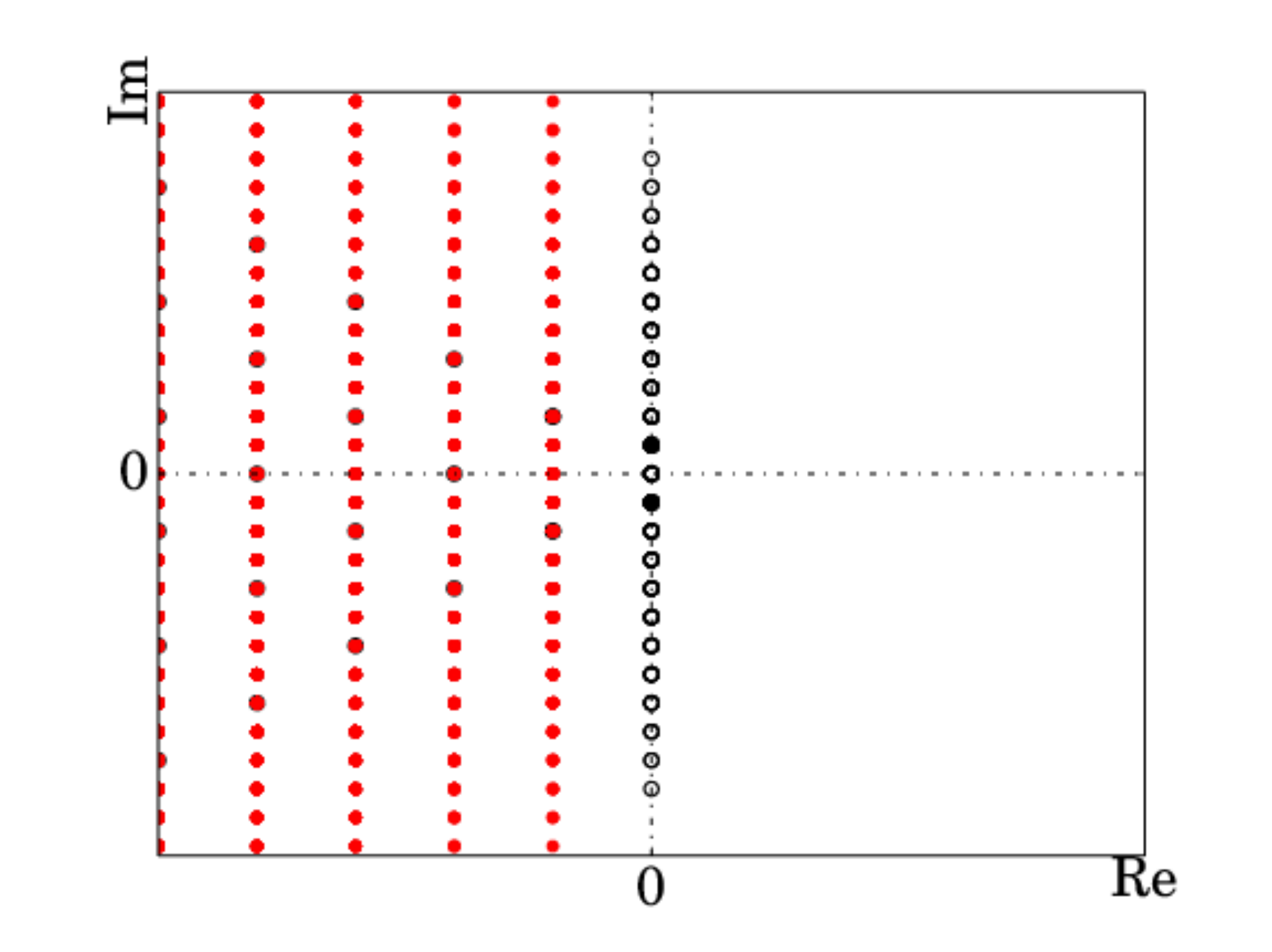} 
\put(110,3){$\sigma$}
\put(5,75){$\omega$}
\end{overpic}
\caption{$-0.2 \pm 0.15j$ and $0.0\pm 0.75j$} \label{fig:resonanced}
\end{subfigure}
\caption{The nonlinear resonance phenomena. Note $0.2 = 2 \times 0.1$ and $5 \times 0.15 = 0.75$}\label{fig:resonance}
\end{figure}

The nonlinear resonant phenomenon is the result of alignment of Koopman spectrums. It can be explained using figure~\ref{fig:resonance}, where all the cross interaction spectrums by two pairs of complex conjugate modes, with spectrums $-0.2\pm 0.75j$ and $0.11\pm 0.71j$, are shown. These two spectrums have no \emph{integer multiple relation} among the real part or the imaginary part. Therefore, the cross interaction spectrums scatter in the complex plane, resulting weak dynamics, and hard to be detected. However, if the two spectrums move towards some integer multiple relation, these scattered Koopman spectrums will aggregate, see figure~\ref{fig:resonanceb} and~\ref{fig:resonancec}, This results in enhanced dynamics. In the case the critical Koopman modes saturate, the above aggregated spectrums will further aggregate, resulting in figure~\ref{fig:resonanced}. This creates even stronger dynamics and can be easily detected. This is the reason that residue criterion picked Floquet modes with integer multiple of base frequency.

It is well-known to disturb some systems, a perturbation of the system's natural frequency is more efficient\citep{rathnasingham1997system,cattafesta1997active}. It is explained similarly since, by the forced linear differential equation~(\ref{eqn:lineardiffforced}), perturbation close to natural frequency can be effectively excited. These excited modes then superimposed on the original modes creating stronger dynamics. Thus efficient perturbations are usually at the same nature frequency or with an integer multiple of it.

\subsubsection{The coherent structure}

Coherent structure in turbulent flow has been discovered for a long time. Leonardo da Vinci first recorded and sketched the repeated patterns in the fluids, and called them the coherent structure~\citep{richter1970notebooks}.~\citet{reynolds1883xxix} in his classic experiments also observed repeated patterns in the turbulent tube with the help of spark light. Moreover, coherent structures are well observed and documented in turbulence research. For instance,~\citet{holmes1996turbulence} successfully identified them using the proper orthogonal decomposition (POD) techniques. However, none of them explained why these coherent structures exist.

In this study, the coherent structure is the result of invariant Koopman modes. In part 1, Koopman modes are found state-independent for many `smooth' systems. They are the reason for coherent structure in fluid systems. The repeated patterns are associated with the complex conjugate Koopman spectrums. Note for fully turbulent flow, the ergodic system defines a unitary operator, which has pure complex spectrums. Therefore, these modes have periodic dynamics.

\begin{figure}
\centering
\begin{subfigure}[b]{0.49\linewidth}
\includegraphics[width=1.0\linewidth, trim={60 50 350 50}, clip]{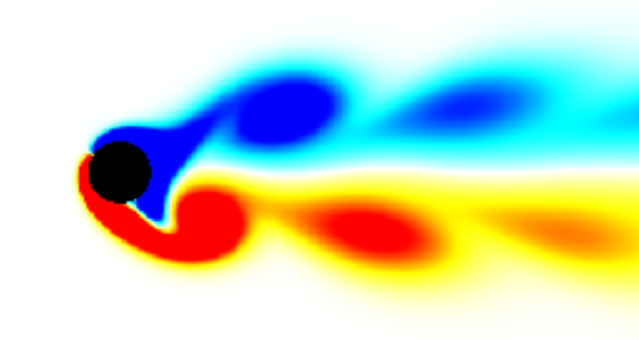}
\caption{DNS} \label{fig:oscDNS}
\end{subfigure}
\begin{subfigure}[b]{0.49\linewidth}
\includegraphics[width=1.0\linewidth, trim={60 50 350 50}, clip]{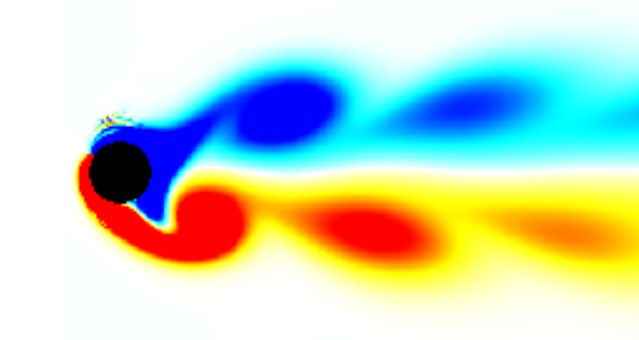}
\caption{Rebuilt with 20 DMD mode} \label{fig:oscROM}
\end{subfigure}\\
\begin{subfigure}[b]{0.7\linewidth}
\centering
\includegraphics[width=0.2\linewidth, trim={100 70 440 70}, clip]{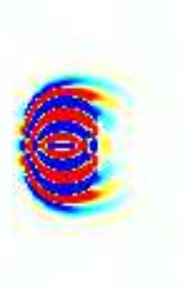}
\includegraphics[width=0.2\linewidth, trim={100 70 440 70}, clip]{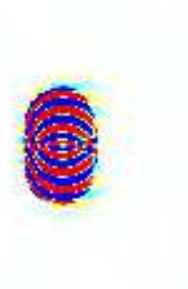}
\includegraphics[width=0.2\linewidth, trim={100 70 440 70}, clip]{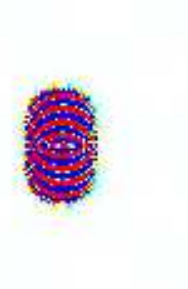}
\caption{Three High-frequency modes around moving boundary.}\label{fig:localDCmodes}
\end{subfigure}
\caption[DMD decomposition captures poor dynamics around the moving solid]{DMD decomposition captures dynamics well far away from the moving solid. (a) DNS simulation of oscillating cylinder at $Re=50$, the oscillation sphere has a dimensionless number of $St=0.2$. (b) DMD modes reconstruction of flow field with 20 DMD modes. (c) Three high-frequency modes appeared around the moving solid.}
\label{fig:movingdmd}
\end{figure}

Even in situation when the spectral continuity requirements are not satisfied, coherent structures may still exist in fluid dynamics. For instance, figure~\ref{fig:oscDNS} shows the numerical simulation of an oscillating cylinder at $Re=50$, with a dimensionless oscillating (up and down) frequency $st=0.2$. Spatial discontinuity at the solid boundary results in unbounded operator, see part 1. However, repeated patterns in the far field is still observed, implying this type of discontinuity may not influence the far wake. It is also observed that 20 DMD modes almost rebuild the DNS results. There is only slight difference around the moving cylinder. Therefore, the repeated patterns in the far wake is also because of the Koopman modes. While around the moving solid boundary, three high-frequency DMD modes are shown in figure~\ref{fig:localDCmodes}. These modes only have significant value around the moving cylinder, reflecting the influence of moving boundary.

\subsubsection{A least-square study for the sub-dynamics of Koopman modes}

The nonlinear dynamics of the Hopf bifurcation was studied by~\citet{stuart1958non}, by considering decomposition
\begin{equation}
\boldsymbol{x}_n = \boldsymbol{x}_0 + \{ \boldsymbol{v}_1 A_1(t) + \boldsymbol{v}_2 A_2(t) + \cdots + c.c. \}
\end{equation}
where $\boldsymbol{x}_0$ is the base flow, $\boldsymbol{v}_i$ are the normal modes of the linearized system at the equilibrium point. $A_i(t)$ are the corresponding coefficients. Let $\boldsymbol{v}_1$ to be the critical normal mode. For the weak nonlinear cases, $A_1(t)$ (or written as $A$) can be approximated by the Stuart-Landau equation~(\ref{eqn:stuartlandau}). For the highly nonlinear system, high order Stuart-Landau equation can be used~\citep[see, chap. 5.3.2]{schmid2012stability}
\begin{equation}
\frac{d A}{d \tau } = \lambda_1 A + \lambda_2 A|A|^2 + \lambda_3 A |A|^4 + \lambda_4 A |A|^6 + \cdots.
\end{equation}

\begin{figure}
\centering
\begin{overpic}[width=0.85\linewidth, left]{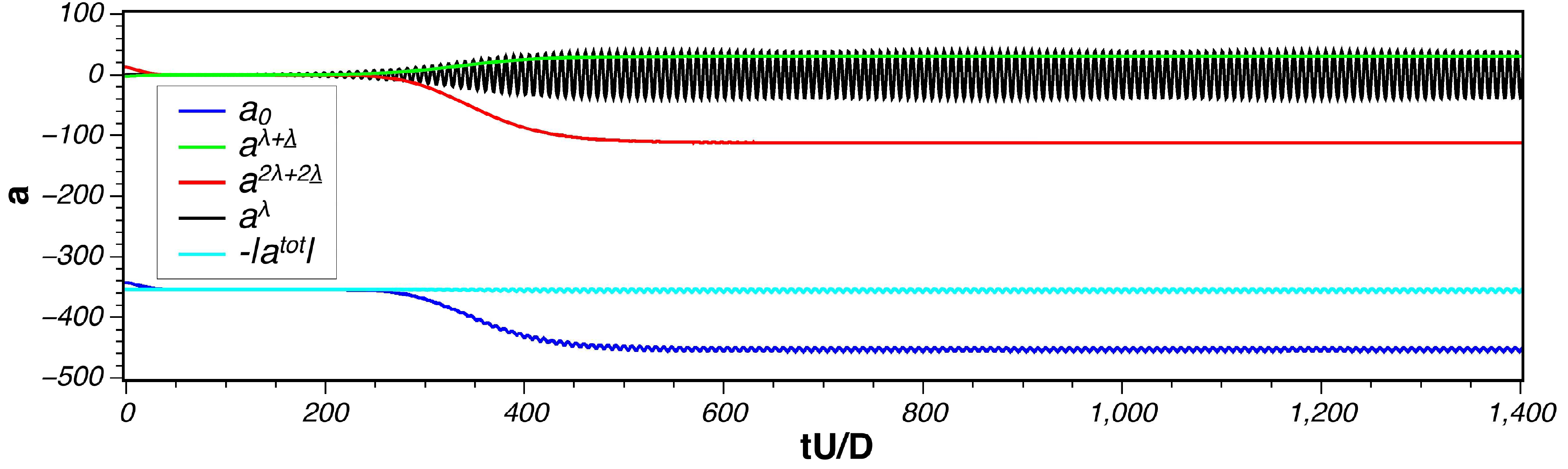}
\put(325.8,0){\includegraphics[height=0.25\textwidth, trim={285 0 40 40}, clip]{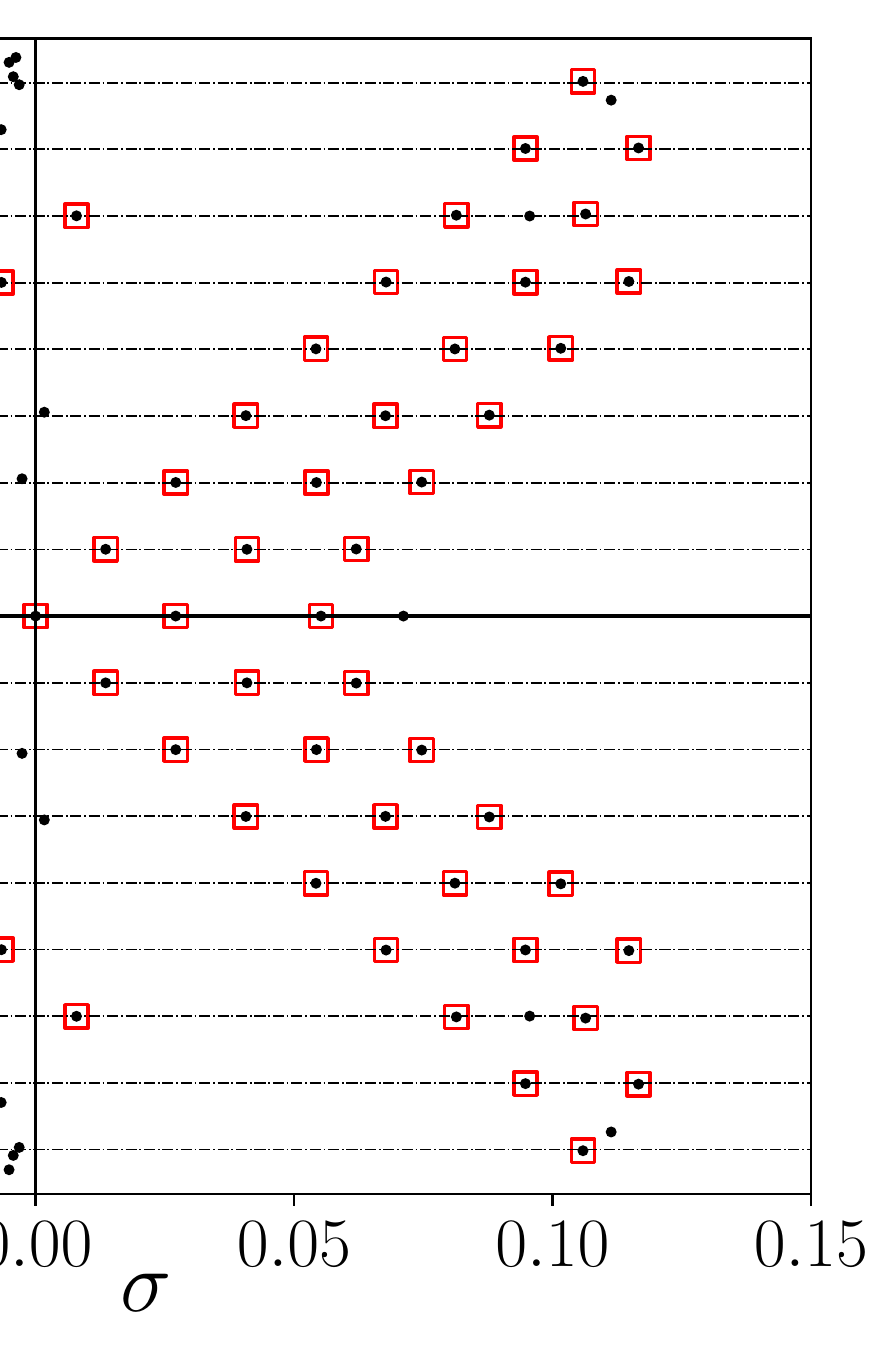}}
\put(309,0){\includegraphics[height=0.25\textwidth, trim={0 0 503 40}, clip]{figure12a}}
\put(324.5,49.5){\color{blue}{O}}
\put(334.5,49.5){\color{green}{O}}
\put(344.5,49.5){\color{red}{O}}
\put(330,54.5){\color{black}{O}}
\end{overpic}
\caption[Time history of different DMD modes]{The time history of modes $\phi_0$, $\phi^{\lambda+\bar{\lambda}}$, $\phi^{2\lambda+2\bar{\lambda}}$, $\phi^{\lambda}$ (only the real part) shown in Fig.~\ref{fig:modulation} and the first-three-term approximation of $\left( \phi_0 \right)^p$ in Eqn.~\ref{eqn:formationmean}. The spectrums are labeled with the same color as the temporal coefficients and are shown in right figure.} \label{fig:temporal4Koopman}
\end{figure}

In our approach, the temporal coefficients of critical Koopman modes are numerically approximated by the least-square solution of the whole bifurcation process using the critical Koopman modes shown in figure~\ref{fig:spectruminit}. Figure~\ref{fig:temporal4Koopman} shows the results of 5 Koopman modes ($\phi_0$, $\phi^{\lambda+\bar{\lambda}}$, $\phi^{2\lambda+2\bar{\lambda}}$, $\phi^{\lambda}$, $\left( \phi_0 \right)^p$). Here are the results.

Firstly, the base flow $\phi_0$ changed magnitude significantly during the nonlinear transition. As seen from figure~\ref{fig:temporal4Koopman}, its magnitude increased during the transition process $200<\frac{tU}{D}<500$. The monotonic modes $\phi^{\lambda + \bar{\lambda}}$ and $\phi^{2\lambda+2\bar{\lambda}}$ increased significantly from trivial. However, the magnitude of the mean flow maintained a constant level as indicated by $|a^{tot}|$
\begin{equation}
|a^{tot}| = \frac{||a_0 \phi_0 + a^{\lambda+\bar{\lambda}} \phi^{\lambda+\bar{\lambda}} + a^{2\lambda+2\bar{\lambda}} \phi^{2\lambda+2\bar{\lambda}} ||}{||\phi_0||}.
\end{equation}
This is because the DMD modes are not orthogonal. It is also noticed even the magnitude does not change, velocity profile has changed, see figure~\ref{fig:modulation}a. The evolution of critical mode $\phi^{\lambda}$ is shown in figure~\ref{fig:temporal4Koopman} by $a^{\lambda} $ (the black line), which follows the prediction of Stuart-Landau equation.

\section{Conclusions}

This article developed a uniform framework for both linear and nonlinear dynamics by transferring dynamics from state space to the infinite-dimensional dual space of the state. By the linearity and completeness of dual space and the local Koopman spectral problem on it, the linear structure and invariant subspaces decomposition is achieved for dynamics, therefore, extending the two important properties of superposition principle and invariant subspaces with exponential dynamics from linear to nonlinear dynamics. Under this framework, the nonlinear dynamics differ with linear ones in two aspects, the locality of exponential dynamics and the infinite-dimensionality, where the later is due to nonlinear interaction and can be conveniently described by the recursive proliferation rule.

Koopman decomposition of linear systems for both linear and nonlinear observables can be explicitly obtained by solving the linear spectral problem. For a nonlinear system, though no explicit formulation is available, the hierarchy structure of Koopman spectrums for nonlinear systems can be employed to understand and compute the Koopman decomposition. The hierarchy is obtained by decomposing the dynamics into base dynamics and nonlinear perturbation. The base dynamics can be studied either analytically or numerically. Furthermore, the perturbation is decomposed into a linear and nonlinear part. The linear part is studied via the linear system. The proliferation rule is applied to obtain spectrums for the nonlinear part. 
GSA technique could be used to compute the Koopman decomposition by assuming small perturbation and disparate time scales. The Koopman decomposition can also be numerically computed by the DMD algorithm. On the other hand, POD provides a numerical algorithm for linear structure via Mercer eigenfunction decomposition, which is the spectral problem of the second-order correlation kernel of dynamics.

Finally, this paper numerically studied the formation of K\'arm\'an vortex of fluids passing a fixed cylinder at $Re=50$, where a Hopf bifurcation process transits from unstable equilibrium to stable limit cycle. During this process, the spectrums of critical Koopman modes asymptotically falls on the imaginary axis. This results in an infinite number of Koopman modes with the same frequencies folding on top of each other, generating the periodic solution. Therefore, the Fourier decomposition of periodic dynamics is explained. The Floquet solution is explained similarly. By folding the cross-interactions of critical Koopman modes with other Koopman modes when the growth rates decrease to 0, Floquet modes, as well as their high-order derived ones are formed. The framework can easily explain some interesting flow phenomena as well. For example, the coherent structure in turbulent flow is because of the invariant Koopman modes. The nonlinear resonance phenomenon is because of superposition of an infinite number of Koopman modes when the frequencies are aligned. From the GSA study, the interaction information transfers directionally. Therefore, energy cascading from low frequencies to high ones, or energy backscattering from high ones to low ones can be explained by the nonlinear interaction between different Koopman modes. It is observed that the wave number of the Koopman modes increases proportionally with frequencies, therefore, the higher frequencies the smaller vortex structures in the Koopman mode. This may explain the fine vortex structures in the complex flows such as turbulence.

\section*{Acknowledgements}

The authors gratefully appreciate the support from the Army Research Lab (ARL) through Micro Autonomous Systems and Technology (MAST) Collaborative Technology Alliance (CTA) under grant number W911NF-08-2-0004. 

\section*{Declaration of interests}

The authors report no conflict of interest.

\bibliographystyle{jfm}
\bibliography{ms}

\end{document}